\documentclass[preprint,3p,onecolumn,numbers,nonatbib]{elsarticle}

\usepackage{amsthm}
\usepackage{amsmath}
\usepackage{verbatim}
\usepackage{float}
\usepackage{bm,bbm}
\usepackage{subcaption}
\usepackage{url}
\usepackage{pifont}
\usepackage{amssymb} 

\usepackage{amsmath}
\usepackage{amsthm}
\usepackage{amssymb}
\usepackage[table,xcdraw]{xcolor}
\usepackage{longtable}
\usepackage{float}
\usepackage{wasysym}
\usepackage[utf8]{inputenc}
\usepackage{hyperref}
\usepackage{oplotsymbl}
\usepackage{graphicx, color}

\usepackage{mathrsfs} 

\usepackage{lipsum}
\usepackage[T1]{fontenc}
\usepackage{comment}
\usepackage{multirow}
\usepackage{longtable}
\usepackage{booktabs}
\usepackage{lscape}
\usepackage{url}
\usepackage{subcaption}
\usepackage{enumitem} 

\usepackage{bbm}
\usepackage{bm}

\usepackage{tcolorbox}

\definecolor{marketblue}{RGB}{102,153,255}

\definecolor{regulatoryred}{RGB}{170,42,63}

\definecolor{opgreen}{RGB}{77,172,101}

\definecolor{techorange}{RGB}{255,165,79}

\definecolor{manipgrey}{RGB}{180,180,180}

\definecolor{interyellow}{RGB}{255,204,51}

\journal{}
\date{}

\begin{document}
\begin{frontmatter}

\title{Mapping Microscopic and Systemic Risks in TradFi and DeFi: a literature review}

\author[1]{Sabrina Aufiero}
\author[1]{Silvia Bartolucci}
\author[1,2]{Fabio Caccioli}
\author[3]{Pierpaolo Vivo}

\affiliation[1]{organization={Department of Computer Science, University College London},
            addressline={66-72 Gower Street}, 
            city={London},
            postcode={WC1E 6BT}, 
            country={UK}}
\affiliation[2]{organization={Systemic Risk Centre, London School of Economics and Political Sciences},
            city={London},
            postcode={WC2A 2AE}, 
            country={UK}}
\affiliation[3]{organization={Department of Mathematics, King’s College London, Strand},
            city={London},
            postcode={WC2R 2LS}, 
            country={UK}}

\begin{abstract}
This work explores the formation and propagation of systemic risks across traditional finance (TradFi) and decentralized finance (DeFi), offering a comparative framework that bridges these two increasingly interconnected ecosystems. We propose a conceptual model for systemic risk formation in TradFi, grounded in well-established mechanisms such as leverage cycles, liquidity crises, and interconnected institutional exposures. Extending this analysis to DeFi, we identify unique structural and technological characteristics -- such as composability, smart contract vulnerabilities, and algorithm-driven mechanisms -- that shape the emergence and transmission of risks within decentralized systems. Through a conceptual mapping, we highlight risks with similar foundations (e.g., trading vulnerabilities, liquidity shocks), while emphasizing how these risks manifest and propagate differently due to the contrasting architectures of TradFi and DeFi.
Furthermore, we introduce the concept of \textit{crosstagion}, a bidirectional process where instability in DeFi can spill over into TradFi, and vice versa. We illustrate how disruptions such as liquidity crises, regulatory actions, or political developments can cascade across these systems, leveraging their growing interdependence. 
By analyzing this mutual dynamics, we highlight the importance of understanding systemic risks not only within TradFi and DeFi individually, but also at their intersection. Our findings contribute to the evolving discourse on risk management in a hybrid financial ecosystem, offering insights for policymakers, regulators, and financial stakeholders navigating this complex landscape.
\end{abstract}

\end{frontmatter}
\newpage
\tableofcontents
\newpage

\section{Introduction} \label{sec:Introduction}

The increasing entanglement between traditional finance (TradFi) and decentralized finance (DeFi) calls for a better understanding of how systemic risks form, propagate, and interact across these two distinct, yet converging, financial ecosystems. While TradFi operates through centralized institutions governed by regulatory oversight, DeFi is characterized by a permissionless, algorithm-driven infrastructure built on distributed technology (e.g., blockchains). Despite their structural differences, both systems are susceptible to similar foundational risks -- such as liquidity shortages, leverage spirals, and correlated exposures -- that may turn into system-wide crises under certain conditions.

This work provides a comprehensive literature review aimed at mapping systemic risk formation in both TradFi and DeFi. We propose a conceptual framework that traces the trajectory from micro-level risks to systemic events, identifying key amplifiers and typical transmission channels along the way. This review draws on established theories from economics and network science already available in the context of TradFi, and extends these insights to the emerging DeFi ecosystem by highlighting its unique technological and organizational make-up.

Importantly, we emphasize that -- while the underlying types of risk are broadly shared between TradFi and DeFi -- their manifestation, amplification, and system-wide impact differ significantly due to the contrasting architectures of these systems. 
Crucially, there are no risks that are entirely exclusive: every risk observed in one system has a counterpart in the other. What differentiates them is not their existence, but rather their likelihood, structural role, and the degree to which they localize or escalate into systemic events. For instance, technological risk is significantly amplified in DeFi due to the reliance on immutable smart contracts and automated execution, whereas in TradFi, it is often mitigated by institutional redundancies and human oversight. 
The differences are therefore contextual: each risk type is embedded within a distinct operational and technological environment, which modulates its behavior, severity, and systemic consequences.

We also introduce and examine the notion of \textit{crosstagion} -- the bidirectional transmission of instability between DeFi and TradFi -- reflecting the increasing interdependence of these domains through stablecoins, tokenized assets, and institutional crypto adoption.

The scope of this paper includes:
\begin{enumerate}[label=(\roman*)]
\item A review of theoretical and data-driven studies on systemic risk in TradFi, incorporating classical equilibrium models, network-based contagion frameworks, and agent-based approaches (in Sec. \ref{sec:TradFi}).
\item An extension of these concepts to DeFi, focusing on smart contract risks, composability, protocol interdependencies, and decentralized governance (in Sec. \ref{sec:DeFi}).
\item A comparative mapping of risk categories across both systems, identifying where they align, diverge, and interact (in Sec. \ref{sec:results}). The main result is shown in Fig. \ref{fig:mapping}, where we propose a two-dimensional risk map that positions key risks observed in TradFi and DeFi according to their speed of contagion and degree of systemic impact. 
\item The introduction of the concept of crosstagion (in Sec. \ref{sec:results2}).
\end{enumerate}

By synthesizing these perspectives, this review aims to offer a structured foundation for future research and policy efforts that seek to understand and mitigate systemic vulnerabilities in both centralized and decentralized finance.

The intersection between TradFi and DeFi is marked by several parallels, particularly in how systemic risks manifest within these domains. Understanding these parallels requires mapping well-known phenomena in TradFi, such as fire sales, asset depreciation, and issues with collateral, onto the DeFi landscape, where they often appear in new and often disguised forms due to the unique characteristics of decentralized systems.

\begin{itemize}

    \item \underline{Fire Sales and Asset Depreciation in DeFi}

    In Traditional Finance, fire sales refer to the rapid liquidation of assets at significantly reduced prices, often occurring during times of financial distress. These sales can trigger cascading effects, further depressing asset prices and exacerbating financial instability. A comparable scenario unfolds in DeFi, especially during market downturns. When the value of crypto assets used as collateral drops below a critical threshold, DeFi protocols automatically trigger liquidation processes to maintain the required collateralization levels. This often leads to a situation where multiple assets are sold off rapidly, mirroring fire sales in TradFi, and causing a sharp decline in asset prices across the market.    
    The automated nature of DeFi liquidations can accelerate these downturns, as smart contracts execute sales without allowing human intervention, potentially leading to even greater volatility and deeper market crashes. This automation highlights a significant risk in DeFi: the lack of discretionary control can exacerbate financial contagion effects, akin to the systemic risks observed in TradFi during periods of financial crisis.

    \item \underline{Collateralization and Overcollateralization}

    Collateralization is a fundamental concept in both TradFi and DeFi. In TradFi, assets are often pledged to secure loans, with the understanding that if the borrower defaults, the collateral can be liquidated to cover the loss. DeFi follows a similar principle but often requires overcollateralization due to the inherent volatility of crypto assets. For example, platforms like MakerDAO require users to lock up more value in collateral than the value of the loan they receive to mitigate the risk of price swings \cite{maker2017whitepaper}. 
    However, this overcollateralization introduces its own set of problems. It locks up a significant amount of liquidity, which can strain the DeFi ecosystem during periods of high demand for liquidity. Additionally, if the value of the collateral falls too quickly, even over-collateralized loans can become under-collateralized, leading to forced liquidations. This mechanism can spiral into a situation where falling asset prices lead to more liquidations, driving prices further down -- a process that closely mirrors the vicious cycles seen in TradFi during crises.

    \item \underline{Systemic Risk and DeFi’s Impact on TradFi}

    The systemic risk in DeFi is amplified by the interconnectedness of various protocols and the reliance on volatile collateral. Unlike TradFi, where central banks and financial institutions can intervene during crises, DeFi operates without centralized oversight, which means that crises can unfold rapidly and with limited mechanisms for containment. As DeFi continues to grow and intertwine with TradFi -- such as through the adoption of stablecoins and other crypto assets by traditional financial institutions -- the potential for systemic risk to spill over into the broader financial system increases.
    This interconnectedness raises concerns about \textit{crosstagion}, where instability in DeFi markets could impact traditional markets, and vice versa. For example, the failure of a major DeFi protocol or a sharp decline in the value of a widely held stablecoin could trigger a broader financial crisis, particularly if traditional financial institutions are exposed to these assets.

    \item \underline{Regulatory and Risk Management Considerations}

    The lack of regulatory oversight in DeFi presents challenges that are distinct from those in TradFi. TradFi benefits from well-established regulatory frameworks and the presence of central banks, which act as lenders of last resort to prevent cascading defaults across the financial system. These centralized interventions are designed to stabilize markets, ensuring that systemic risks do not spiral out of control. However, this safety net in TradFi can also lead to some form of ``moral hazard'', as market participants may engage in riskier practices under the assumption that central authorities will eventually intervene to stabilize the system in the event of a crisis. In contrast, DeFi's decentralized nature makes it difficult to implement such safety nets, leaving the system vulnerable to the kind of cascading defaults that are rarely observed in TradFi due to regulatory interventions and built-in safeguards.
    Moreover, the dynamic and highly interconnected interactions between various DeFi components can amplify risks beyond the scope of individual protocols, creating systemic challenges that are difficult to manage without a centralized authority. 
\end{itemize}   

DeFi presents significant innovations and opportunities for financial systems, but it also inherits and amplifies certain risks common in TradFi. As DeFi continues to evolve, understanding these risks and developing appropriate regulatory and risk-management frameworks will be crucial for mitigating their impact on both DeFi and TradFi ecosystems.

\section{Traditional Finance}\label{sec:TradFi}

In the traditional economic perspective, equilibrium stands as the central guiding principle, portraying markets as self-correcting and inherently stable, even when confronted with external shocks. The traditional economic equilibrium approach is a foundational framework, and analyzes how supply and demand interact within markets to reach a state of balance, or equilibrium. In this state, market forces of supply and demand are balanced, and in the absence of external influences, the values of economic variables will not change. This approach assumes that markets are efficient, agents are rational, and all available information is reflected in prices, leading to a natural gravitation toward stability even after experiencing shocks.
A central concept in this framework is the General Equilibrium Theory, which was formalized by economists Arrow and Debreu (1954). Their seminal work demonstrated mathematically that under certain conditions, all markets in an economy will simultaneously reach equilibrium \cite{arrow1954existence}. 
A key concept is the Efficient Market Hypothesis (EMH), introduced by Fama (1970) \cite{fama1970efficient}. The EMH posits that asset prices fully reflect all available information, which implies that it is impossible to consistently achieve returns that exceed average market returns on a risk-adjusted basis. This hypothesis supports the notion that financial markets are inherently stable and self-correcting, as prices adjust to new information instantaneously.
The theory of Rational Expectations, developed by Muth (1961) \cite{muth1961rational}, further underpins the traditional equilibrium approach by suggesting that individuals make forecasting decisions based on all available information and past experiences, leading to outcomes that, on average, align with the predictions of economic models. This implies that systemic risks are anticipated and mitigated through rational decision-making processes of market participants.

However, this approach has been criticized for oversimplifying the complexities of financial systems. It often overlooks the intricate network of interdependencies and the potential for feedback loops that can amplify shocks, leading to systemic crises.

The focus on systemic risk in financial markets has increased significantly since the 2007 US mortgage crisis. This crisis highlighted the weaknesses in the financial system, particularly when the collapse of Lehman Brothers in September 2008 exposed these vulnerabilities, which were further exacerbated by the subsequent Eurozone debt crisis. Such events lead to a chain reaction that undermines the essential confidence needed for the financial system to function effectively. These crises are particularly severe given the increasing size of the financial sector as a proportion of national economies, along with its growing global integration \cite{silva2017analysis}. In addition, advances in technology have accelerated trading speeds, contributing to this volatility. As noted by Grilli et al. (2014) \cite{grilli2015markets}, the shift of resources from productive sectors to the financial sector over the past decades has heightened financial instability. The financialization of the economy has, according to their research, played a significant role in making financial crises more frequent and intense.

\textit{Systemic risk} is an elusive concept and difficult to pin down with a single definition. Many definitions of systemic risks have been introduced over the years, and they can be defined in various ways depending on the scope and focus \cite{rochet1996interbank, allen2000financial}. 
Oort (1990) \cite{oort1990banks} identifies several factors that make banking systems vulnerable: (i) the risk of widespread financial crises due to interconnected bank failures, (ii) the systemic risks associated with new financial products, and (iii) the effects of external shocks such as debt crises, market rate shifts, and deregulation.
De-Bandt and Hartmann (2000) \cite{de2000systemic} suggest that any definition should encompass widespread disruptions in the banking and financial sectors. According to Summer (2003) \cite{summer2003banking}, there is no universally agreed-upon definition of systemic financial risk. Lehar (2005) \cite{lehar2005measuring} describes systemic risk as the possibility of multiple financial institutions going bankrupt simultaneously. The European Central Bank (ECB, 2009) \cite{ecb2009fsr} defines systemic risk as the possibility of one institution’s failure prompting the collapse of other participants due to liquidity and credit constraints. This jeopardizes the overall stability of the financial system. Adrian and Brunnermeier (2010) \cite{95a16e2e-027f-345c-a203-76b01010c3b9} take a more focused approach, identifying systemic risk as the widespread malfunction of institutions, leading to disruptions in credit and capital supply across the economy. Billio et al. (2012) \cite{billio2012econometric} add that systemic risk often manifests in abrupt shifts in regime, where economies oscillate between periods of low volatility during growth and high volatility during downturns. Patro et al. (2013) \cite{patro2013simple} offer a more modern perspective, describing systemic risk as a scenario in which the entire financial system faces simultaneous stress, leading to credit and liquidity crises. They argue that systemic risk not only affects financial markets but also has a profound impact on the economy by reducing capital supply and intensifying capital losses. Additionally, the authors highlight systemic risk as the likelihood of a significant financial decline triggered by large, pervasive events.
De Bandt et al. (2012) \cite{bandt2012systemic} provides a straight-forward description of systemic events. Narrow systemic events affect one or a few institutions,  leading to adverse effects on other banks, while broad systemic events occur when correlated shocks hit many institutions simultaneously, having highly correlated exposures. Systemic risk is also classified as strong if it leads to the failure of previously solvent institutions, and weak if it does not cause failures. Contagion arises from strong, narrow events, and systemic risk is defined as the risk of experiencing a strong systemic event. Thus, according to the De Bandt taxonomy, the measurement of systemic risk is an attempt to quantify the occurrence of contagion or a widespread macroeconomic shock resulting in the failure of several financial institutions. 
Benoit et al. (2017)  \cite{benoit2017risks} introduce a definition (and an approach) with a high-level abstraction, allowing them to encompass a wide range of theoretical models and perspectives on systemic risk. The authors define it as the risk that many market participants are simultaneously affected by severe losses, which then spread through the system. 

In this context, different models are distinguished by their specific definitions of the proportion of exposure that each financial institution has to a systematic risk factor. Such factor is then described as a key determinant in the overall systemic risk of the financial system. The available models vary in how they conceptualize the risk factor, depending on the features of the risk being considered, such as market-wide shocks or institution-specific vulnerabilities, and this variance is crucial in understanding how systemic events are modeled across different frameworks.

While older models attempt to understand systemic risk with a traditional economic equilibrium approach, subsequent works on empirical networks and agent-based models challenge the notion that financial markets gravitate towards a steady-state of stability. These models explore how financial systems evolve and respond to shocks, emphasizing the complexity and potential instability inherent in modern financial networks. 

A network-based approach offers a clearer picture of risk propagation and its potential consequences, allowing for more accurate assessments of systemic vulnerabilities \cite{colander2009financial, bardoscia2017pathways, haldane2011systemic}.

The expansion of the complex networks literature and analytical tools in recent decades has greatly enhanced our ability to study systemic risk, by revealing how various types of financial vulnerabilities emerge from—and propagate through—the structure of interconnections among institutions \cite{jackson2021systemic}.
The financial system as a whole consists of financial actors (referred to as banks, though the term broadly applies to any financial entity whose actions impact others: institutions, pension funds, companies or households), markets (e.g. the stock or the bond market), contracts (e.g. the ownership of a stock, or a loan between two banks, or from a bank to a firm, or from a bank to a household), and regulatory bodies (e.g. financial supervisors and central banks). A network is thus a natural and general description of the financial system at different scales. Nodes represent financial institutions and edges stand for a financial relationship, such as interbank lending or asset exposure.
The most direct connections occur via financial contracts, such as lending, borrowing, joint investments, and asset transactions. These interactions allow institutions to manage liquidity and risk -- see Allen and Gale (2000) \cite{gale2007financial}; Eisenberg and Noe (2001) \cite{eisenberg2001systemic}; Elliott et al. (2014) \cite{elliott2014financial}; Corsi et al. (2018) \cite{corsi2018measuring}; Forer et al. (2025) \cite{forer2025financial}. Additionally, even in the absence of direct transactions, institutions sharing similar exposures develop correlations in their portfolio performances 
-- see Acharya and Yorulmazer (2007) \cite{acharya2007too}; Allen et al., (2012) \cite{allen2012asset}; Diebold and Yilmaz (2014) \cite{diebold2014network}. 
The network is characterized by a matrix \( \textbf{G} = \{ g_{ij} \} \), where \( g_{ij} \) represents the strength of the connection between bank \(i\) and bank \(j\). A connection, \( g_{ij} \), can have multiple components, reflecting various types of financial contracts that exist between the institutions.

The different types of contracts can interact, as shown by Bardoscia et al. (2019) \cite{bardoscia2019multiplex} for UK bank data.
A vector \(\textbf{V} = \{ V_i \}\) captures the values associated with each institution \(i\), taking into account all its assets and liabilities, including any defaults and potential bankruptcy costs.
Since banks are interlinked, their values are interdependent, meaning that \( V_i \) is influenced not only by the financial status of \(i\) but also by that of other banks connected to it \cite{jackson2021systemic}. The value of a bank is therefore expressed as a function of the values of all other banks in the system, denoted as:
\begin{equation}
    V_i = F_i(\textbf{V} \mid \textbf{G})  \ \text{.}
\end{equation}
Banks' values are determined by a system of \(n\) equations with \(n\) unknowns, written as:
\begin{equation}\label{eq:1}
\textbf{V} = \textbf{F}(\textbf{V} \mid \textbf{G}) \text{.}
\end{equation}

Under certain conditions\footnote{When the function \(\textbf{F}(\cdot \mid \textbf{G})\) is non-decreasing and bounded in \textbf{V}, Tarski’s fixed point theorem applies, and there exists an equilibrium.} the system admits an equilibrium state. 

This equilibrium is defined as a self‑consistent configuration of variables that simultaneously satisfies \eqref{eq:1}, ensuring that every balance sheet identity and constraint is met. Importantly, the notion of equilibrium here denotes internal coherence among stocks and flows rather than a fixed point of strategic best responses or the outcome of a dynamic adjustment process.
These stationary points may include both maximum and minimum values, which correspond to the highest and lowest possible stable outcomes for all banks.
Network models typically use simple behavioral assumptions: these models reflect global interaction, where each actor’s behavior depends on others, as described in Bargigli and Tedeschi (2014) \cite{bargigli2014interaction}. 
In a financial network the relationships change over time: as a result, a complete description of the financial system requires a temporal multiplex network\footnote{A multiplex network is a collection of networks (also called layers) with the same set of nodes, but with different links.}, where each layer is associated with a specific type of relationship (e.g. interbank loans of a given maturity). However, in many cases, one focuses on individual processes, the timescale of which is much shorter than the timescale over which those relationships change. This simplification makes it possible to represent the financial system as a single-layer static network \cite{bardoscia2021physics}. 
Interbank networks display several notable features and stylized facts:
\begin{itemize}
    \item Heavy-tailed degree distributions: in the context of interbank networks, a node’s degree corresponds to the number of counterparties it has. These networks are characterized by a small number of highly connected hubs -- often systemically important institutions -- and a large number of nodes with low connectivity.
    \item High clustering: clustering in network theory refers to the interconnectedness of a node’s neighbors, often quantified using the clustering coefficient. Interbank networks exhibit a high level of clustering, meaning that banks tend to have densely connected counterparts, forming numerous closed loops (triangles). This distinguishes interbank networks from random networks, where clustering is typically lower.
    \item Negative assortativity: assortativity indicates the tendency of nodes to connect with others of similar degree. Unlike social networks, where high-degree nodes are often connected to other high-degree nodes (positive assortativity), interbank networks exhibit negative assortativity. In these systems, highly connected nodes are more likely to link with less-connected nodes, a pattern similar to that observed in technological networks like the internet.
    \item Core-periphery structure: these networks typically feature a central core of densely interconnected nodes, representing systemically important institutions, surrounded by a less connected periphery of smaller institutions. This structural organization highlights the critical role of the core in maintaining network stability.
\end{itemize}

Although financial networks models differ in the details, they all share a common trait: financial interconnections inherently create or enhance systemic risks. A formal model of financial networks is thus useful to measure, predict, and trace the sources of systemic risk.
The models used are essentially static, where a network is given, a shock is applied, and the system evolves according to a deterministic dynamic. These models are highly useful for understanding how the structural constraints of the network and simple behavioral rules of financial institutions can lead to the endogenous amplification of shocks. However, to fully understand how a system can endogenously transition from a normal state to a crisis regime, we need dynamic models in which investments and connections evolve over time. This is where agent-based modeling becomes particularly valuable. In agent-based models, individual financial institutions (agents) interact with one another based on a set of behavioral rules, allowing for the emergence of complex system-wide phenomena \cite{axtell2025agent, thurner2011systemic, samanidou2007agent}. By simulating the evolution of investments, network structures, and decision-making processes, agent-based models can offer a more realistic representation of how systemic risk develops and propagates over time.
Agent-based models go further by incorporating local feedback mechanisms and adaptive behavior, allowing the interactions within the network to directly influence the actions of economic agents. 
The absence of comprehensive data means that fully understanding the process of financial network formation and evolution may remain elusive for now. Generally, there is insufficient data on the dynamic linkages across firm balance sheets which has led to a continued reliance on theoretical and simulation-based methods. 

Today, the study of systemic risk intersects disciplines such as banking, macroeconomics, econometrics, and network theory, offering many insights into the phenomena that threaten financial stability on a global scale.

\subsection*{The process of systemic risk formation}

\begin{figure}
    \centering
    \includegraphics[width=\linewidth]{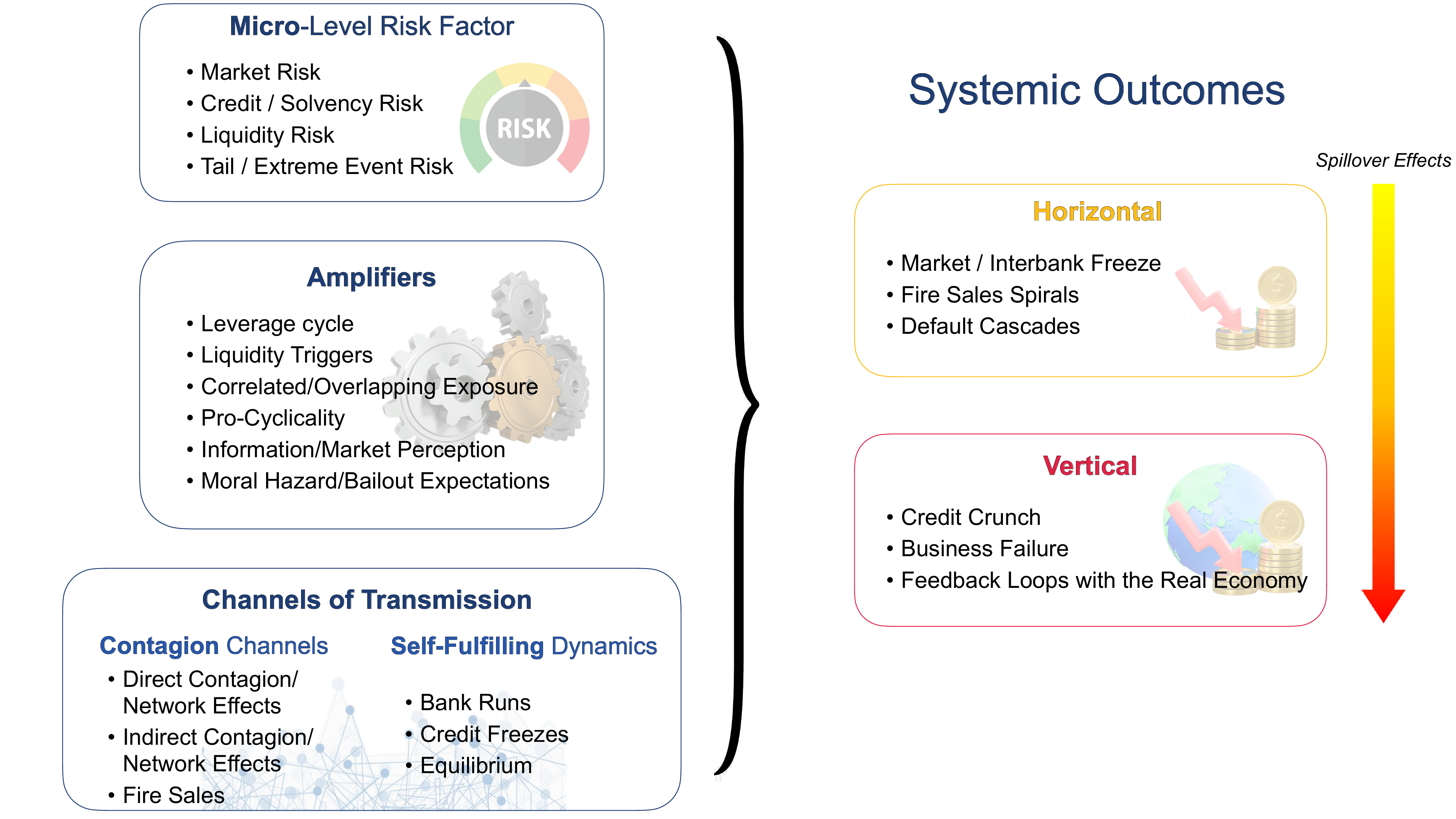}
    \caption{A schematic representation of the TradFi systemic risks formation process, illustrating the progression from micro-level risks to amplifiers, transmission channels, and systemic outcomes. The outcomes are categorized into horizontal systemic risks, which remain confined within the financial sector, and vertical systemic risks, which spill over into the real economy.}
    \label{fig:schema tradfi}
\end{figure}

We now propose a classification that examines the process of systemic risk formation, starting from individual (or microscopic) risks, progressing through amplifiers and transmission channels, and culminating in systemic risks.
Individual risks primarily affect specific institutions or portfolios. While these risks may initially appear contained, they have the potential to escalate into systemic issues when they interact with amplifiers. Amplifiers are factors that magnify the impact of adverse events, transforming localized problems into broader crises. These mechanisms boost the effects of individual shocks, enabling them to propagate across the financial system. Transmission channels are the mechanisms through which individual shocks propagate through the financial system. Inspired by Jackson et al. (2021) \cite{jackson2021systemic}, we classify these channels into two main categories: contagion channels -- which create externalities among financial institutions, allowing shocks to spread directly or indirectly -- and self-fulfilling prophecies and feedback effects, where individual fears or behaviors collectively trigger systemic outcomes. 
Systemic risk represents the final stage of this process, where localized disturbances evolve into widespread financial instability. Building on the framework of De Bandt et al. (2012) \cite{bandt2012systemic}, we identify two distinct dimensions of systemic risk. The first, horizontal systemic risk, remains contained within the financial sector; in contrast, vertical systemic risk spills over into the real economy, causing broader disruptions.
This classification offers a structured framework for analyzing how localized risks can escalate into systemic crises: the schematic representation is provided in Fig. \ref{fig:schema tradfi}.

\textbf{Microscopic risks} (top-left corner of Fig. \ref{fig:schema tradfi}), or individual risks, affect individual institutions or specific portfolios, often appearing confined in scope. Within this category fall several distinct types of risk, each examined in detail in Sec. \ref{subsec:microscopic-tradfi}. 
Market risk arises from price fluctuations and volatility across various assets, including equities, bonds, and foreign exchange markets. 
Credit or solvency risk, on the other hand, reflects the potential for borrower defaults, credit downgrades, or bankruptcies, which can erode the capital base of financial institutions. 
Liquidity risk can manifest in two forms: funding liquidity, where institutions are unable to roll over short-term debt or secure new funding, and market liquidity, where assets cannot be sold quickly without incurring significant price discounts. 
Tail risks, or extreme event risks, stem from low-probability, high-impact shocks, such as market crashes or sudden liquidity freezes. 
Additionally, operational risk, while less central to traditional systemic contagion, can sometimes play a significant role, particularly in cases of cyber-attacks or major internal process failures. Together, these risks form the foundational vulnerabilities that, if amplified, may escalate into broader systemic crises.

\textbf{Amplifiers} (mid-left in Fig. \ref{fig:schema tradfi}) are mechanisms that magnify the impact of events, transforming localized issues into broader systemic crises. 
Within this group are multiple amplification mechanisms, discussed in detail in Sec. \ref{subsec:amplifier-tradfi}. One such amplifier is leverage, which amplifies both returns and losses. When losses occur, margin calls and forced asset sales can trigger fire-sale spirals, exacerbating financial instability. 
Liquidity triggers, such as margin calls, can force asset sales at depressed prices, further amplifying financial vulnerabilities.
Correlated or overlapping exposures further intensify risks, as many institutions often hold the same or similarly risky assets, leading to widespread losses if asset prices decline.
Pro-cyclicality is another critical amplifier, as financial institutions tend to lend and invest aggressively during economic booms but retrench during busts, reinforcing cyclical fluctuations.

Information and market perceptions also play a significant role. Rumors or sudden loss of confidence can drive investors and depositors to withdraw funds en masse, intensifying financial distress. 
Finally, moral hazard, fueled by expectations of government bailouts, can encourage excessive risk-taking, creating larger crises when such support is absent or inadequate. 
Together, these amplifiers transform initial shocks into systemic disruptions.

\textbf{Channels of transmission} (bottom-left corner of Fig. \ref{fig:schema tradfi}) are the mechanisms through which individual shocks propagate throughout the financial system. 
We subdivide these channels into contagion channels, which transmit shocks from one balance sheet to another, and self‑fulfilling dynamics, where feedback loops amplify the stress; both classes are discussed in detail in Sec. \ref{subsec:channel-tradfi}.
Contagion channels create externalities among institutions. 
For example, direct counterparty exposure occurs when a default or severe distress at one institution directly impacts its counterparties through mechanisms such as derivatives or loans. 
Indirect contagion arises when multiple institutions hold the same asset; in such cases, a fire sale or a price shock by one institution simultaneously affects all holders. 
Similarly, fire sales -- rapid asset sell-offs at depressed prices -- can trigger further losses for other institutions with comparable positions. 
Network effects further amplify these dynamics, as interlocking claims in areas like securities lending, or payment systems propagate distress from one node to another.
In addition to contagion channels, self-fulfilling dynamics play a significant role in transmission. 
Bank runs, where depositors or short-term creditors withdraw funds en masse due to fear, create liquidity shortages that validate and intensify the panic. 
Funding runs or credit freezes occur when wholesale lenders refuse to roll over financing, forcing institutions to deleverage rapidly. 
Equilibrium multiplicities further complicate these dynamics, as the financial system can shift between a stable ``good'' equilibrium -- characterized by confidence and accessible funding -- and a ``bad'' equilibrium marked by panic and a lack of funding. These transmission channels collectively enable individual shocks to escalate into broader systemic disruptions.

\textbf{Systemic outcomes} (right panels in Fig. \ref{fig:schema tradfi}) can be categorized as either horizontal or vertical, depending on the extent of their impact. 
Both types of outcomes are examined in depth in Sec. \ref{subsec:systemic outcomes-tradfi}. Horizontal systemic events remain confined within the financial sector. 
For instance, default cascades occur when the failure of one institution undermines its counterparties, potentially triggering a domino-like chain of further defaults. 
Similarly, market freezes or interbank freezes arise when banks and other intermediaries refuse to lend or trade, effectively paralyzing liquidity and price discovery within the financial system. 
Fire-sale spirals further exacerbate horizontal crises, as forced liquidations and plummeting asset prices create widespread losses across multiple institutions. While the damage in these scenarios is significant, it remains largely contained within the boundaries of the financial sector, affecting banks, insurers, etc.
In contrast, vertical systemic events involve spillovers into the real economy, leading to broader disruptions. 
Credit crunches are a key example, as banks reduce lending to households and firms, stifling investment, consumption, and economic growth. This, in turn, can result in widespread business failures and rising unemployment, as firms face financing constraints and the broader economy contracts. Moreover, feedback loops between the financial sector and the real economy amplify these effects; declining corporate profits and increasing defaults feed back into further bank losses, intensifying the crisis. 
This two-dimensional perspective on systemic crises (horizontal vs. vertical) and (contagion vs. self-fulfilling) captures \textit{how} and \textit{why} financial sector turmoil can escalate into full-blown economic disruptions.
Of course, several mechanisms can be combined.

\subsection{Microscopic Level Risks}\label{subsec:microscopic-tradfi}
From the perspective of an individual bank, several interrelated threats can arise long before they manifest as systemic crises.  
For instance, market risk emerges when fluctuations in asset prices threaten the value of a bank’s holdings and reduce its capital buffer.  
Credit (or solvency) risk stems from the possibility that borrowers will default on loans, directly jeopardizing the bank’s balance sheet and ability to meet obligations. 

\begin{itemize}    
\item 
   \textbf{Market risk} is the risk of losses in positions arising from movements in market variables like prices and volatility \cite{cpmi2003glossary}. Market risk can manifest in various forms, such as: equity risk, which arises from fluctuations in stock prices or their implied volatility; interest rate risk, driven by changes in interest rates or their volatility; currency risk, stemming from shifts in foreign exchange rates or their volatility (e.g. EUR/USD, EUR/GBP, etc.); commodity risk, associated with price movements or volatility in commodities (e.g. corn, crude oil).
   
\item 
    \textbf{Credit risk} is the risk that a borrower will not be able to repay the debt to a bank. In this sense, the debtor may be the receiver of a bank loan, the issuer of a debt security, or even another bank borrowing in the interbank market \cite{almarzoqi2015does}. Losses can occur in various scenarios, such as when a borrower defaults on a loan, a company fails to meet its debt obligations, a business or individual delays payments on trade invoices, or an insolvent entity -- such as a bank, insurance company, or bond issuer -- is unable to fulfill its financial commitments, often exacerbated by bankruptcy protections granted by governments. On the other hand, the \textbf{solvency risk} defines the risk that a bank cannot meet maturing obligations because it has a negative net worth; that is, the value of its assets is smaller than the amount of its liabilities. This may happen when a bank suffers some losses from its assets because of the write-offs on securities, loans, or other bank activities, but then the capital base of the institution is not sufficient to cover those losses. In such a case, the bank unable to meet its obligations defaults and loses its franchise value. In order to avoid such risk, banks need to keep an adequate buffer of capital, so that in case of losses, the bank can reduce capital accordingly and remain solvent \cite{almarzoqi2015does}.

\item  
    \textbf{Liquidity risk} is the risk that a bank will not be able to meet its short-term payment obligations, either because it is not able to accrue enough funding on the wholesale market (funding liquidity), or because its securities or investments cannot be sold quickly enough to get the right market price (market liquidity) \cite{almarzoqi2015does}.
    Liquidity risk often materializes when each bank over-invests in illiquid assets. As Bhattacharya and Gale (1987) \cite{bhattacharya1985preference} explain, banks tend to over-invest in illiquid assets, leaving the entire banking sector vulnerable to liquidity shortages. This occurs because, while it would be optimal for all banks to hold liquid assets, individual banks often rely on others to do so, leading to underinvestment in liquidity overall. When a liquidity shortage is expected, banks without liquid assets borrow from those that do have liquidity, typically through the interbank market. The lenders then sell some of their liquid assets to provide the necessary funds. This creates an issue, where banks are incentivized to hold illiquid assets, assuming that they can always rely on other institutions for liquidity during times of stress. As a result, in equilibrium, banks tend to over-invest in illiquid assets, which makes them all more susceptible when a liquidity shock occurs.

\item
    \textbf{Tail risk} In its canonical usage, tail risk describes the chance that asset returns deviate far beyond the range predicted by a normal distribution (typically, movements exceeding three standard deviations from the mean). Classical portfolio theory (under the Gaussian assumption) assigns only $0.3\%$ probability to such events, implying that they should occur roughly three times in a thousand trading days \cite{kelly2014tail}. Empirical evidence, however, shows that return distributions are skewed and exhibit fat tails: large negative moves happen far more frequently than the Gaussian benchmark suggests, producing black swan losses that conventional risk models systematically understate \cite{bhansali2008tail}.  
    An influential paper on tail risk by Thurner et al. (2012) \cite{thurner2012leverage} develops a model to explore how leveraged asset purchases with margin calls can lead to extreme market fluctuations. The authors show that leveraging amplifies downward price movements during market downturns, as margin calls force funds to sell into falling markets, triggering clustered volatility and power law tails in asset returns. They find that these dynamics arise from the interplay of leverage, wealth dynamics, and market behavior, rather than irrational trends. Moreover, they highlight that common risk control policies, such as basing leverage limits on volatility, can unintentionally exacerbate market instability by increasing leverage during low-volatility periods and triggering sharper contractions during high-volatility periods.
    
    The term \textit{tail risk} is also used more expansively to denote the probability of rare, catastrophic events, and shocks whose origins may lie outside purely financial mechanisms. In this sense tail risk encompasses operational‑type disasters such as pandemics or abrupt geopolitical crises. McRandal and Rozanov (2012) \cite{mcrandal2012primer} highlight at least seven significant tail events between the late 1980s and early 2010s, illustrating how infrequent but high impact episodes can trigger systemic stress.

    Thus, whether viewed through abnormal price moves or catastrophes, tail risk captures the vulnerability of financial systems to extreme shocks.
\end{itemize}

\subsection{Amplifiers}\label{subsec:amplifier-tradfi}
A number of mechanisms have been proposed to explain why relatively small shocks can lead to large aggregate impacts, in particular when they simultaneously affect many institutions.

\begin{itemize}
\item 
    A type of risk-amplifier involves banks increasing their risk exposures in a correlated manner. One well-explored mechanism for this behavior in macro-finance literature is the \textbf{leverage} cycle. The concept emphasizes that, due to limited collateralization options, borrowers -- whether households, firms, or financial institutions -- can borrow more when asset values are high and less during downturns, exacerbating the business cycle. Influential models explaining this process include works by Bernanke and Gertler (1989) \cite{bernanke1986agency}, Kiyotaki and Moore (1997) \cite{kiyotaki1997credit}, and Holmstrom and Tirole (1997) \cite{holmstrom1997financial}.
    Although the economic dynamics differ, the formation of bubbles is closely related to leverage cycles, as financial institutions often take large positions in the same assets, raising the risk of a future value decline. Bubbles can be linked to leverage cycles: leveraged investors have strong incentives to bid up asset prices \cite{allen2000bubbles}. If losses grow too large, the inability to repay debts leads to bubbles bursting, which contributes to crises. Allen and Carletti (2013) \cite{allen2013systemic} extend this model to explore real estate bubbles, which were a significant factor in the 2008 financial crisis. 
    Adrian and Shin (2009) \cite{adrian2010liquidity} investigate the relationship between leverage and liquidity, particularly in environments dominated by mark-to-market accounting. They find that as asset prices adjust, these changes are immediately reflected in institutions' net worth, prompting them to modify the size of their financial positions.
    Caccioli et al. (2009) \cite{caccioli2009eroding} highlight how the unchecked expansion of financial instruments can lead to significant market fluctuations and instability. These instruments often create a scenario where trading volumes rapidly increase, saturating investor demand. While this makes the market appear arbitrage-free, efficient, and seemingly complete, it compromises the system’s overall stability. Such dynamics are typical of leverage cycles and financial bubbles, where short-term efficiency conceals the long-term risks of collapse. 
    Brock at al. (2009) \cite{brock2009more} found that more hedging instruments may destabilize markets when traders have heterogeneous expectations and adapt their behavior according to performance-based reinforcement learning. 
    
\item 
    \textbf{Liquidity crises} are an ideal example of how amplification occurs in financial markets.
    Liquidity can refer to market liquidity, which describes the ease with which an asset can be converted into cash; funding liquidity, which refers to the ability of borrowers to secure external financing; or accounting liquidity, which assesses the strength of an institution’s balance sheet based on its cash-like assets \cite{amihud2012market}. Allen and Gale (2004) \cite{allen2004financial} suggest that this cycle is driven by market incompleteness: in the absence of contingent securities, liquidity must be found by selling assets, leading to further declines in prices. Some economists also define a liquid market as one that can accommodate ``liquidity trades'' -- the sale of securities by investors to address sudden cash needs -- without significant price fluctuations. A liquidity shortage may result from factors such as asset prices falling below their fundamental long-term values, worsening external financing conditions, a decline in market participants, or difficulties in trading assets \cite{krishnamurthy2010amplification}. These factors often reinforce one another during a liquidity crisis. Participants seeking cash may struggle to find buyers for their assets, either due to reduced market activity or a general shortage of cash among market participants. As a result, asset holders may be compelled to sell at prices below their fundamental values. Borrowers, in turn, face increased loan costs, stricter collateral requirements, and limited access to unsecured debt. Furthermore, the interbank lending market typically experiences significant disruptions during such crises. A liquidity crisis can be exacerbated by mechanisms that link asset market liquidity and funding liquidity, amplifying the effects of a small economic shock. This feedback loop can lead to a severe liquidity shortage, ultimately triggering a full-scale crisis \cite{krishnamurthy2010amplification}.
    Coordination failures and bank runs exemplify how liquidity shortages can rapidly escalate small shocks into crises. Banks and financial institutions are inherently vulnerable to liquidity pressures due to coordination challenges with their creditors. The foundational works of Bryant (1980) \cite{bryant1980model} and Diamond and Dybvig (1983) \cite{diamond1983bank} illustrate how fear of insolvency can drive depositors and short-term creditors to withdraw funds simultaneously, creating a sudden and severe funding liquidity crunch. This withdrawal of liquidity forces institutions to liquidate assets at depressed prices, further exacerbating the crisis. Recent research highlights that modern financial systems, heavily reliant on short-term funding, are even more susceptible to liquidity-driven instability, amplifying the potential for cascading failures throughout the system.

    The balance sheet mechanism is one way in which a small negative economic shock can be amplified. When asset prices fall due to a financial shock, the capital of financial institutions is eroded, worsening their balance sheets. This triggers two self-reinforcing liquidity spirals. First, to maintain their leverage ratios, institutions are forced to sell assets precisely when prices are depressed, further driving prices down and eroding net worth -- a process referred to as the loss spiral by Brunnermeier and Pedersen (2008) \cite{brunnermeier2009market}. Simultaneously, tighter lending standards and increased margins create a margin spiral, where borrowers face stricter conditions, leading to fire sales, lower asset prices, and worsening external financing conditions.
    In addition to the balance sheet mechanism, liquidity can also dry up for reasons unrelated to the borrower's creditworthiness. For example, banks may hoard funds as a precaution against uncertain future access to capital markets, reducing the funds available in the economy and slowing economic activity \cite{acharya2013precautionary}.  

\item 
    Financial institutions face similar risks due to \textbf{correlated exposure} when they invest in the same types of assets, a phenomenon that is extensively analyzed in the risk literature. This body of research explores why financial institutions opt to take on similar risk exposures, thereby reinforcing amplification mechanisms, and why they assume large risk positions, increasing their vulnerability to default and exposing their counterparties to contagion.
    In Acharya (2009) \cite{acharya2009theory}, it is noted that the failure of one bank results in a lower overall level of risky investment, which drives up the return on safe assets in equilibrium and squeezes the profits of surviving banks. The bank failure creates negative externalities impacting other institutions. To mitigate this, banks often have incentives to invest in the same assets, ensuring that they either fail or survive together.
    A similar idea is explored in Acharya and Yorulmazer (2008) \cite{acharya2008information}, where creditors view one bank's default as a signal that other banks could fail in the future, contributing to a herding behavior. 
    Farhi and Tirole (2012) \cite{tirole2012overcoming} suggest that when bailouts involve fixed costs for the government (like keeping interest rates low), they tend to benefit the entire economy. Thus, bailouts are more efficient when many banks fail simultaneously.
    Anginer et al. (2014) \cite{anginer2014does} find that greater competition among banks reduces risk by encouraging diversification, which makes the financial system more resilient to shocks. However, in environments with weak supervision, a lack of competition can increase fragility. Cubillas and González (2014) \cite{cubillas2014financial} add that financial liberalization can heighten risk-taking, particularly in developing countries where new opportunities emerge. In developed nations, stronger competition tends to drive risk-taking behavior, demonstrating that the relationship between competition and risk is complex and dependent on institutional environments.
    
\item 
    \textbf{Pro-cyclicality} refers to the tendency of financial institutions, market participants, and regulatory frameworks to reinforce fluctuations in the business cycle, rather than counteracting them. During an economic upswing, rising asset values and optimistic risk assessments encourage banks to expand credit and leverage, pushing asset prices even higher and intensifying the boom. Conversely, in a downturn, risk aversion increases, credit shrinks, and institutions may engage in forced asset sales, driving values lower and deepening the bust \cite{adrian2010liquidity}. This feedback loop amplifies both the peaks and troughs of financial cycles, turning what might have been a contained adjustment into a more severe crisis -- a key reason why pro-cyclicality is considered a major amplifier of systemic risk.   

\item 
    \textbf{Informational amplification} occurs when information (or perception) about the health of one bank amplifies risks across other institutions. This process is driven by changes in market perceptions regarding the creditworthiness of specific institutions and the valuation of their assets, which can escalate and create a broader crisis of confidence. For instance, if depositors and investors interpret the failure of bank \textit{j} as a signal of potential trouble at bank  \textit{i}, this informational link amplifies the initial shock, increasing the risk. Chen (1999) \cite{chen1999banking} illustrates that correlated returns between banks can lead to bank runs, where panic spreads across institutions, even if the initial failure was unrelated to the others. Dasgupta (2004) \cite{dasgupta2004financial} examines how cross-deposits among banks amplify this effect, meaning that negative information about one institution can trigger a run on linked banks. Similarly, Cespa and Foucault (2014) \cite{cespa2014illiquidity} highlight how market illiquidity can be contagious, with liquidity drops in one asset leading to a domino effect across correlated assets, further exacerbating financial instability.

\item 
    \textbf{Moral hazard} arises when individuals or institutions adjust their behavior knowing that they are insulated from the negative consequences of excessive risk-taking -- often due to explicit or implicit guarantees from governments or other backstops. In the banking context, \textbf{bailout expectations} amplify this moral hazard because banks anticipate that if their positions sour, they can rely on external support or rescue funds rather than bearing the full brunt of their own losses. As a result, they may engage in riskier lending, higher leverage, and weaker liquidity management, in the belief that safety nets will shield them from failure \cite{farhi2012collective}. This dynamic amplifies systemic vulnerabilities by concentrating more fragility in the financial system: once trouble emerges, the cost of interventions grows larger, and contagion risks spread more swiftly if bailouts prove insufficient or are delayed.
\end{itemize}

\subsection{Channels of Transmission}\label{subsec:channel-tradfi}

The channels of transmission describe the mechanisms through which localized shocks propagate across the financial system, transforming initial disturbances into broader systemic disruptions.

\begin{itemize}
\item 
    \textbf{Direct contagion} occurs when losses incurred by one financial institution spread to others interconnected with it, causing cascading defaults and insolvencies across multiple banks. Financial contagion models typically concentrate on the interconnections between financial institutions primarily interacting through their balance sheets. Each balance sheet reflects two sides: assets (including loans, investments, and liquid funds) and liabilities (debts and other obligations). 
    A liquidity contagion occurs when a bank cannot meet its payment obligations in full due to a shortage of liquid assets, leading other banks to also default on their payments. This creates a domino effect, where payments are either delayed or significantly reduced, placing further pressure on the financial system. Such shocks originate from markets, leading to contagion among banks unable to meet their own obligations due to non-payments from counterparties \cite{bardoscia2019full}.
    Solvency contagion, on the other hand, happens when a bank's equity becomes insufficient to cover its liabilities. When a bank’s equity falls to zero, its creditors write off their interbank assets, discounting these as lost. This reduction in asset values triggers a chain of defaults, where one bank's failure to pay leads to subsequent failures of connected institutions.

    Financial systems exhibit the characteristics of complex systems, where interdependencies between multiple actors and counterparties play a central role. Following a shock, the behavior of asset prices and financial aggregates often deviates from predictable patterns, displaying irregular volatility, non-linearities, and abrupt discontinuities (such as liquidity freezes). Once destabilized, these systems may transition from linear dynamics to non-linear behavior, characterized by path dependence, sustained oscillations, and even regime shifts. The \textbf{network structure} of financial systems -- shaped by evolving relationships, financial innovation, and regulatory arbitrage -- further amplifies these dynamics. 
    A crucial component of this framework is interbank liabilities, where one bank's debt to another represents its interbank obligations. Such liabilities are directly tied to another bank’s assets and are represented as links in the network that tracks financial interactions.
    These bilateral interbank relationships are modeled as a system of directed, weighted obligations that are connected through transactions like loans, payments, or other forms of financial engagement. When a financial institution faces failure, a common approach to modeling contagion involves simulating a shock that reduces the external assets of one or more banks. This shock then propagates across the network, much like an epidemic, affecting other banks that are connected by these financial liabilities. Formally, this process is governed by dynamic equations that model how a bank's balance sheet evolves as its assets and liabilities are impacted \cite{bardoscia2021physics}. 
    While these links can certainly trigger domino effects and spread bank defaults, the overall risk of a systemic event depends on the structure of the entire matrix of connections, which presents a more nuanced challenge. 
    Allen and Gale (2000) \cite{allen2000financial} argue that interbank markets allow for risk-sharing, which can reduce the likelihood of any one bank defaulting: they show that more complete markets with increased diversification and interconnection always have a positive impact on the economy. However, this also introduces a trade-off between minimizing individual defaults and increasing the potential for contagion.  
    They found that complete networks are more resilient compared to incomplete networks -- incomplete networks allow for contagion to spread even when institutions are not directly connected.
    Freixas, Parigi, and Rochet (2000) \cite{freixas2000systemic} demonstrate that a circular arrangement of interbank connections is less stable than a fully connected network. They also explored how the formation of these links can introduce risks, particularly around coordination failures. Similarly, Allen, Babus, and Carletti (2012) \cite{allen2012asset} suggest that clustering banks into smaller groups can reduce contagion compared to a fully connected network, though it might also reduce the incentives for banks to roll over short-term debt.
    Important results in this literature include, for instance, the idea that networks are “robust yet fragile” \cite{gai2010contagion, acemoglu2015systemic}: connected networks, in which all institutions are connected to each other (at least indirectly) are more robust to shocks, because of risk-sharing, but are more likely to see all institutions fail conditionally on a large shock. 
    More highly connected networks have a low probability of a contagious event, due to risk sharing. Yet, when a contagious event occurs the outcome will be more damaging and widespread due to the same connectivity that helps prevent some events. 
    In \cite{gai2010contagion}, Gai and Kapadia (2010) theoretically studied contagion on Erd\H{o}s–R\'enyi random networks and derived conditions under which the initial default of a randomly selected bank could lead to a global cascade of defaults. We highlight this paper because the model's behavior can be understood in terms of a percolation problem, making it an excellent example of how complex systems contribute to the understanding of systemic risk.

\item
    \textbf{Indirect contagion} happens when shocks propagate between banks even in the absence of direct links: it can due, for instance, to common asset holdings, leading to correlated losses. 
    The 2008 financial crisis is a prime example of institutions being overly exposed to the same markets, particularly the mortgage and subprime mortgage markets. Since then, several studies have examined this sort of correlation explicitly. For instance, Elliott, Georg, and Hazell (2018) \cite{elliott2021systemic} find that German banks tend to lend to other banks with portfolios similar to their own, showing an increase in lending by 31\% when portfolio similarities rise from the 25th to the 75th percentile. This heightened similarity makes the institutions more vulnerable to shared risks. 
    Several key factors contribute to such correlation \cite{jackson2021systemic}. First, competition pushes institutions to pursue similar investments, as seen in the savings and loan crises, where banks became highly exposed to fixed-rate mortgages. Second, regulatory frameworks often dictate the types of assets banks must hold, such as government bonds, which lead to similar risk exposures across institutions. For example, during the 2010 Greek debt crisis, many European banks were found holding large amounts of Greek debt, amplifying shared vulnerabilities. Third, banks have strong incentives to align their portfolios with counterparties they transact with. A bank’s solvency is interdependent on the solvency of its counterparties, making correlated portfolios a strategic decision. Lastly, large economic shocks, such as the Covid-19 pandemic, simultaneously impact the entire financial sector, creating widespread correlation across portfolios. This results in systemic vulnerabilities even in the absence of direct exposure between institutions.

    The dynamics of these interactions can be mapped through a \textbf{network structure} where banks are linked to assets in their portfolios: the network is a bipartite one, with two types of nodes representing banks and assets, and links connect banks to the owned assets. 
    To understand the role of overlapping portfolios in absorbing or amplifying external shocks, it is crucial to assess how the properties of the overlapping portfolio network
    influence its stability \cite{caccioli2014stability, duarte2021fire, cont2017fire, barucca2021common}. This involves specifying how banks adjust their portfolios to mitigate losses and how asset prices react during trading. 
    If one institution is under stress and needs to sell part of its assets, those assets may be devalued due to market impact the tendency of prices to respond to trading activity. This devaluation leads to mark-to-market losses for other institutions holding the same assets. In turn, these institutions may be forced to liquidate their own assets in response to the losses they experience, further depressing prices and potentially triggering a fire sale. In this structure, banks interact indirectly through their overlapping portfolios, i.e., the common assets they hold.
    Asset price reactions are typically modeled through a market impact function, where the more an asset is liquidated, the greater its devaluation \cite{bouchaud2009markets}. While some studies assume linear or logarithmic relationships, others incorporate scenarios where heavily devalued assets are bought by investors at lower prices.
    The simplest bank dynamic often modeled is the linear threshold approach, where a bank remains passive until its losses exceed a certain threshold, typically set at equity level \cite{watts2002simple}. Once crossed, the bank liquidates its entire portfolio. Under this assumption, it is possible to derive analytical results for the case of random networks: diversification, while reducing individual risk, does not always improve systemic stability. 
    Although passive investment strategies provide a useful baseline, in reality, banks react to changing market conditions, especially during crises, by rebalancing portfolios. This behavior is driven by internal risk management procedures or regulatory constraints \cite{adrian2010liquidity}. Active risk management is often modeled through leverage targeting, where banks sell assets to maintain their leverage ratios. This strategy optimizes their expected return on equity while satisfying Value at Risk (VaR) or Expected Shortfall (ES)\footnote{An increasing number of systemic risk measures adopt a holistic approach, not focusing on a specific source or transmission channel of risk. These measures often rely on real-time market data, such as securities prices or derivatives linked to financial institutions, allowing them to detect shifts in systemic risk quickly. However, a common criticism is their lack of theoretical grounding, making it difficult to pinpoint specific risk sources. Notable examples of central metrics in systemic risk research include Value at Risk, which describes the maximum loss incurred in a predefined period of time and confidence level, and Expected Shortfall, which measures the portfolio’s loss when it exceeds the limit set by VaR.} requirements. Some hybrid models suggest that banks adopt leverage targeting after reaching a specific loss threshold, before which they remain passive \cite{cont2017fire}.

\item 
    \textbf{Fire sales} are another mechanism for contagion that operate without network linkages, while reinforcing network effects. The downward spiral in asset prices that results from the dumping of illiquid assets in response to liquidity shocks can spread losses through markets, amplifying a cascade of defaults -- when a bank becomes insolvent, it often has to sell prematurely significant amounts of assets. Such dumping depresses prices for those assets, reducing the portfolio values of other banks holding similar assets. This can lead other banks to default and drive their asset sales into a downward spiral -- see Capponi and Larsson (2015) \cite{capponi2015price}, Greenwood et al. (2015) \cite{greenwood2015vulnerable}. 
    Several factors influence the severity of this contagion. One issue is market depth -- financial markets are often not deep enough to absorb large-scale liquidations without significant price impacts. Asymmetric information also plays a role, with market participants interpreting fire sales as a signal of underlying weakness, which exacerbates the price declines. 
    Many studies consider the effects of fire sales by assuming the existence of one asset that is common to all banks and is liquidated when banks default, but their focus remains mainly on understanding how the topology of the interbank exposures network affects its stability \cite{gai2010contagion, cifuentes2005liquidity}. More recent research, using data on Austrian interbank exposures, suggests that the interaction between contagion channels, including overlapping portfolios, contributes to aggregate losses (Caccioli et al. (2015) \cite{caccioli2015overlapping}). This conclusion was further supported by studies on Mexican banks, using detailed data on both direct exposures and shared assets, indicating the amplified effects of these interconnected risks \cite{poledna2021quantification}. Caccioli et al. (2024) \cite{caccioli2024modelling} examine the impact of fire sales on the UK financial system through commonly held assets across different financial sectors. In particular, they model indirect contagion via fire sales across UK banks and non-banks subject to different types of constraints. They find that performing a stress simulation that does not account for common asset holdings across multiple sectors can severely underestimate the fire sale losses in the financial system. 

\item 
    A risk can be amplified even in the absence of any change in fundamental values, through \textbf{self fulfilling dynamics} and feedback effects. 
    This phenomenon underlies events such as bank runs, where self-fulfilling panic leads depositors to withdraw funds, pushing an otherwise solvent institution into insolvency \cite{reinhart2009aftermath}. Similarly, credit freezes occur when uncertainty about the economic outlook causes banks to pull back lending, further tightening liquidity. This often leads to a downward spiral, as lending dries up \cite{bebchuk2011self}.
    Recent literature has emphasized the critical role cycles play in creating multiple equilibria within financial networks (Jackson and Pernoud, 2020 \cite{jackson2024credit}). As soon as a financial network allows for multiple equilibria, a shift in beliefs can move the system discontinuously from one equilibrium to another. In a network of interbank obligations, cycles enhance counterparty risk without creating value in the financial system; clearing cycles hence reduce the risk of default cascades without affecting the values of banks \cite{roukny2018interconnectedness}. However, in practice, these cycles are extremely complex and intricate, often involving numerous banks with varying debt maturities and contract types, making coordination failures more likely \cite{corsi2016micro}.
    We finally highlight that commonalities in asset holdings and fire sales can also generate multiple equilibria and hence a self-fulfilling worsening of the financial system. For example, if two banks hold the same asset and one is forced to sell, the price drop may push both banks into financial instability. In such cases, the self-fulfilling nature of the crisis can become evident, as the fire sale itself triggers further negative outcomes, including default. 
    Naturally, all these forms of risk interact, and are often at play at the same time.
\end{itemize}

\subsection{Systemic Outcomes}\label{subsec:systemic outcomes-tradfi}

Systemic risk can arise from interdependencies in values, where a change in one bank’s value directly affects others, potentially creating a ripple effect throughout the financial network. For instance, a shift in the value of bank \(i\) can influence the values of all interconnected banks, propagating across the network and leading to significant consequences. 
Additionally, systemic risk can emerge from the presence of multiple equilibria. Even in the absence of a crisis or changes in real economic fundamentals, the interconnected nature of financial institutions can generate contagion effects and feedback loops, leading to substantial shifts in the financial system -- see May et al. (2008) \cite{may2008ecology} and Bormetti et al. (2015) \cite{bormetti2015modelling}. 
The first example reflects the direct impact of changes in financial conditions, while the second highlights structural fragility within the system. Such fragility can result in instability through sudden shifts between equilibria, as seen in events like market freezes. These are often driven by increased uncertainty, which erodes trust among banks and counterparties, exacerbating systemic vulnerabilities.
Systemic risk can be categorized as horizontal or vertical depending on the severity of its effects \cite{bandt2012systemic}. The horizontal perspective confines its focus to events within the financial sector, such as bankruptcies of financial intermediaries or market crashes. In contrast, the vertical perspective gauges the severity of systemic risk based on its broader impact on economic output, with greater effects indicating a more severe systemic event.

\begin{itemize}
\item 
    \textbf{Market freezes} are an extreme manifestation of illiquidity, where institutions are unable to access liquidity due to adverse selection. During the 2008 financial crisis, this phenomenon was evident in interbank markets, where safe banks avoided lending to risky ones. Flannery (1996) \cite{flannery1996financial} points out that this adverse selection problem can lead to illiquid banks losing access to funding, requiring central bank intervention.
    Acharya, Gale, and Yorulmazer (2011) \cite{acharya2011rollover} show how repo markets can collapse when investors lose confidence in the value of collateral, triggering a systemic crisis.
    Caballero and Simsek (2013) \cite{caballero2013fire} explore how market freezes are exacerbated when banks know about shocks to their direct counterparties but have little information about those further along the counterparty chain. This lack of transparency, particularly during large shocks like the Lehman Brothers collapse, causes banks to stop lending. 

\item
    Greenwood, Landier, and Thesmar (2015) \cite{greenwood2015vulnerable} create a structural model that examines systemic risk arising from \textbf{fire-sale spillovers}. When banks are forced to sell assets, this impacts prices, triggering negative effects on other similarly exposed institutions. Their model shows how leverage distribution across banks contributes to systemic risk and calibrates this on data from European and U.S. banks during financial crises. 
    Brunnermeier, Gorton, and Krishnamurthy (2014) \cite{brunnermeier2013liquidity} propose the Liquidity Mismatch Index (LMI) to measure liquidity risk by comparing the future cash-equivalent values of assets and liabilities under stress scenarios. The LMI helps assess the potential liquidity shortfalls and joint liquidity events. 

\item 
    A liquidity (or solvency) contagion brings to a cascading failure, known as a domino effect or \textbf{default cascade} -- mathematically modeled through interbank obligations. As each bank defaults, the remaining equity of creditors is wiped out, leading to more insolvencies. This chain reaction continues until the entire network is either stabilized or the defaults spread across the system, overwhelming it. 
    Jackson and Pernoud (2020) \cite{jackson2024credit} examine self-fulfilling cascades and freezes in great depth, showing how the presence of cycles and asset portfolios in banks can influence insolvency events. They suggest that cascades are less likely if certain banks have enough capital buffers, preventing the chain reaction. Moreover, the study explains that specific cycles involving key debts can lead to multiple equilibria.

\item
    Battiston et al. (2012) \cite{battiston2012default} assess that while diversification of individual credit risk can provide benefits by mitigating default impacts, it may also have unintended systemic effects, especially during \textbf{credit crises}. The increased number of counterparties involved through diversification could expose agents to a higher likelihood of credit runs. They argue that the structure of financial interconnections and differences in the robustness of financial institutions are crucial factors to consider when designing policies to enhance the stability of the financial market.

\end{itemize}

Lehar (2005) \cite{lehar2005measuring} was among the first to empirically estimate the probability of a systemic crisis by analyzing correlations in bank portfolios. His portfolio-based approach allows for the calculation of the likelihood of simultaneous defaults across multiple banks, as well as identifying how much each individual bank contributes to the overall risk within the system.
Cai, Saunders, and Steffen (2018) \cite{cai2018syndication} develop another method for measuring interconnectedness by assessing similarities between two portfolios of loans using Euclidean distances.
In another approach, De Nicolò and Lucchetta (2011) \cite{de2011systemic} analyze the relationship between a systemic real risk indicator, based on GDP Value at Risk (GDP VaR), which measures the risk of extreme declines in a country's GDP, and a systemic financial risk measure, defined as VaR (Value at Risk), which quantifies the potential loss in a large portfolio of financial firms over a specific time period under normal market conditions. They develop a model to predict tail risks in real activity across multiple countries, offering a practical tool for risk surveillance.
Glasserman and Young (2014) \cite{glasserman2015likely} emphasize that the interconnected nature of the modern financial system plays a crucial role in understanding the propagation of crises. Due to the intricate network of connections between financial institutions, critical disruptions in one part of the system can have far-reaching effects on others, potentially destabilizing the entire system. Examples include the collapse of Lehman Brothers, the failure of AIG, and the exposure of European banks to sovereign default risks. 
Arinaminpathy et al. (2012) \cite{arinaminpathy2012size} highlight the critical role that large, well-connected banks play in financial stability. Their failure can have far-reaching effects, shaking trust in the market and causing widespread disruptions. Tougher capital requirements for such large institutions could enhance the system's resilience, especially in less diluted systems. Banulescu and Dumitrescu (2014) \cite{banulescu2015sifis} define Systemically Important Financial Institutions (SIFIs) as those whose failure, due to their size, complexity, and interconnectivity, would significantly harm the financial system and the economy. Kaufman (2014) \cite{kaufman2014too} further discusses the public policy debate around \textit{Too Big to Fail} (TBTF) institutions, noting that varying definitions make it difficult to conclude how such firms should be managed to avoid collateral damage in case of failure.
Early studies that empirically analyzed contagion within financial networks include the work of Upper and Worms (2004) \cite{upper2004estimating} and Elsinger, Lehar, and Summer (2006) \cite{elsinger2006risk}. They explored how the insolvency of a bank could spread to others, using real data on credit connections between Austrian banks. Acemoglu, Ozdaglar, and Tahbaz-Salehi (2015) \cite{acemoglu2015systemic} introduce a notion of distance over the financial network that captures the propensity of a bank to be in distress when another
bank is in distress. Their model shows that under certain conditions, increased interconnections within the financial system can stabilize it, while beyond a certain threshold, it increases fragility. Drehmann and Tarashev (2011) \cite{drehmann2011systemic} introduced two measures of financial interconnection and contagion: one assessing a bank's ability to propagate shocks through the system and another examining its vulnerability to shocks originating from other banks. These measures provide a way to quantify system-wide risk contributions.
Iyer and Peydró (2011) \cite{iyer2011interbank} highlighted how interbank linkages and bank runs act as contagion channels, showing, through a natural experiment, that the likelihood of contagion grows when weaker banks are connected to those exposed to significant withdrawals.
In particular, Jamilov et al. (2024) \cite{jamilov2024two} find that even when the fundamentals of the banking sector -- like capital and liquidity -- seem strong, systemic bank runs still happen and contribute to economic downturns. This suggests that disruptions on the liability side (like problems with short-term borrowing or withdrawals) play a significant role in causing the severe and long-lasting recessions that often follow banking crises.
Amiti et al. (2011) \cite{amini2016resilience}
studies the propagation of balance-sheet or cash-flow insolvencies across financial institutions and how they may be modeled as a cascade process on a network representing their mutual exposures. These models share mathematical similarities with linear threshold models, making them good for analytical techniques, and allowing researchers to derive key metrics -- such as the scale of a default cascade. Beyond the linear threshold framework, more comprehensive models account for write-offs triggered by both defaults and rising probabilities of default. This approach is central to the DebtRank family of models \cite{bardoscia2015debtrank, bardoscia2016distress}, alongside empirical models and valuation-based approaches that seek to measure and predict cascading defaults within the financial system \cite{fischer2014no, barucca2020network}.
Additionally, the topology of the financial network plays a crucial role in determining the system's resilience to shocks. Studies have shown that more diversified and interconnected networks can be more resistant to small shocks but are less resilient to large systemic events. Bardoscia et al. \cite{bardoscia2017pathways} study the stability of interbank networks under the DebtRank
framework as these networks become more interconnected. They observe that, on average, the system tends to become more unstable as the network's connectivity increases. The result provides a clear example of how an action that, according to standard models, should reduce individual risk by increasing diversification can, in fact, inadvertently lead to greater systemic risk.
This robust-yet-fragile nature of financial networks indicates that while diversification reduces the probability of widespread contagion in normal times, it may exacerbate the impact when a significant shock occurs.
A core-periphery structure within the banking sector exemplifies this dynamic. The core, composed of highly interconnected large national or international banks, contrasts with the more sparsely connected periphery of regional banks, which primarily maintain links with core institutions. Empirical research, such as the work by Soramäki et al. (2007) \cite{soramaki2007topology}, confirms the prevalence of such a structure in interbank lending. For instance, Soramäki et al. document a fully connected core of 25 banks, where large exposures exist, and similar findings are reported for the U.S. \cite{bech2010topology}, Germany \cite{craig2014interbank}, and the Netherlands \cite{blasques2018dynamic}.
The core-periphery framework has clear implications for systemic risk. The dense connections within the core increase the likelihood of contagion, as shown by Elliott, Golub, and Jackson (2014) \cite{elliott2014financial}. A core-centered network is more susceptible to default cascades across a broader range of transactions than more balanced structures.

To conclude this section, we mention the paper of Glasserman and Young (2016) \cite{glasserman2016contagion}: they suggest that the relationship between interconnectedness in financial networks and overall financial stability remains a complex topic with no straightforward answers. They point out that while more interbank connections could theoretically promote stability through risk-sharing mechanisms, they may also heighten fragility by creating additional pathways for contagion. The complexity of this relationship, they argue, is further influenced by factors like leverage levels and the varying sizes of institutions, which together shape the dynamics of financial contagion in ways that are not yet fully understood, even at the theoretical level.

\section{Decentralized Finance}\label{sec:DeFi}

On November 1, 2008, the pseudonym Satoshi Nakamoto published a white paper proposing a new payment infrastructure based on blockchain technology, placing the mechanism of cryptographic proof at the center of the intermediation between different parties, instead of relying on trust in existing estabilshed institutions and centralised intermediaries \cite{antonopoulos2021mastering, nakamoto2008bitcoin}. 

A typical financial system can be represented, at an abstract level, as a collection of states and transactions that describe the transition from one state to another. Over the last decade, advances in technology have enabled an alternative architecture for storing and managing information where no single entity has full control over all the states and transactions or any subset of them. Instead, multiple parties (so-called validators) hold their own copies of states and jointly decide which transactions are admissible. This architecture became known as distributed ledger technology (DLT). A blockchain is a form of DLT in which all transactions are recorded and organized in blocks that are linked together using cryptographic protocols to form a chain. Bitcoin was the first and remains one of the most famous application of blockchain technology for peer-to-peer transactions \cite{makarov2022cryptocurrencies}. One of the main advantages of DLT is the elimination of a central point of failure. Since multiple copies of records exist, the corruption of a single node or a single copy has no effect on the security of the blockchain.

The decentralization and distributed nature of the new infrastructure has implications for the scalability of the network. Intuitively, as the network becomes more decentralized more resources need to be spent to achieve the protocol consensus and to secure the network. This trade-off between decentralization, security, and scalability was famously formulated by Vitalik Buterin, a cofounder of Ethereum \cite{buterin2013ethereum} -- a decentralized, open-source blockchain platform --, and became known as the scalability trilemma \cite{buterin2022proof}. A large number of new blockchain-based solutions have been introduced to try to
achieve the three goals simultaneously, such as sidechains \cite{singh2020sidechain} and layer-2 solutions such as the lightning networks \cite{antonopoulos2021mastering, bartolucci2020percolation}.

Crypto assets now span a wide range of classes -- such as stablecoins, utility tokens, CBDCs, and cryptocurrencies -- each characterized by differing levels of technological security and governance stability. These features significantly influence their adoption prospects and long-term viability, as investor preferences are shaped not only by perceived robustness, but also by expected future use cases and adoption benefits \cite{bartolucci2020model}. Empirical evidence further supports that investment patterns in crypto markets reflect these underlying features, with co-investment behavior and token returns closely tied to functional characteristics and utility narratives \cite{mungo2024cryptocurrency}.

Decentralized Finance (DeFi) refers to a form of financial market that operates without central instances and intermediaries, such as (central) banks, centralized exchanges or other components known from existing financial systems.
Replicating existing financial services in a more open and transparent way, DeFi has gained a lot of attraction: the volume traded on decentralized exchanges (DEX) in 2024 reached $\$285$ billion \cite{binance2025post}. In particular, DeFi protocols do not rely on intermediaries and centralized institutions; instead, it is based on open protocols and decentralized applications (DApps). Agreements are enforced algorithmically, transactions are executed in a secure and verifiable way, and legitimate state changes persist on a public blockchain. Thus, this architecture can create an immutable and highly interoperable financial system with unprecedented transparency, equal access rights, and little need for custodians, as most of these roles can be assumed by “smart contracts” \cite{schar2021decentralized}.
DeFi provides a range of applications, for example allowing users to engage in activities such as buying USD-pegged stablecoins on decentralized exchanges, earning interest on lending platforms, and investing in liquidity pools or on-chain funds. 

The backbone of all DeFi protocols and applications is composed of so-called smart contracts. \textbf{Smart contracts} generally refer to small applications stored on a blockchain and executed in parallel by a large set of validators. Their advantage is a high level of security: smart contracts will always be executed as specified and allow anyone to verify the resulting state changes independently. Their transparency aims to minimize manipulation risks, as anyone can verify the encoded rules and allowed actions, ensuring trust in decentralized systems.
The sector has experienced ha strong growth, with the value of funds that are locked in DeFi-related smart contracts recently crossed $80$ billion USD. 

DeFi makes use of \textbf{crypto assets}, such as cryptocurrencies and tokens, as a means of exchanging and storing wealth. Crypto assets provide DeFi products with higher liquidity, flexibility, increased speed and accessibility than conventional assets, allowing DeFi to offer innovative financial products and services.

DeFi is also an umbrella term for a large number of other related protocols in closely related fields -- such as applications prediction markets, gambling, exchanges and governance. Examples include DAO, Oracles, Stablecoins, Liquidity Pools, Lending Pools, Digital Exchanges (DEXs), and Automated Market Makers (AMMs) \cite{weingartner2023deciphering, ibba2023preliminary, aufiero2024dapps}.

Many DeFi apps, in their quest to avoid placing trust in any actor or institution, have experimented with new organizational forms, so-called decentralized autonomous organization (\textbf{DAO}). The basic idea of DAO is to spread control over decisions among all interested stakeholders by issuing special ``governance” tokens that give their holders the power to propose changes to the protocol and vote on them.

While the blockchain infrastructure tries to remove the reliance on third-party enforcement, smart contracts often need to access data stored outside the main blockchain ledger and network if they want to interact with external protocols or entities. Consider, for example, the submission of a limit order in a decentralized exchange, where a user sets a smart contract to automatically sell a token of Bitcoin when the price hits a certain target level. For this contract to work, the contract needs to access up-to-date Bitcoin prices from external exchanges. If data is not obtained in an accurate and timely fashion, a smart trader could reap large gains by taking advantage of stale or wrong prices. \textbf{Oracles} effectively function as intermediaries that deliver external data to smart contracts operating on a blockchain \cite{caldarelli2020understanding}. Since smart contracts are inherently confined to the blockchain environment and cannot fetch data from outside sources, oracles are crucial for supplying this external information. These tools bridge the gap between the blockchain-based distributed ecosystem and the traditional finance centralized providers (e.g., exchanges) by feeding the smart contracts with necessary data, enabling them to execute based on conditions tied to external inputs.

\subsection{Core DeFi solutions}
While DeFi aspires to create a parallel and independent financial system based on algorithmic automation via smart contracts, key components of the DeFi system rely, in practice, on traditional financial market infrastructure. The most critical nexus between the two systems is via \textbf{stablecoins} \cite{carter2021defi}. These digital assets consist of dollar-denominated tokens exchanged on public blockchains and, in principle, backed by fiat dollars locked as reserves at financial institutions.
Stablecoins are intended to offer greater predictability and stability than other cryptocurrencies or tokens, whose prices are frequently volatile and susceptible to changes. Stablecoins typically serve as a medium of exchange, unit of account, store of value, and collateral for loans, among other functions.
The growth of stablecoins has increased significantly since mid-2020, coinciding with the rise of DeFi activities. By late 2021, the total value of major stablecoins in circulation had reached $ \$120$ billion. As of late 2024, that number has grown to $ \$174$ billion, showing sustained demand \cite{forbes_stablecoins_2024}. For comparison, the largest money market fund, the Vanguard Federal Money Market Fund (VMFXX), held $ \$310.4$ billion in assets as of September 2024. 
In particular, USD Tether has gained substantial scale as a \textit{vehicle currency} for investors seeking to trade in and out of cryptoassets. Being the first stablecoin, its growth has benefited from a user base built up early on, which has attracted new adopters seeking ease of trading \cite{aramonte2021defi}.
Stablecoin issuers receive assets (collateral) in exchange for their own liabilities (stablecoins). While this mechanism looks superficially similar to how banks operate, there are fundamental differences. Issuers lack public backstops, such as deposit insurance, and rely on private backstops (collateral) to ensure that stablecoins maintain a steady value and are suitable as mediums of exchange. As such, the expansion of the balance sheets of stablecoin issuers, at least currently, is driven more by the appetite of investors to hold the stablecoins than by any desire of the issuers to acquire more assets. In other words, this growth is liability-driven, while the expansion of bank balance sheets is commonly asset-driven \cite{mcleay2014money}.

A \textbf{liquidity pool} represents a collection of digital assets in a smart contract that is used to facilitate trades between the assets on a \textbf{decentralized exchange} (\textbf{DEX}). Financial specialists argue that one of the main issues with current decentralized exchanges is the impact of low market liquidity, which can cause significant price fluctuations during trades. This challenge arises when there is not enough liquidity to smoothly handle large transactions, leading to less favorable pricing for users. 
Unlike traditional buyers and sellers markets, many DeFi platforms use Automated Market Makers ({AMMs}), which allow digital assets to be traded in an automatic and permissionless manner through the use of liquidity pools. AMMs determine prices algorithmically based on transaction volumes and users' demand.
As an essential part of DEXes, liquidity pools provide the liquidity that is necessary for these exchanges to function. They are created when users lock their cryptocurrency into smart contracts, which then enable tokens to be traded. In exchange for the liquidity provision, those who fund this pool earn a percentage of transaction fees for each interaction by users. Without liquidity, AMMs would not be able to match buyers and sellers of assets on a DEX, and the whole system would grind to a halt. 

\textbf{Automated Market Makers} (\textbf{AMMs}) are decentralized protocols that allow the creation and oversight of liquidity pools within a decentralized exchange (DEX). AMMs utilize algorithms and smart contracts to automatically connect buyers and sellers, as well as to decide the prices and quantities of assets in the liquidity pool. They offer fair, transparent, and secure pricing on the DEX market. AMMs serve multiple functions, including providing liquidity to the DEX market, letting users trade assets swiftly and easily, and allowing liquidity providers to make
returns on their assets.

Similar to DEXs, lending and borrowing in DeFi are managed by smart contracts. \textbf{Lending protocols} provide the rules, algorithms, and incentives governing the lending process within lending pools (LPs). A lending pool must have adequate liquidity in order to issue loans. Thus, they often depend on liquidity pools to provide the necessary liquidity. Most DeFi loans are over-collateralized, using other cryptocurrencies as collateral, mainly to create leveraged trading positions. A common transaction involves borrowing stablecoins with Ethereum or Bitcoin as collateral. Since crypto values can fluctuate, the risk arises if the collateral’s value drops below the borrowed amount. To address this, smart contracts use oracles to fetch real-time prices and trigger liquidations if the loan-to-value (LTV) ratio falls below a set threshold, typically between 50\% and 80\%, depending on the riskiness of the collateral. Borrowers pay interest on loans and receive a rate on their collateral, while protocols collect fees, which are distributed among token holders. Borrowing rates depend on market conditions and are set to maximize fund utilization. The rates vary across assets, and the protocol adjusts them to respond to changing market dynamics \cite{makarov2022cryptocurrencies}.
The lending space is dominated by a few large players such as Aave, Anchor, and Compound protocols. Most protocols operate on a few chains; for example, Aave is built on Ethereum, Avalanche, and Polygon, Anchor uses only Terra, and Compound only Ethereum. Outstanding loans on the major lending platforms have increased rapidly, reaching a maximum of $\$20$ billion in late 2021 \cite{aramonte2021defi}. 

\subsection{DeFi risks}

Decentralized finance presents unique and novel financial risks for a variety of reasons. First, DeFi, being a nascent and rapidly growing sector of the financial industry, faces significant uncertainty and volatility. Several DeFi protocols and methods are experimental in nature, and there is considerable doubt about their long-term feasibility and sustainability. Second, since DeFi relies on blockchain technology and smart contracts, it introduces additional risks and challenges not typically found in traditional financial systems. Third, since DeFi frequently involves the usage of digital assets, such as cryptocurrencies and tokens, the offered financial services can be extremely volatile and subject to substantial price swings.
DeFi vulnerabilities have continued to cause significant losses to investors in recent years. In 2022, total losses due to hacks and exploits in DeFi reached record levels. According to a Chainalysis report, more than $ \$3.1$ billion was stolen from DeFi protocols in 2022 alone \cite{chainalysis2023crypto}. One of the most significant incidents was the Ronin Network hack in March 2022, where attackers stole approximately $\$625$ million in cryptocurrency, marking one of the largest hacks in the history of DeFi \cite{ronin2022hack}. Being unregulated, DeFi protocols also lack consumer safeguards. According to a report by CertiK, losses due to scams and exploits in DeFi exceeded $ \$2$ billion in the first half of 2022 alone \cite{certik2022h1}.

The key issue for systemic risk within traditional financial markets is the degree of interconnectedness of different institutions or, more generally, of parties involved in a financial network, which can become problematic in the context of cascading effects in liquidity crises or similar atypical scenarios. The vast majority of the aspects mentioned relate in their inherent characteristics primarily to financial systemic risk, i.e., primarily liquidity-related issues and, for example, balance sheet dependencies. In DeFi, however, it is not only financial risks that can manifest themselves as systemic risks, but many more \cite{bekemeier2021deceptive}.
In traditional finance, an attack on the IT infrastructure primarily threatens the digital representations of real assets, such as those used in online banking, transactions, or clearing systems. However, the actual physical assets, like cash or paper securities, remain unaffected. Historically, financial markets operated without reliance on IT, with transactions often carried out using physical certificates \cite{bekemeier2021deceptive}. DeFi, on the other hand, operates under the principle of \textit{code is law}, where all rules and operations are embedded within smart contract ecosystems. In the event of a malfunction or hack, the processes and cryptographic mechanisms governing DeFi cannot easily be replicated or managed in an analog format. This highlights the need for heightened technological considerations when assessing systemic risks in DeFi.

In recent years, there has been a growing body of research addressing the risks associated with DeFi, and many studies have focused primarily on the technological risks, indeed. The first systematic review of DeFi research was conducted by Meyer et al. (2022) \cite{meyer2022decentralized}, while Werner et al. (2021) \cite{werner2022sok} provided a general overview and focused on technical risks. Zhou et al. (2023) \cite{zhou2023sok} addresses DeFi attacks, and Schär (2021) \cite{schar2021decentralized} examined the architecture, market mechanisms, and associated risks. Li et al. (2022) \cite{li2022survey} also centered their work on technical risks but included governance risks as well. 
Qin et al. (2021) \cite{qin2021attacking} examined new types of market manipulation such as those enabled by flash loans. 
Comprehensive risk assessments which also include economical and market
risks are rarely found, with a few exceptions like Carter and Jeng (2021) \cite{carter2021defi} and Chang et al. (2022) \cite{chang2022risk}, who proposed a more structured approach to categorizing risks, identifying six major categories: smart contract risks, cybersecurity operational risks, blockchain infrastructure risks, social risks, financial risks, and societal risks. Another accurate classification of the risks has been done by Weingärtner et al. (2023) \cite{weingartner2023deciphering}: they categorize, through a risk wheel, risks associated with DeFi into systematic and unsystematic categories, the first one being those risks that can be managed or influenced by the DeFi product creator. Kaur at al. (2023) \cite{kaur2023risk}, through an extensive literature survey, classified DeFi risks in five categories and computed the relative importance of each criterion and sub-criteria. According to them, the priority ranking of DeFi risks is in the following order:  (1) Technical Risks, (2) Legal and Regulatory Risks (3) Financial Risks (4) Operational Risks.

\subsection{The process of systemic risk formation}

While systemic risk is a well-researched field within the traditional financial markets literature, systemic risk in DeFi has not yet been widely explored. The vast majority of the aspects studied for TradFi are mainly related to \textit{ financial} systemic risk. In DeFi, however, systemic risk can arise not only from traditional financial exposures, but also from non-financial sources intrinsic to the technology itself.

Bekemeier (2021) \cite{bekemeier2021deceptive} presents a framework that categorizes systemic risks in DeFi into two broad types: financial and non-financial systemic risks. \textit{Financial risks} essentially comprise forms of market risks, as well as additional application-specific risks arising from the particular circumstances of decentralized liquidity management.  \textit{Non-financial risks} comprise all those risks that, compared with the classic understanding of systemic risk, could represent additional triggers for systemic events in DeFi. The framework further differentiates between different risk types, such as organizational and technological risks.

In traditional finance, a risk is considered \textit{systemic} when a localized disturbance -- such as the default of a financial institution -- is amplified through mechanisms like leverage, pro-cyclicality, or interconnected exposures, and transmitted across the system via channels such as common asset holdings or liquidity freezes, ultimately threatening the stability of the broader financial system.
\begin{tcolorbox}[colback=cyan!10, colframe=cyan!80!black, boxrule=0.5pt, arc=4pt, left=6pt, right=6pt, top=6pt, bottom=6pt]
\textbf{Definition (Systemic Risk in DeFi)} \vspace{0.5cm}\\
A risk is considered \textit{systemic} in decentralized finance (DeFi) when it originates within a specific protocol, asset class, or technological component -- such as a smart contract vulnerability, oracle failure, or the collapse of a stablecoin -- and propagates across the DeFi ecosystem through highly interoperable structures such as liquidity pools, yield aggregators, or cross-chain bridges. A risk becomes systemic when it compromises the functionality, liquidity, or trust in multiple protocols, potentially triggering cascading liquidations and liquidity freezes, or widespread users' losses that threaten the stability of the ecosystem as a whole.
\end{tcolorbox}

A DeFi risk is, therefore, systemic if it has the capacity to cascade beyond the boundaries of a single protocol or blockchain platform, disrupting the broader functionality and trust within the ecosystem.

\begin{figure}
    \centering
    \includegraphics[width=\linewidth]{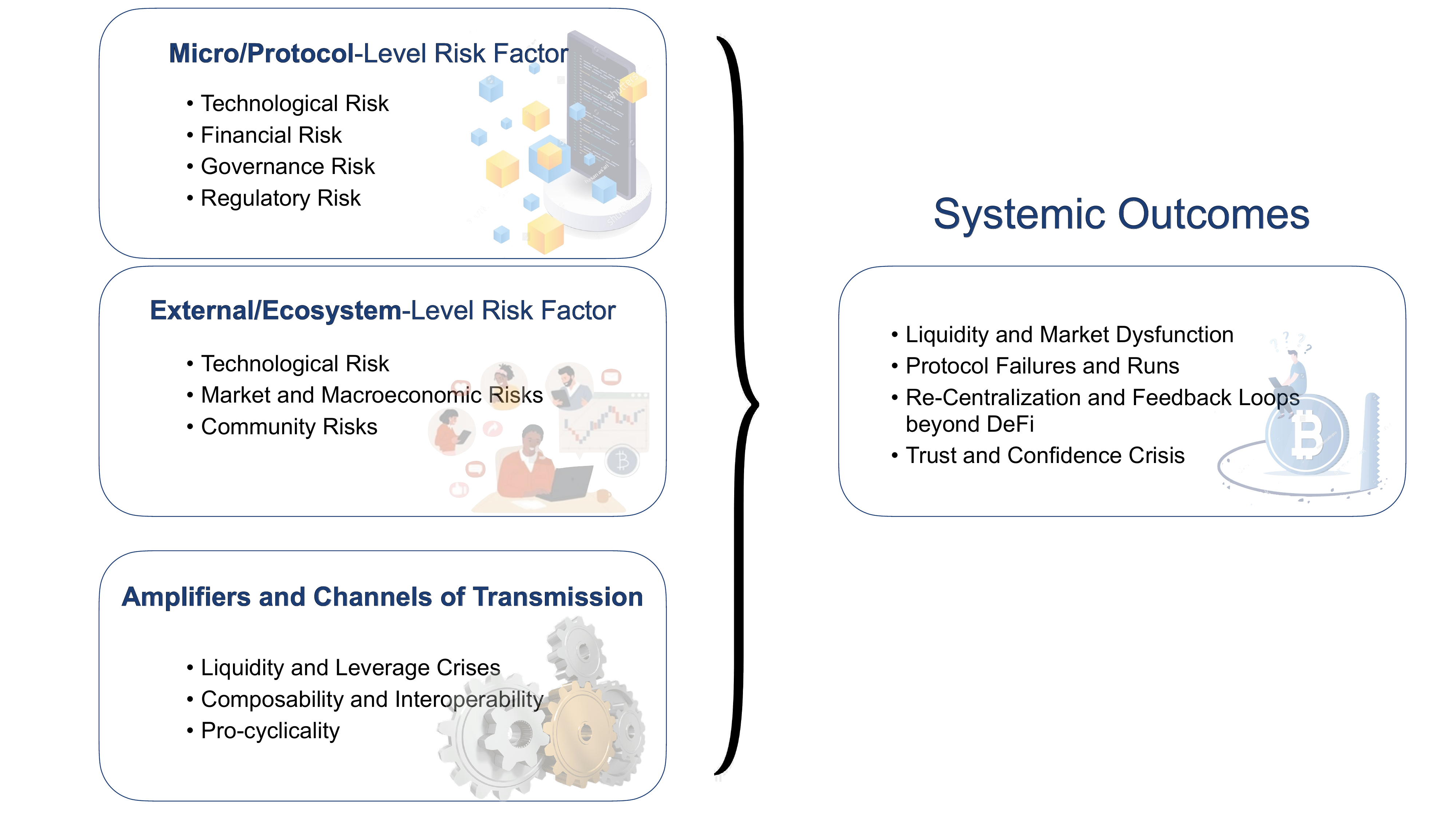}
    \caption{A schematic representation of the DeFi systemic risks formation process, illustrating the progression from micro/protocol-level and external/ecosystem-level risks, to amplifiers and transmission channels, and finally systemic outcomes.}
    \label{fig:schema defi}
\end{figure}

The following taxonomy builds upon the previous framework in Fig. ~\ref{fig:schema tradfi}, while adapting it to the specific context of DeFi. It distinguishes between Micro-/Protocol-Level Risks, External/Ecosystem-Level Risks, Amplifiers and Transmission Channels that propagate or intensify these risks, and the resulting Systemic Outcomes that may undermine the stability of the broader DeFi ecosystem. A schematic overview is provided in Fig.~\ref{fig:schema defi}. It is worth emphasizing that, in DeFi, systemic risk is not confined to financial phenomena alone. Due to the code-dependent and fully digital architecture of the ecosystem, technological risks -- such as smart contract failures, oracle manipulation, or governance exploits -- play a particularly prominent role in shaping systemic vulnerabilities.

\textbf{Micro/Protocol-Level Risks} pertain to individual protocols, platforms, or user interactions. They can arise from internal code, poor governance, or inadequate operational practices within a single DeFi application. Each category is examined in detail in Sec. \ref{subsec:microscopic-defi}.
Technological risks are a prominent category of micro-level risks, arising from vulnerabilities in the code or infrastructure of DeFi protocols. These risks include vulnerabilities in smart contracts, such as bugs, logic flaws, or backdoors that attackers can exploit, as well as dependency risks, which occur when protocols heavily rely on external code libraries or a single development team, thereby creating central points of failure. Cyberattacks targeting front-end interfaces, key theft, or phishing scams are also significant threats, as is inappropriate key management, such as the loss of private keys or compromised administrative privileges. Cryptographic failures, such as flaws in hashing or signature schemes, further increase these technological vulnerabilities. Cyber risks, while predominantly micro-level threats, can have severe implications for the whole system. Danielsson (2016) \cite{danielsson2016cyber} states that cyber risks can evolve into systemic threats, not by themselves but as triggers, particularly if the timing aligns with broader vulnerabilities. 
Financial risks specific to individual protocols are another major concern. Liquidity risks emerge when lending pools or decentralized exchanges (DEXs) lack sufficient liquidity to accommodate large withdrawals or trades. Credit and counterparty risks are present in undercollateralized or partially collateralized lending systems, which can lead to defaults or liquidation shortfalls. 
Governance risks also play a critical role. One prominent concern is governance capture, where large token holders or malicious actors exploit their voting power to seize control of decentralized autonomous organization (DAO) proposals. This can lead to decisions that prioritize the interests of a few stakeholders over the broader community, undermining the decentralized ethos of the protocol.
Lastly, regulatory and compliance risks are increasingly relevant at the protocol level. Some protocols may face exposure to regulations, such as stablecoin wrappers targeted by local authorities, or ambiguous legal statuses that could result in forced shutdowns or re-centralization. These regulatory uncertainties add another layer of complexity to managing micro-level risks in DeFi. 

\textbf{External/Ecosystem-Level Risks} in DeFi arise from factors outside individual protocols but have the potential to impact multiple platforms due to their reliance on shared infrastructure or resources. They are discussed in detail in Sec. \ref{subsec:external-defi}. 
One major category of external risks is technological in nature. These risks often originate from the underlying blockchain infrastructure. Issues such as network congestion, network attacks (e.g., DoS attacks) \cite{antonopoulos2021mastering}, or significant bugs in the blockchain protocol layer can disrupt the functionality of multiple DeFi protocols simultaneously. Additionally, the reliance on oracles for price feeds introduces vulnerabilities, as incorrect or manipulated data can severely impact multiple decentralized exchanges (DEXs) and lending platforms simultaneously. Scalability constraints, such as high gas fees or block-size limitations, further exacerbate transaction inefficiencies, possibly limiting DeFi’s growth and accessibility.
Market and macroeconomic risks also play a crucial role in ecosystem-level vulnerabilities. They arise from rapid price fluctuations in governance tokens or collateral assets, as well as interest rate risks in algorithmic lending pools. General crypto market volatility, such as large price swings in BTC or ETH, often correlates with movements in DeFi assets, creating widespread instability. Foreign exchange and currency risks, particularly when stablecoins lose their peg \footnote{Stablecoin depegging occurs when the market price of a stablecoin deviates significantly from its intended fixed value—typically one unit of fiat currency (e.g., 1 USD). This can signal a loss of trust in the stablecoin's backing mechanism or collateral, and may trigger redemptions, liquidity shocks, or broader contagion within DeFi systems that rely on the stablecoin as a stable unit of account or collateral.}, can result in the simultaneous failure of multiple DeFi applications. Similarly, interest rate shocks in variable-rate borrowing and lending protocols can trigger abrupt spikes or collapses, creating ripple effects across the ecosystem.
Finally, community risks can destabilize the ecosystem. Disagreements among developers or stakeholders on protocol upgrades can lead to fragmentation or forks, weakening the cohesion of the community. Additionally, social engineering attacks, such as hacks on official communication channels like Discord or Twitter, pose risks to large user bases, potentially undermining trust and participation in the ecosystem.

\textbf{Amplifiers and channels of transmission} are the mechanisms that enable localized risk events to escalate into system-wide disturbances, magnifying their impact across the DeFi ecosystem; they are discussed in detail in Sec. \ref{subsec:amplifier and channel-defi}.
One key amplifier is leverage, which can lead to liquidation cascades. High levels of leverage in lending protocols make the system particularly vulnerable to price drops. A decline in token values can trigger forced liquidations, further depressing prices in a feedback loop that intensifies market instability. The composability and interoperability of DeFi protocols also act as critical transmission channels. Protocols are deeply interconnected, meaning that a governance exploit, liquidity shock, or failure in one protocol can cascade through yield aggregators, cross-chain bridges, or dashboard platforms, affecting the broader ecosystem. Pro-cyclicality amplifies these dynamics by exacerbating market trends. During bull markets, over-optimism leads to excessive leverage and risk-taking. In contrast, during bear markets, widespread unwinding of positions accelerates price declines, amplifying crashes. 

When micro-level and ecosystem-level risks are amplified and transmitted through interconnected channels, they can culminate in \textbf{systemic outcomes} that threaten the stability of the entire DeFi ecosystem, resulting in a systemic crisis. These outcomes represent the most severe consequences of cascading failures within DeFi, and they are examined in depth in Sec. \ref{subsec:systemic outcomes-defi}.
One potential systemic outcome is a global liquidity lock-up, where key stablecoins lose their peg or cross-chain bridges fail. Such events trigger widespread liquidity shortages, disrupting the functionality of multiple DeFi platforms and hindering users’ ability to access or move funds. A credit freeze or market freeze can also arise during systemic crises. Borrowing becomes either prohibitively expensive or entirely unavailable, leading to empty lending pools. Similarly, massive contagion and protocol failures occur when the collapse of a major lending platform or decentralized exchange (DEX) causes forced liquidations and solvency crises across interconnected protocols, propagating distress throughout the ecosystem. Spirals of liquidations and bank runs represent another critical risk, where rapid races to the exit from staking or farming positions drive token prices even lower. This downward spiral not only erodes market stability but also amplifies the distress across multiple protocols. Systemic outcomes can also result in re-centralization. In such cases, large entities -- such as centralized exchanges, influential decentralized autonomous organizations (DAOs), or regulators -- intervene to rescue failing protocols or impose control. While these actions may stabilize individual components of the ecosystem, they erode DeFi’s core principle of decentralization. Moreover, feedback loops beyond DeFi represent a significant systemic threat. If major crypto assets or stablecoins fail, the contagion can extend to centralized exchanges or even traditional finance, especially if large institutions hold these assets. The interconnection between DeFi, centralized platforms, and TradFi further amplifies the potential scale of such crises. Finally, a systemic confidence crisis can emerge, as users lose trust in the viability of DeFi. This leads to widespread capital flight, with users moving their funds to traditional finance or centralized platforms. The resulting collapse in user participation and capital inflows can effectively lock up the DeFi ecosystem, stalling its growth and functionality.

Defining a risk as systemic in DeFi thus hinges on whether its impact can cascade beyond one protocol to threaten the broader network of smart contracts, liquidity pools, and ultimately user confidence. Much like in TradFi, the critical difference lies in how open, automated, and composable DeFi is, causing potential disruptions to propagate faster and possibly more broadly than in traditional markets.

\subsubsection{Microscopic/Protocol Level Risks}\label{subsec:microscopic-defi}

At the level of an individual DeFi platform, a range of vulnerabilities -- such as smart contract exploits, liquidity mismatches, or governance failures -- can threaten both user funds and the protocol’s core functionality.
 
\begin{itemize}
    \item The complexity of \textbf{technological risks} in DeFi is high.
    DeFi products operate on a distributed ledger system; therefore, they heavily rely on the blockchain’s underlying protocol. Potential protocol errors in blockchain systems may arise from changes to the protocol, such as hard forks, soft forks, or protocol upgrades \cite{antonopoulos2021mastering}. A hard fork involves changes that are not backward-compatible, requiring all participants to update their software; whereas soft forks are backward-compatible and do not require universal updates. Some blockchain networks, such as Ethereum, implement upgrades through governance mechanisms where users can propose and vote on changes. These modifications, whether executed via a hard fork, soft fork, or any allowed another method, carry risks, including possible fraud or undesirable outcomes, as explored by Barrera and Hurder (2018) \cite{barrera2018blockchain}. At the protocol level, these risks lay the foundation for vulnerabilities that extend to smart contracts, as these applications are directly built on top of the underlying blockchain infrastructure. Issues with protocol upgrades or scalability can directly affect the execution and security of smart contracts, exacerbating their inherent risks.
    
    Individual developers or teams are responsible for the creation and deployment of contracts on various blockchain platforms. Translating business processes in smart contract using specific programming languages (e.g., Solidity), can introduce exploitable vulnerabilities or errors \cite{aufiero2024dapps}, e.g. re-entrancy vulnerability \cite{tikhomirov2018smartcheck}.  Contracts need to be carefully checked before deployment on blockchain platforms to spot potential bugs that can be exploited by malicious behaviors \cite{luu2016making}. In DeFi, the open-source nature of protocols and the low barriers to entry mean that virtually anyone -- including non-specialist developers -- can write and deploy smart contracts with minimal oversight, facilitating rapid innovation, but also increasing the risk of poorly designed or insecure protocols and the proliferation of contracts within the ecosystem.
    
    Most current smart contract and blockchain platforms lack privacy-preserving mechanisms, especially concerning transactional privacy \cite{ron2013quantitative}. The consequences of such vulnerabilities can be severe. Malicious actors may leverage these weaknesses to misappropriate funds or manipulate the protocol's intended functionality. This risk is particularly acute given the autonomous nature of these systems, where transactions execute automatically based on predetermined conditions. Technical exploits are common: Werner et al. (2021) \cite{werner2022sok} identified 21 such attacks on DeFi protocols between February to December 2020, costing users an aggregate of $\$144.3$ million -- although in some cases, funds were returned by attackers. These exploits are varied in their approach, taking advantage of reentrancy bugs, transaction sandwiches, logical bugs, and governance fragilities.

    Dependency risk arises when a protocol relies on external components -- such as third-party smart contract libraries, or front-end services -- that may themselves be vulnerable or subject to failures. If a critical dependency experiences an outage, security breach, or update incompatibility, it can cause disruptions in the ecosystem. Mitigating dependency risk requires contingency planning -- such as employing redundant infrastructure -- so that a single failure point does not threaten the protocol’s core functionality \cite{aufiero2024dapps}.

    Moreover, the accessibility of DeFi and blockchain protocols and algorithms openly through the internet raises various security concerns. A first threat lies in cybersecurity, where these systems are vulnerable to hacking and theft. Attackers can compromise users' private keys or exploit vulnerabilities within the smart contracts that govern blockchain networks. Once access is gained, hackers may steal assets or manipulate transactions, posing significant risks to users’ financial holdings. Securing private keys and ensuring robust smart contract design are crucial to safeguarding assets within these ecosystems (Destefanis et al., 2018 \cite{destefanis2018smart}; Ibba et al., 2024 \cite{ibba2024curated}). Private keys are a critical tool, acting as a unique cryptographic signature that grants access to digital assets and allows users to execute transactions on the blockchain. If product administrators or users mishandle these keys -- whether through inadequate storage, sharing, or other means -- the consequences could be severe. Unauthorized parties could gain access to wallets, leading to theft of assets or execution of fraudulent transactions. Since DeFi lacks centralized control, there is no recourse for recovering lost or stolen funds caused by compromised keys.

\item 
    DeFi services also face \textbf{financial risks}, which primarily revolve around the depletion of funds due to user transactions involving digital assets. These risks typically concerns liquidity and credit risks.
    Liquidity risk refers to the possibility of lacking sufficient funds to meet the value of a financial asset. DeFi platforms often incentivize market makers to liquidate under-collateralized loans, acting like foreclosure auctions. However, if liquidation incentives fail, the original counterparties and liquidity providers face unexpected default risk. Unlike centralized exchanges, decentralized services might not offer last-resort remedies, such as taking trading offline during flash crashes.
    Credit risk, or counterparty risk, refers to the chance that counterparties might default on their financial obligations. In DeFi, this risk is heightened by the volatility of underlying assets, excessive leverage, and potential inaccuracies in algorithmic interest rates calculations. DeFi loan protocols such as Compound demonstrate that loan durations are generally short, with volatile loan rates \cite{cryptoslate2021aave}. This volatility reflects the emerging market dynamics of DeFi, which is significantly different from the more stable conditions in traditional banking. Additionally, the presence of leveraged investment strategies among DeFi users may create new forms of systemic risk \cite{saengchote2021defi}. The anonymous nature of DeFi networks also complicates the assessment of creditworthiness, adding to the overall challenge of managing credit risk \cite{werner2022sok}.

\item  
    \textbf{Governance risks} arise from the decision-making structures within DeFi, where the lack of transparency or accountability may expose users to risks. When decision-making power is concentrated in the hands of a small group, it can undermine the broader community's interests and reduce the legitimacy of collective outcomes. Consensus protocols such as the so-called Proof-of-Work (PoW) or Proof-of-Stake (PoS) mechanism can be vulnerable to centralization, where a small number of participants control a significant portion of the network's validation resources. In Proof-of-Stake (PoS) protocols, consensus is maintained by a limited set of validators; if these actors are compromised or behave maliciously, they can jeopardize network integrity and expose users to significant risks. Additional risks tied to these protocols include so-called  selfish mining and pool hopping attacks \cite{gencer2018decentralization}. DeFi products can also be affected by the speed and scalability of a consensus mechanism. If the protocol cannot manage a sufficiently large amount of transactions or has a high latency, it might cause DeFi users to experience delays in transactions' confirmation and potential financial losses linked to the delayed executions.

\item 
    Decentralized financial systems operate within a rapidly evolving \textbf{regulatory} landscape that often falls outside the scope of traditional financial oversight frameworks. This creates an environment fraught with legal ambiguities and diverse regulatory challenges. Approaches to regulating DeFi vary significantly across jurisdictions, reflecting differing priorities and concerns about innovation, consumer protection, and financial stability. One key concern is the legal status of smart contracts, which, being self-executing, often lack recourse if misused.
    In the United States, regulatory oversight of DeFi is largely driven by enforcement actions from agencies such as the Securities and Exchange Commission (SEC). However, the absence of a comprehensive legislative framework has resulted in a patchwork of precedents rather than a unified set of rules \cite{sec_enforcement_2022, gensler2021remarks}. 
    By contrast, the European Union has taken a more structured approach, with emerging regulatory frameworks such as the  Markets in Crypto-Assets (MiCA) regulation designed to provide clearer guidance for operators and users in the DeFi ecosystem \cite{mica_proposal_2022, ecb_2022_crypto}.
    Asian jurisdictions present a spectrum of regulatory stances. On one end, some countries impose stringent restrictions or outright bans on DeFi activities, while others adopt more flexible frameworks aimed at fostering innovation while ensuring consumer protection. In certain cases, jurisdictions are actively crafting new regulations to integrate DeFi into their broader financial systems, highlighting a willingness to balance experimentation with oversight \cite{mas2022guidelines, china2021ban}.
\end{itemize}

\subsubsection{External/Ecosystem Level Risks}\label{subsec:external-defi}

Beyond the boundaries of any single application, DeFi participants also face broader threats arising from shared infrastructures, uncertain regulatory environments, and interdependencies across multiple chains and protocols.

\begin{itemize}
\item 
    One major category of external risks in DeFi is \textbf{technological} in nature, stemming from weaknesses in the underlying blockchain infrastructure and related systems. Network congestion and high transaction fees, for example, can drastically slow or even temporarily halt critical operations across multiple DeFi applications \cite{werner2022sok, fsb2022risks}. More severe threats include the possibility of a 51\% attack on a Proof-of-Work-based chain, where a single entity gains sufficient hash power to reorder transactions or censor blocks \cite{antonopoulos2021mastering}. Significant bugs at the protocol layer -- such as consensus flaws -- also pose broad-based hazards, as even minor disruptions can trigger chain reorganizations or forks, jeopardizing the security of smart contracts and user funds. 
    Furthermore, DeFi protocols rely heavily on oracles for reliable price data. Oracles are critical for DeFi, but their implementation poses several risks, because they do not provide the robust security properties of native blockchain protocols. Therefore, they can be manipulated by either technical or social vulnerabilities \cite{li2020survey}. Ensuring data reliability is paramount, as inaccuracies or delays can disrupt protocol functionality. Oracles are also vulnerable to data manipulation, requiring robust security measures such as the reliance on decentralized networks and cryptographic proofs to enhance resistance. Scalability is another challenge, as oracles must accommodate growing data demands without sacrificing performance. Additionally, the high costs associated with oracle services may deter smaller DeFi projects, potentially limiting broader adoption. In November 2022, an attack on the Solend lending platform incurred a debt of $ \$1.26$ million. The hacker exploited a weakness in the platform’s price data oracle, stole funds, and increased the value of the assets. This attack impacted three loan pools that held stablecoins.
    Lastly, scalability constraints, such as block-size limits, exacerbate inefficiencies, raising gas costs and undermining the broader accessibility and growth of DeFi ecosystems.

\item 
    \textbf{Market risk} involves the decline in asset value due to investor behaviour, new information, or market conditions. The ease of transferring funds and the complexity of DeFi instruments can enable protocol creators, exchange operators, or third-party manipulators to abuse the system. Assets' high volatility is a central feature of the ecosystem that attracts investors, while simultaneously requiring robust risk management strategies. The interconnectedness of DeFi platforms with foundational cryptocurrencies, such as Ethereum, further amplifies these risks. Price fluctuations in these base cryptocurrencies can create ripple effects across the entire ecosystem, jeopardizing stability and functionality.
    Manipulation of stablecoin prices could trigger liquidations, exacerbated by the lack of standardized price discovery mechanisms, leading to volatility and valuation swings \cite{deshmukh2021decentralized}. To navigate these challenges, DeFi participants employ various risk management strategies. Limit orders allow investors to set predetermined price points for transactions, offering control in volatile markets, while stop-loss orders enable automatic asset sales if prices fall below a specified threshold, mitigating potential losses. 

\item 
    Finally, \textbf{community risks} can significantly destabilize a DeFi ecosystem by undermining its social and governance foundations. Disagreements among core developers, token holders, and other stakeholders about protocol upgrades can lead to contentious splits or hard forks, effectively fracturing the network into competing versions \cite{tasca2017taxonomy}. Such forks dilute the project’s user base and developer resources, reducing the overall synergy essential for ensuring consistent protocol maintenance and innovation. Moreover, when influential figures within the ecosystem publicly oppose certain changes -- such as fee adjustments, algorithmic parameters, or governance rules -- factions can emerge, exacerbating political tensions that further erode community cohesion \cite{de2020blockchain}.
    Sophisticated social engineering attacks represent another community-level threat: malicious actors can exploit official communication channels (e.g., Discord or Twitter) to impersonate developers or moderators, direct users to phishing sites, and seize login credentials or private keys. Such breaches can sow panic, trigger mass withdrawals, and catalyze rumor-driven bank runs if users lose faith in the project’s leadership or technical competence. The outcome is a potential downward spiral of trust -- especially in a permissionless environment where rapid exit is feasible with minimal friction \cite{fsb2022risks}. 
    Ultimately, mitigating these community risks requires robust mechanisms for dispute resolution, transparent governance frameworks, and continuous vigilance over security practices within communication channels. Without these safeguards, even technically sound DeFi protocols may succumb to fractious politics and exploit-based disinformation, eroding the very network effects on which DeFi’s decentralization relies.
\end{itemize}

\subsubsection{Amplifiers and Channels of Transmission}\label{subsec:amplifier and channel-defi}

Even relatively isolated shocks can escalate into systemic disruptions once they pass through DeFi’s high interconnectivity -- where leverage, protocol's composability, and liquidity feedback loops combine to magnify and rapidly transmit local stress to the broader ecosystem.

\begin{itemize}
\item 
    One key amplifier within the DeFi ecosystem is \textbf{leverage}, particularly in lending protocols, where users pledge collateral to borrow additional assets. 
    Lending pools enable users to lend and borrow crypto assets without needing mutual trust. Loan parameters such as interest rates, maturities, and token prices are set by smart contracts, incentivizing proper behavior. However, like smart contracts, lending pools are complex to design, especially because of their intricate economic incentive systems. This makes it challenging to determine whether a lending pool achieves its intended financial goals. 
    Excessive leverage increases both gains and losses: when token prices decline, heavily leveraged positions can quickly fall below collateral requirements, prompting automated \textbf{liquidations}. These forced sell-offs generate additional downward pressure on prices, which can trigger liquidations in other similarly leveraged positions -- a so-called liquidation cascade. The resulting spiral amplifies market volatility, as the forced sales often occur at depressed prices, further reducing collateral values and triggering yet more margin calls. Thus, leverage brings liquidation risks, as seen when the collapse of FTX led to missing payments on $\$17.7$ million of loans defaults across lending pools, affecting liquidity providers \cite{vidal2023ftx}.
    Liquidity risk in decentralized finance becomes particularly significant under specific conditions. Thin markets, characterized by a limited number of buyers and sellers, make executing trades challenging without causing substantial price impacts. Similarly, large transactions in markets with insufficient depth can result in significant price slippage, where the executed price deviates markedly from the initial quote. During periods of market stress, such as high volatility or turbulence, liquidity can quickly diminish, amplifying risks for all participants. Additionally, some DeFi protocols impose built-in constraints on transaction sizes or frequencies, further limiting liquidity in certain scenarios. 
    Additionally, a few primary depositors provide most of the liquidity in lending pools, while a small number of borrowers take most of the loans \cite{gudgeon2020defi}. When users who play dual roles (both lenders and borrowers) withdraw their stablecoin deposits without repaying loans, it definitely triggers liquidity crises. 
    Impermanent loss is another critical aspect of liquidity risk, particularly in DeFi, due to the unique structure of liquidity pools and the automated market maker (AMM) algorithms that govern them.  It occurs when liquidity providers deposit assets into a liquidity pool, and the prices of those assets change relative to their initial value at the time of deposit. This price divergence can lead to a temporary reduction in the value of the provider's holdings compared to simply holding the assets outside the pool. This phenomenon is peculiar to DeFi because liquidity pools constantly rebalance the ratio of assets in response to market activity, which magnifies the impact of price fluctuations. While this loss is termed impermanent because it may reverse if asset prices return to their original levels, it becomes permanent if the liquidity is withdrawn while the price difference persists. The reliance on AMM mechanisms and the absence of such rebalancing dynamics in traditional finance make impermanent loss a distinct risk in the DeFi ecosystem. This risk is particularly pronounced for more volatile assets, but it can be partially offset by the trading fees or rewards earned from participating in the pool \cite{coinbase_impermanent_loss}.
    Finally, we mention the counterparty risk: it occurs when one party in a transaction cannot meet its obligations. A reentrancy attack can exploit this by draining liquidity \cite{weingartner2023deciphering}.

\item 
    \textbf{Composability and Interoperability} have emerged as defining features of modern DeFi, transforming discrete financial applications into interlinked ecosystems while simultaneously amplifying the scope and severity of potential disruptions. By design, DeFi protocols are often built on modular smart contracts that communicate within each other, enabling seamless integration of services such as lending or trading \cite{aufiero2024dapps}. This high degree of interconnectivity fuels rapid innovation and capital reallocation -- traders and automated yield aggregators can swiftly migrate collateral among platforms offering more attractive rates or incentives. Yet these benefits come at the cost of deeper systemic entanglement: an exploit or price manipulation in a single contract can quickly reverberate through multiple layers of composable contracts \cite{gudgeon2020defi}. 
    For instance, if a core stablecoin experiences a temporary loss of peg due to erroneous oracle data, not only do borrowers and lenders who rely on that stablecoin face immediate liquidation risks, but also AMMs and derivatives platforms may become imbalanced, triggering sudden shifts in trading volumes and token valuations. At the same time, cross-chain solutions compound this risk by extending vulnerabilities across multiple blockchains: an exploit in one layer can invalidate collateral assumptions or freeze liquidity in a distant portion of the DeFi network \cite{christidis2016blockchains}.
    Furthermore, because governance mechanisms are themselves composable -- often allowing tokens from one protocol to vote on upgrades in another -- unforeseen governance captures or quorum manipulations can cascade swiftly and undermine the stability of multiple platforms \cite{werner2022sok}. 

\item 
    \textbf{Pro-cyclicality} refers to the tendency of financial behavior to reinforce boom-and-bust cycles, amplifying both market expansions and contractions. In DeFi, this phenomenon manifests when rising token prices and optimistic sentiment encourage participants to increase leverage, overextend credit, or concentrate investments in a narrow set of protocols -- thereby fueling further price appreciation and liquidity inflows. Once prices begin to fall, however, risk aversion spreads quickly: borrowers face margin calls, investors rush to withdraw liquidity, and lending platforms impose stricter collateral requirements. These defensive maneuvers can trigger a feedback loop of forced liquidations, further depressing asset prices and sapping market confidence. Pro-cyclical dynamics thus act as a powerful amplifier and channel of transmission, since the very strategies that multiply gains during bullish phases exacerbate system-wide stress in a downturn. In the absence of stabilizing forces -- such as circuit breakers, dynamic collateralization rules, or robust governance protocols -- this cyclical swing can intensify volatility, leading to more severe and systemic liquidity shortfalls across interconnected DeFi applications
\end{itemize}

\subsubsection{Systemic Outcomes}\label{subsec:systemic outcomes-defi}
This section examines how localized disruptions -- once magnified by amplifiers and transmitted across interlinked protocols -- ultimately culminate in widespread losses, liquidity lockups, and confidence crises that threaten the entire decentralized ecosystem.

\begin{itemize}
\item 
    One of the gravest systemic outcomes in DeFi arises when core sources of \textbf{liquidity} become inaccessible -- often triggered by stablecoins losing their peg or by failures in cross-chain bridges that lock up user funds \cite{ezzat2022blockchain}. A sudden de-pegging event can send shockwaves through lending markets, forcing protocols to rapidly reassess collateral requirements or declare a state of emergency, thereby constraining user withdrawals and trades. In parallel, cross-chain bridge exploits may render assets on certain blockchains effectively unusable, causing a breakdown of trust and a  ``flight to safety'' into more established tokens \cite{zhang2024security}. These disruptions worsen global liquidity shortages, making it difficult for platforms that rely on stable collateral or cross-chain assets to process redemptions. This can result in sharply rising interest rates or even the collapse of lending pools. During periods of acute distress, a credit freeze can ensue, wherein borrowing becomes prohibitively expensive or is halted altogether, effectively emptying the liquidity pools that underpin money markets. Markets may seize up further if liquidity providers rush to withdraw funds, exacerbating a vicious cycle in which plummeting token prices fuel additional margin calls and panic selling \cite{meyer2022decentralized}. Ultimately, this \textbf{market dysfunction} reverberates across interconnected DeFi services, with once-stable platforms exposed to widespread contagion.

\item 
    \textbf{Protocol Failures} is another form of systemic outcome. The collapse of a major lending protocol or decentralized exchange can trigger a chain reaction of forced liquidations, rapidly spilling over into interconnected DeFi services. In such scenarios, once-trusted platforms may suddenly face severe solvency challenges if their collateral bases plummet in value or become locked by liquidity constraints, prompting automated smart contracts to liquidate user positions at a fraction of their original worth. This phenomenon is further amplified when staking and farming participants sense incoming turmoil and rush to exit their positions -- effectively replicating the classic bank \textbf{run} dynamics observed in traditional finance, but at far greater speed due to the permissionless nature of DeFi \cite{auer2020rise}. 
    Large holders rush to redeem stablecoins for safer assets or to withdraw liquidity from DEX pools, creating a mass exodus that drains capital reserves and compounds systemic stress. In many DeFi protocols, capital is locked in smart contracts and reliant on secondary markets for price discovery, so widespread redemptions can leave pools illiquid or force the protocol to mark down collateral precipitously. This, in turn, sows panic among remaining participants, who may stampede to exit before liquidity disappears \cite{werner2022sok}. The combination of leveraged positions facing margin calls and a rush to redeem or exit unstable pools can initiate feedback loops where declining asset prices accelerate forced liquidations, deepening price falls and entrenching systemic fragility.
    In addition, yield aggregators automatically rebalance assets across multiple platforms, often amplifying losses if one leg of the strategy collapses. These interdependencies mean that a local failure -- initially confined to a single protocol -- can morph into a system-wide contagion event that affects DEXs, lending pools, stablecoins, and cross-chain bridges alike. The ultimate outcome is a feedback loop of declining token prices, margin calls, and exacerbated liquidity drains, illustrating how DeFi’s hallmark composability can become a powerful amplifier of stress when market sentiment turns negative. Composability, indeed,  acts as a potent channel, intensifying feedback loops throughout the ecosystem. 

\item 
    Another dimension of systemic distress in DeFi arises from the way governance structures and the broader financial landscape can precipitate a \textbf{return to centralization}, particularly during crises. When a critical protocol faces insolvency or a governance exploit, large stakeholders -- whether centralized exchanges, influential DAOs, or government regulators -- may intervene, effectively assuming control to prevent a total collapse. While these interventions may temporarily stabilize lending pools or restore pegged assets, they undermine the decentralized ethos that originally defined many DeFi initiatives, leaving them subject to top-down directives. Such rescues often come with conditions that centralize governance authority (for instance, consolidated voting power in a single DAO or partial acquisition by a centralized exchange), reinforcing a hierarchical power structure that contradicts the open, permissionless ideals of DeFi. 
    Early precedents include the 2016 DAO hack, where a large portion of stolen Ether was rescued via an Ethereum hard fork -- an action that raised questions about the supposedly immutable nature of smart contracts \cite{dupont2017experiments}. A similar realignment unfolded in 2020 when the pseudonymous founder of SushiSwap withdrew significant development funds and spurred intervention by prominent figures, including centralized exchange operators, effectively seizing control \cite{qin2021attacking}. Even MakerDAO’s emergency governance votes during the 2020 Black Thursday crash, while designed to protect the protocol, illustrated how rapidly decision-making could revert to centralized measures in times of stress \cite{werner2022sok}. More recent, the collapse of TerraUSD (UST) in 2022 prompted Terraform Labs and other large entities to exercise top-down authority -- suspending the chain, injecting reserves, or modifying key parameters -- in a last-ditch bid to defend the peg \cite{aramonte2021defi}. 
    Although these maneuvers can restore a measure of stability, they also highlight that DeFi ecosystems, under pressure, often default to structures resembling traditional, centralized models \cite{gudgeon2020defi}.
    
    Moreover, \textbf{feedback loops beyond DeFi} add a structural layer of fragility. If major crypto assets or stablecoins fail, the contagion can easily spread to centralized exchanges, which often hold substantial amounts of these assets on behalf of retail and institutional clients. A widespread loss of confidence in stablecoins, for instance, could force centralized exchanges to freeze user withdrawals or delist affected tokens, rippling into the realm of traditional finance if large banks or investment funds have exposure to these crypto instruments. Consequently, the intricate interconnections between DeFi protocols, centralized platforms, and TradFi institutions amplify the scale of potential crises: a local DeFi failure can escalate into a multifaceted meltdown that challenges the stability of both on-chain and off-chain financial systems. Ultimately, these governance and structural risks highlight the paradox of DeFi in practice: while seeking to liberate finance from centralized oversight, it may create novel pathways for systemic contagion -- requiring precisely the kind of top-down interventions it hoped to avoid.

\item 
    Finally, \textbf{trust and confidence crisis} are present as well, as platforms must continuously adapt to technological shifts to avoid obsolescence, maintain user trust to mitigate reputational damage, and prevent user exits that could diminish value or cause platform failure. Such a crisis of confidence can arise from a series of high-profile exploits, sudden stablecoin de-pegs, or governance breakdowns, all of which erode the perception that DeFi is secure and transparent. Faced with uncertainty about collateral security or protocol solvency, large liquidity providers withdraw funds en masse, while ordinary users may follow suit out of fear of being last in line \cite{werner2022sok}. By undermining user confidence -- arguably DeFi’s most crucial asset -- this type of systemic crisis can prove more damaging than any individual exploit, with repercussions that ripple outward to centralized crypto exchanges and potentially even institutional investors holding DeFi-related assets.
\end{itemize}

\section{Mapping Risks in TradFi and DeFi}\label{sec:results}

TradFi and DeFi embody two fundamentally different approaches to delivering financial services. TradFi relies on a centralized framework, where banks and institutions act as intermediaries, controlling access and managing assets. By contrast, DeFi leverages blockchain technology and smart contracts to create a decentralized ecosystem in which peer-to-peer transactions are facilitated without reliance on central authorities \cite{yawrisk}.

\textit{Accessibility} is one of the most prominent differences between the two systems. Traditional financial services are often constrained by geographic, regulatory, and institutional barriers, including limited operating hours, creditworthiness requirements, and minimum balance thresholds. In contrast, DeFi offers permissionless access to anyone with an internet connection, operating 24/7 and eliminating many of the gatekeeping mechanisms found in TradFi.

\textit{Transaction speed and cost} also diverge significantly between traditional finance and DeFi. Traditional financial systems are often slow, with international transfers sometimes requiring several days to process and involving significant fees. DeFi enables near-instant settlement and lower transaction costs, particularly for cross-border transfers, through blockchain-based execution.

\textit{Transparency} further distinguishes the two. In TradFi, internal processes, risk exposures, and decision-making structures are often opaque to the public. DeFi, by contrast, operates on public blockchains, where transaction histories and smart contract logic are openly accessible, enabling greater accountability -- but at the expense of user privacy.

\textit{Innovation and adaptability} are likewise shaped by their respective architectures.
Product innovation and development in traditional finance is frequently constrained by regulatory barriers and institutional resistance to change. In DeFi, open-source protocols and composable infrastructures allow for rapid experimentation and the creation of novel financial instruments.

\textit{Regulatory oversight} marks another fundamental distinction.  Traditional finance operates within strict regulatory frameworks, which provide consumer protections, systemic stability and recourse mechanisms, but may also hinder innovation. DeFi, however, largely operates in a regulatory grey zone. While this grants greater flexibility and lowers entry barriers, it also exposes users to higher risks due to the absence of formal oversight, enforceable protections, or standardized disclosure requirements.
\textit{Interoperability and composability} are central to DeFi’s design. Traditional financial systems often operate in isolation, with limited integration between platforms. In contrast, DeFi's composability enables seamless interactions between different protocols, allowing users to combine different services (e.g., lending, trading, insurance) into complex, automated financial workflows.
\textit{Governance structures} also differ markedly. TradFi institutions typically rely on centralized decision-making processes managed by executives and regulators. Many DeFi protocols adopt decentralized governance models, where stakeholders -- typically token holders -- participate directly in protocol decisions, such as upgrades, fee structures, or risk parameters.

Despite its many advantages, DeFi introduces a new set of systemic and operational risks. Smart contract vulnerabilities, oracle manipulation, and governance attacks can have widespread consequences due to the tightly interconnected nature of DeFi protocols. Moreover, the low barriers to participation -- while democratizing access -- can allow malicious actors to exploit weakly designed contracts or mislead uninformed users.

These differences in design philosophy and infrastructure are reshaping the drivers, channels, and manifestations of systemic risk. 
The decentralization inherent in DeFi reduces some risks associated with centralized control (e.g., single points of failure), but introduces new challenges related to governance, coordination, and accountability. Transparency in DeFi transactions, facilitated by public blockchains, reduces information asymmetry but can raise privacy concerns for users. The immutable nature of smart contracts contrasts with the flexibility in traditional finance to rectify errors through centralized intervention. Additionally, while DeFi aims to provide inclusive financial services without barriers, this openness can attract malicious actors and requires users to possess a certain level of technical understanding to participate safely.

To address these evolving risks, DeFi platforms are gradually adopting risk mitigation mechanisms inspired by traditional finance. Examples include decentralized insurance protocols that cover losses from hacks or contract failures, governance tokens that enable stakeholder participation akin to corporate voting rights, and routine smart contract audits that mirror internal controls and compliance functions in TradFi to mitigate technological risks.

As Bekemeier (2021) \cite{bekemeier2021deceptive} notes, a nuanced comparison of systemic risk across DeFi and TradFi reveals multiple overlapping dimensions -- technological, financial, and governance-related -- each of which can amplify or contain risk depending on the architecture and assumptions of the underlying system.

\begin{itemize}

\item 
    The \textit{object} of concern changes. Whereas the traditional perspective tends to look at the contribution of financial institutions, which in themselves form the aggregate of many individual risks, the object of concern in DeFi are primarily individuals interacting with the protocol, developers and applications, which leads to a potentially broader distribution of entities contributing to systemic risk, and, consequently, a multiplication of the level of interconnectedness that is described in the traditional literature as one of the main triggers of systemic risk. 

\item 
        The \textit{point} of concern in traditional financial systems is primarily concerned with the overall stability of the monetary system. In DeFi, the main concern is to ensure the stability of the ecosystem on \textit{all technological layers} (infrastructure and protocols), and all further layers on top, such as the application layer \cite{schar2021decentralized}.

\item 
    The \textit{authority} of concern is different. Traditional financial systems are guarded by regulatory frameworks and supervisory authorities, which are external. Furthermore, regulation can be found on national, as well as supranational levels, such as regulation of the European Union. Financial market legislation and regulation act as a social construct, and they are not necessarily embedded in any computer code or software, although certain regulatory requirements may have an impact on the design of IT systems. In contrast, DeFi does not include regulatory or supervisory authorities within the ecosystem. All rules and regulations are embedded in smart contract code and are defined by the developers and partly the community.

\item 
    Concerning \textit{regulatory implementation} significant differences occur. In traditional finance, regulations are established ex ante through legal frameworks, approved by standardized legislative procedures, and can be adjusted ex post, particularly after crises, through political processes. In contrast, DeFi defines its governance and rules ex ante via smart contracts. However, ex post changes are often technologically impossible due to the immutable nature of smart contracts. 

\item 
    Finally, for measuring systemic risk, in the DeFi ecosystem on-chain data can be used to assess systemic risk, following the structure of the definitions in the corresponding protocols and delivering a single point of truth. In the traditional financial systems, there are numerous data sources and country-specific interpretations, leading to significant difficulties relating to data aggregation for the final assessment of systemic risk \cite{bisias2012survey}.
\end{itemize}

Now, we turn our attention to the core issues where risks in TradFi and DeFi share similar foundations, yet manifest through distinct mechanisms. The logic behind these risks is consistent across both systems, but the specific vulnerabilities differ due to the technological structures and operational frameworks involved.

\begin{figure}
    \centering
    \includegraphics[width=\linewidth]{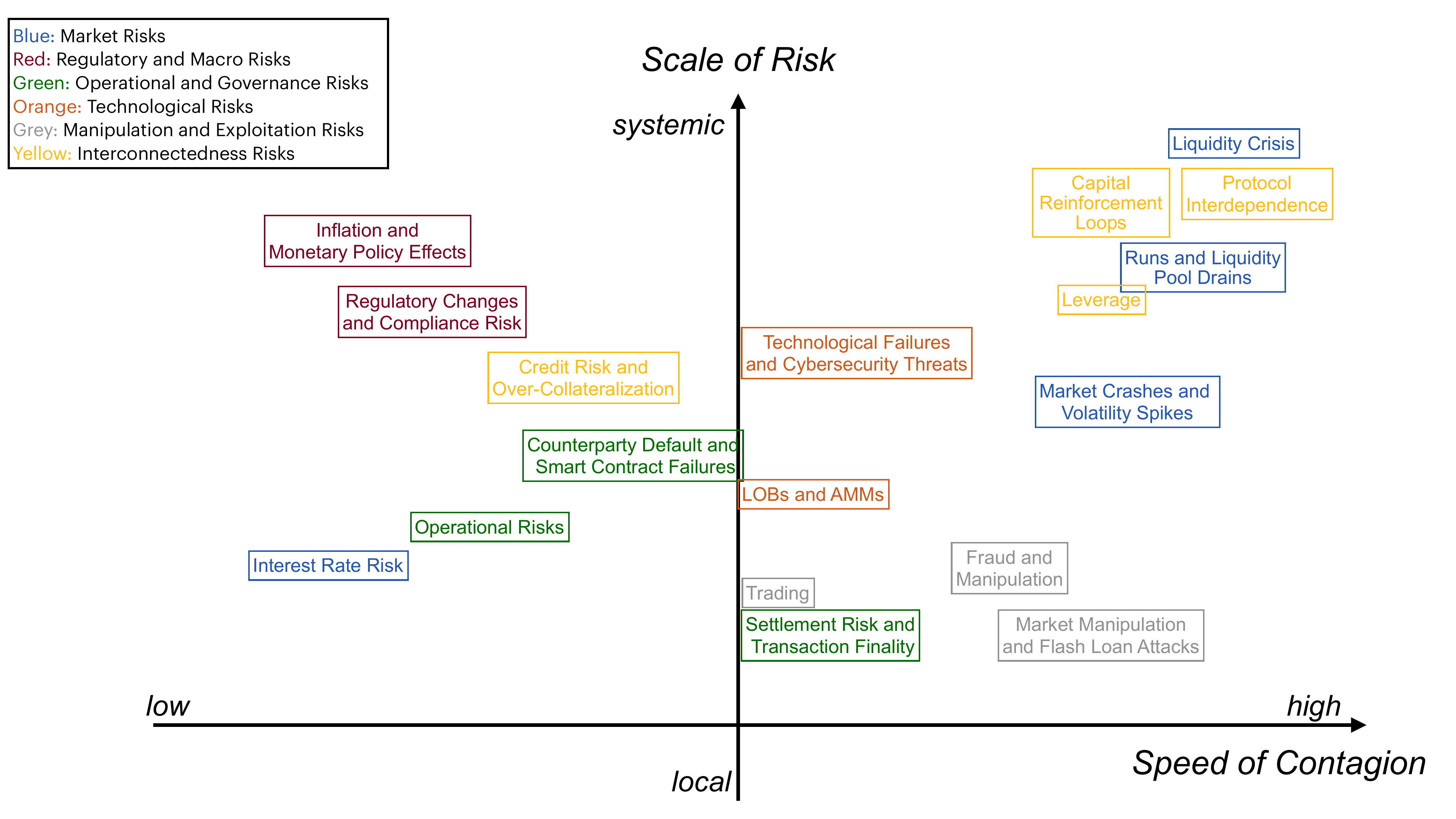}
    \caption{A two-dimensional plot categorizing risks in DeFi and TradFi with similar foundations. The horizontal axis represents the speed of contagion, from slowly spreading risks (left) to rapidly propagating risks (right). The vertical axis captures the scale of risk, ranging from localized, micro-level risks (bottom) to system-wide, macro-level risks (top). Risks are positioned based on these dimensions, and the color-coding corresponds to the legend (on the up left corner), classifying risks into six categories: blue for market risks, green for operational and governance risks, orange for technological risks, red for regulatory and macro risks, grey for manipulation and exploitation risks, and yellow for interconnectedness risks.}
    \label{fig:mapping}
\end{figure}

In the plot in Fig. \ref{fig:mapping}, we propose a map of these key risks in TradFi and DeFi, using a framework inspired by the two-layer approach used in the previous Sections that distinguishes between micro-level and system-wide risks. The horizontal axis represents the speed of contagion, ranging from risks that spread slowly and remain relatively contained (far left) to those that propagate rapidly across the ecosystem (far right). This dimension is particularly relevant in DeFi, where the interconnected and algorithm-driven nature of the system can cause some risks -- such as flash crashes or oracle manipulations -- to cascade almost instantaneously across protocols. The vertical axis captures the progression from localized, protocol-level risks (at the bottom) to broader, system-wide phenomena (at the top). 
The risks are positioned on the map based on their categorization along the two axes. The colors indicate categories as defined in the legend (on the up left corner), where we classify the risks into distinct groups: market risks (blue), regulatory and macro risks (red), operational and governance risks (green), technological risks (orange), manipulation and exploitation risks (grey), and interconnectedness risks (yellow).

Moving from left to right, we will now analyze each risk category individually.

\begin{tcolorbox}[colback=marketblue!10, colframe=marketblue!80!black, boxrule=0.5pt, arc=4pt, left=6pt, right=6pt, top=6pt, bottom=6pt]\textbf{Interest Rate Risk}
\end{tcolorbox}

Interest rate changes in traditional finance, often influenced by central bank policies, affect borrowing costs and investment returns. In DeFi, interest rates on lending platforms are typically dynamically adjusted based on supply and demand curves. Sudden shifts in these rates can impact borrowers and lenders, influencing the attractiveness and stability of lending protocols. The absence of a central authority means that rate adjustments are more market-driven but can also be more volatile.

In TradFi, the 2022 global interest rate hikes by central banks, particularly the U.S. Federal Reserve, led to significant losses in bond portfolios, exemplified by the collapse of Silicon Valley Bank, which was forced to sell long-term securities at a loss, triggering a liquidity crisis \cite{lowenstein2023svb}. In DeFi,
similar pressures emerged during the 2021 crypto bull market, when borrowing demand surged. Protocols like Aave and Compound responded with significant rate increases \cite{cryptoslate2021aave}, making borrowing prohibitively expensive and destabilizing collateralized positions as users struggled to meet rising repayment obligations.

\textit{Interest rate risk typically originates in a localized manner and exhibits a relatively slow speed of contagion in both TradFi and DeFi. Nevertheless, it exerts a profound influence on broader market dynamics over time, shaping investment behavior, liquidity conditions, and the stability of financial protocols.}

\begin{tcolorbox}[colback=regulatoryred!10, colframe=regulatoryred!80!black, boxrule=0.5pt, arc=4pt, left=6pt, right=3pt, top=3pt, bottom=3pt]\textbf{Inflation and Monetary Policy Effects}
\end{tcolorbox}
    In traditional finance, central banks control the money supply, influencing inflation and currency values through monetary policy. DeFi projects may have tokens with fixed supplies or built-in inflationary mechanisms. Changes such as protocol upgrades or adjustments in tokenomics can affect the supply and value of tokens. Unlike traditional currencies, these changes are often governed by decentralized communities, introducing novel dynamics in monetary policy effects.

   \textit{ Inflation and monetary policy effects are generally slower-moving risks, driven by structural adjustments in monetary policy or tokenomics, but their eventual impact can be system-wide, especially in TradFi due to the central role of fiat currencies and in DeFi when stablecoins or widely-used tokens are affected.}

    \begin{tcolorbox}[colback=regulatoryred!10, colframe=regulatoryred!80!black, boxrule=0.5pt, arc=4pt, left=6pt, right=3pt, top=3pt, bottom=3pt]\textbf{Regulatory Changes and Compliance Risk}
\end{tcolorbox}

    Traditional financial institutions are subject to regulatory oversight, and shifts in regulations can significantly impact their operations, introducing compliance costs and altering market dynamics. DeFi currently operates in a relatively unregulated environment, but this lack of regulation poses risks. Sudden regulatory actions could restrict DeFi operations, limit access in certain jurisdictions, or impose new compliance requirements, affecting both platforms and users. The uncertainty surrounding future regulations adds an additional layer of risk to the DeFi ecosystem.

 \textit{   Regulatory changes and compliance risks are typically slower-moving risks too, as they involve the development, announcement, and implementation of legal or policy changes. However, their impact can be highly systemic, particularly if regulations target key components of the financial or DeFi ecosystem, such as stablecoins issuance or cross-border transactions.}

    \begin{tcolorbox}[colback=opgreen!10, colframe=opgreen!80!black, boxrule=0.5pt, arc=4pt, left=6pt, right=3pt, top=3pt, bottom=3pt]\textbf{Operational risks}
\end{tcolorbox}
    Operational risks in traditional finance stem from human errors, process failures, or external events that can lead to financial losses. In DeFi, these risks may manifest through coding errors in smart contracts, mismanagement of private keys by users, and exploitable flaws in decentralized governance processes. The complexity of smart contracts and the technical knowledge required to interact with DeFi platforms can increase the likelihood of operational problems, emphasizing the need for users' education and intuitive platform designs.
    
\textit{    Operational risks are typically localized events that arise from human errors, process failures, or governance mismanagement. However, in both TradFi and DeFi, they can propagate indirectly if they compromise trust or create vulnerabilities in critical components of the systems.
}

 \begin{tcolorbox}[colback=interyellow!10, colframe=interyellow!80!black, boxrule=0.5pt, arc=4pt, left=6pt, right=3pt, top=3pt, bottom=3pt]\textbf{Credit Risk and Over-Collateralization}
\end{tcolorbox}
    In traditional lending, credit risk arises from the possibility that borrowers may default on their loans. Financial institutions manage this risk through credit assessments, relying on credit scores and financial histories. By contrast, DeFi platforms, lacking access to such data, mitigate credit risk through over-collateralization. Borrowers must lock up assets that exceed the value of the loan, thereby protecting lenders from defaults.  While this approach secures lenders' interests, this mechanism can be capital-inefficient, limiting borrower participation and constraining the expansion of lending markets in DeFi.
    Building upon this model, DeFi operates similarly to narrow banking, where over-collateralization is the cornerstone for all lending activities. However, this method introduces additional systemic risks beyond credit defaults. One significant risk stems from investors taking on highly leveraged positions and the potential for runs on stablecoins. Stablecoins play a crucial role in the DeFi ecosystem by providing liquidity and a medium of exchange pegged to stable assets such as the USD. Yet, they are vulnerable to runs if investors doubt the quality or liquidity of the assets backing them. For stablecoins backed by traditional assets, a lack of transparency or timely information about reserves can prompt investors to redeem their holdings all together, similar to bank runs or money market fund crises, unless these stablecoins are fully backed by liquid assets like cash or short-term government securities. Proposed solutions to mitigate these risks include issuing stablecoins through insured banks, requiring one-to-one backing with safe assets, or establishing central bank digital currencies (CBDCs) \cite{gorton2023taming}.
    Furthermore, the ability to establish highly leveraged positions within DeFi introduces another layer of systemic risk. The crypto ecosystem is known for offering high leverage -- sometimes up to $100$ times -- for various derivative products. Such extreme leverage can amplify market volatility and has been linked to severe deleveraging cycles and sharp declines in cryptocurrency prices \cite{makarov2022cryptocurrencies}. DeFi exacerbates this issue by facilitating leverage through protocols that accept tokens from other protocols as collateral. 

    In TradFi, credit risk became evident during the Eurozone debt crisis (2010-2012), when countries like Greece struggled to repay their debts, leading to significant credit risk exposure for banks holding sovereign bonds. In DeFi the collapse of Celsius in 2022 highlighted the risks of over-collateralized crypto loans during market downturns \cite{pontem2022celsius}.

 \textit{Credit risk and over-collateralization occupy a middle-to-high scale of systemic importance -- especially if large stablecoins or heavily leveraged positions are involved. However, they typically do not escalate as instantly as a bank run or a sudden liquidity crisis. Over-collateralized lending and stablecoin backing failures can unfold over hours or days (e.g., liquidations triggered by price drops, runs on partially collateralized stablecoins).}

 \begin{tcolorbox}[colback=opgreen!10, colframe=opgreen!80!black, boxrule=0.5pt, arc=4pt, left=6pt, right=3pt, top=3pt, bottom=3pt]\textbf{Counterparty Default and Smart Contract Failures}
\end{tcolorbox}
    Counterparty default risk in traditional finance arises when a party involved in a financial transaction fails to meet their obligations, leading to potential losses for the other party. Lenders often mitigate this risk through due diligence and legal contracts. In contrast, DeFi aims to eliminate counterparty risk by leveraging decentralized protocols and smart contracts. To address the risk of user defaults, DeFi protocols typically implement automated liquidation mechanisms. If a borrower's collateral falls below a predefined threshold, smart contracts trigger a ``liquidation call'', enabling the sale of collateral to cover outstanding debts. This process ensures that creditors are repaid without the need for legal enforcement or arbitration. However, extreme market conditions, such as rapid price declines, can lead to insufficient collateral recovery, resulting in systemic risks for the protocol. To mitigate such scenarios, some DeFi platforms incorporate internal insurance mechanisms, such as reserve funds or liquidity pools, designed to absorb losses arising from defaults or protocol vulnerabilities. These insurance pools are funded by transaction fees, user contributions, or protocol treasury allocations and act as a buffer to protect users from catastrophic losses. 
    While this approach removes reliance on human discretion, it introduces a new form of risk: smart contract failure. Bugs, vulnerabilities, or malicious manipulations -- such as oracle attacks or poorly designed liquidation mechanisms -- can result in significant financial losses, mimicking the impact of a traditional counterparty failing to fulfill their obligations. 

  \textit{Counterparty default and smart contract failures typically originate as localized events, but their speed of contagion and scale of risk can vary based on the interconnectedness of the protocols involved. }

 \begin{tcolorbox}[colback=techorange!10, colframe=techorange!80!black, boxrule=0.5pt, arc=4pt, left=6pt, right=3pt, top=3pt, bottom=3pt]\textbf{Technological Failures and Cybersecurity Threats}
\end{tcolorbox}

    System outages, data breaches, and cyber-attacks pose significant risks in traditional finance, disrupting services and compromising sensitive information. DeFi faces similar challenges, compounded by the immutable nature of smart contracts. Once deployed, smart contracts cannot be easily altered, and any vulnerabilities can be exploited by hackers. Additionally, issues like network congestion on blockchains can delay transactions, impacting users' experience and trust. The reliance on technology in DeFi makes robust cybersecurity measures and efficient network infrastructure critical.

  \textit{  Technological failures and cybersecurity threats are often localized events that affect specific institutions or platforms. However, in highly interconnected systems like DeFi, a major exploit or outage can propagate quickly through composable protocols, making their speed of contagion moderate to high. The scale of risk varies but can escalate to systemic levels if critical infrastructure is compromised.}

 \begin{tcolorbox}[colback=techorange!10, colframe=techorange!80!black, boxrule=0.5pt, arc=4pt, left=6pt, right=3pt, top=3pt, bottom=3pt]\textbf{Limit Order Books (LOBs) and Automated Market Makers (AMMs)}
\end{tcolorbox}
    LOBs and AMMs are two distinct mechanisms used to facilitate trading in financial markets. A Limit Order Book is a traditional system employed by centralized exchanges where buy and sell orders are listed with specified prices and quantities. Traders place limit orders, and the exchange matches these orders based on price and time priority, allowing for direct transactions between buyers and sellers at agreed-upon prices \cite{briola2024deep}. 
    Blockchain-based decentralized exchanges fall into two broad categories: decentralized limit order books where an order is a smart contract registered on the blockchain, and swap ex-changes where prices are set by a deterministic automated market making rule.
    Automated Market Makers rely on liquidity pools funded by users who deposit their assets, and trades are executed against these pools using mathematical -- deterministic -- formulas to determine asset prices. The most common form of the latter is the constant product rule where relative prices of crypto assets are determined by iso-liquidity curves. Park (2021) \cite{park2021conceptual} found that, although this pricing rule is simple, its use could give rise to persistent arbitrage opportunities when there are multiple competing trading systems.
    While both systems aim to provide liquidity and efficient price discovery, they differ fundamentally in operation: LOBs match individual orders to facilitate trades, whereas AMMs use algorithms and pooled liquidity to enable continuous trading without the need for direct order matching.

   \textit{ Limit Order Books (LOBs) and Automated Market Makers (AMMs) have risks that are generally localized but can escalate under specific circumstances (e.g., arbitrage exploits, liquidity drains). Their speed of contagion is moderate to high in DeFi due to the automation and transparency of AMMs, while in TradFi, LOBs operate at a slower pace due to centralized oversight and time-based order matching.}

     \begin{tcolorbox}[colback=gray!10, colframe=gray!80!black, boxrule=0.5pt, arc=4pt, left=6pt, right=3pt, top=3pt, bottom=3pt]\textbf{Trading}
\end{tcolorbox}
The idea of using software code to represent and execute contracts is not a recent innovation. For instance, when using an online brokerage platform, a limit order to buy stocks at a certain price is executed automatically by a software system. Such automated agreements dominate financial and e-commerce sectors due to their speed and efficiency. However, in traditional systems, parties can still seek legal redress in case of disputes, as for instance if incorrect information was used in the transaction.
    The economic distinction between traditional electronic contracts and blockchain-based smart contracts lies in the execution and enforcement mechanisms. In a blockchain-based system, once a contract is deployed and executed, the distributed verification mechanism it relies on prevents any unilateral reversal unless explicitly programmed in advance. Moreover, due to the (pseudo)anonymity of public blockchain networks, parties involved in a transaction are often unidentifiable, making it virtually impossible to assign legal accountability or serve formal legal notice.

    Smart contracts autonomously execute the transaction once predefined conditions are met, reducing the likelihood of a party reneging on the agreement post-execution. This characteristic limits flexibility, but ensures adherence. However, this rigidity removes the possibility for efficient breach, such as in the case of a mutual mistake. For example, if a seller finds the asset's value to be far greater than agreed just before delivery, a traditional contract might allow renegotiation. In contrast, smart contracts will execute regardless, as they cannot account for unplanned contingencies.
    As an example, one significant risk that spans both traditional and decentralized finance is front-running. In TradFi, front running involves profiting from advanced knowledge of a future trade. For example, if a broker knows that a large buy order is about to move the market, they might place their own order just ahead of it to take advantage of the price movement \cite{cai2003there}. This concept has found a parallel in DeFi, where front-running occurs when a trader monitors an impending transaction on the blockchain and places their own trade ahead of it. This is often possible by exploiting the transparency of the transaction pool (mempool), where trades are temporarily stored before being confirmed. Since the mempool is publicly accessible, traders with the right tools can anticipate transactions and manipulate the order to profit from price changes, potentially harming the original trade.

    Malinova et al. (2024) \cite{malinova2024learning} propose a framework to evaluate the viability of automated market makers (AMMs) in equity markets and assess their potential to outperform traditional market-making mechanisms. Using empirical data from U.S. equity trading, they estimate that well-designed AMMs could save U.S. investors billions of dollars annually. These gains stem from the distinctive way AMMs manage risk: unlike traditional market makers, who require compensation for a broad array of inventory and informational risks, AMM liquidity providers are primarily exposed to short-term intraday risks and can be compensated accordingly. In addition, AMMs improve capital efficiency by continuously deploying locked funds to provide liquidity -- capital that would otherwise remain idle in traditional brokerage accounts. Notably, this model offers particular advantages for smaller firms, as it lowers barriers to entry for liquidity provision and helps attract a broader base of investors through a more efficient and accessible infrastructure.

  \textit{  In DeFi, mempool transparency and automated market markers lead to faster risk escalation, while TradFi’s institution-driven execution slows contagion. These risks are typically localized to specific markets or protocols, but can spill over into connected markets if a critical entity fails, amplifying their impact. Trading risks are best positioned in the mid-right quadrant of the plot, reflecting a moderate scale of risk and a moderate to high speed of contagion, particularly in the automated and interconnected DeFi ecosystem.}

\begin{tcolorbox}[colback=opgreen!10, colframe=opgreen!80!black, boxrule=0.5pt, arc=4pt, left=6pt, right=3pt, top=3pt, bottom=3pt]\textbf{Settlement Risk and Transaction Finality}
\end{tcolorbox}

    In traditional finance, settlement delays can expose counterparties to price volatility and credit risk. In DeFi, transactions are settled directly on-chain and require network confirmations to achieve finality. However, issues such as blockchain reorganizations -- i.e., temporary forks in the chain where recent blocks are replaced by an alternative version of the ledger -- can delay or reverse recent transactions. Combined with network congestion or latency, these events can create settlement uncertainty or lead to failed trades. Ensuring timely and secure settlement is, therefore, essential to maintaining trust, reliability, and efficiency in decentralized financial markets.

    \textit{Settlement risk and transaction finality are generally localized, tied to specific transactions or protocols, but can escalate if they affect critical infrastructure, such as major blockchains or stablecoin networks. In DeFi, issues such as blockchain congestion or reorganization can propagate more quickly in automated environments, while TradFi settlement risks spread more slowly due to manual processes and safeguards as clearinghouses. These risks typically have a low to moderate scale, remaining localized unless they disrupt widely used systems such as Ethereum or SWIFT, which can cause broader ripple effects. }

\begin{tcolorbox}[colback=gray!10, colframe=gray!80!black, boxrule=0.5pt, arc=4pt, left=3pt, right=3pt, top=3pt, bottom=3pt]\textbf{Fraud and Manipulation}
\end{tcolorbox}
    Traditional finance is no stranger to fraudulent activities, including insider trading, Ponzi schemes, and manipulation of financial statements. DeFi, while decentralized, is also vulnerable to fraudulent practices. Common scams include rug pulls, where developers abandon a project and abscond with investors' funds, and pump-and-dump schemes involving low-liquidity tokens \cite{xu2019anatomy}. Additionally, DeFi is susceptible to oracle manipulation attacks, where malicious actors exploit price feeds to manipulate asset prices to their advantage. These forms of fraud highlight the need for transparency, a stricter regulation, and investors' education within the DeFi space.

     In TradFi, Enron Corporation, an American energy, commodities, and services company, manipulated its financial statements to hide billions in debt in 2001, leading to one of the largest corporate collapses in history. Similarly, Bernie Madoff’s $\$65$ billion Ponzi scheme during the 2008 financial crisis illustrates how insider manipulation and deception can devastate investors and undermine market trust. In DeFi, newer schemes such as rug pulls and pump-and-dump tactics exploit the trust of retail investors. An example of rug pull is the \textit{Squid Game Token} scam in 2021: developers launched a token inspired by the popular Netflix series, which gained significant media attention. Once the token's value skyrocketed, the developers sold their holdings, crashing the price to near zero and fleeing with over $\$3$ million. A pump-and-dump scheme happened with \textit{SaveTheKids Token} (2021): promoted by influencers as a charitable project, the token's price was first artificially inflated through coordinated buys and then dumped, leaving retail investors with significant losses.

    \textit{Fraud and manipulation risks can spread rapidly in DeFi due to the automated nature of smart contracts and the interconnectedness of protocols. In TradFi, these risks propagate more slowly due to regulatory oversight and manual processes. Most incidents remain localized to specific projects or protocols, but when they involve systemically important assets -- such as major stablecoins or liquidity providers -- the fallout can ripple across the ecosystem. The risk is typically low to moderate in scale but can be amplified by leverage, liquidity crises, or high protocol dependency.}

\begin{tcolorbox}[colback=gray!10, colframe=gray!80!black, boxrule=0.5pt, arc=4pt, left=3pt, right=3pt, top=3pt, bottom=3pt]\textbf{Market Manipulation and Flash Loan Attacks}
\end{tcolorbox}

Market manipulation in traditional finance often involves influencing asset prices through large trades, coordinated activity, or the spread of false information. In DeFi, manipulation can take novel forms, such as via so-called flash loan attacks -- a mechanism unique to blockchain-based systems.

First introduced in 2020 on the lending protocol Aave, a flash loan allows a borrower to access large amounts of liquidity (up to the size of the lending pool) without collateral, provided the loan is repaid within the same blockchain transaction. If repayment does not occur, the transaction is automatically rolled back: this feature eliminates risk for the lender because if the loan is not repaid as intended, it is as if the transaction never happened. Protocols offering flash loans earn revenue by charging small fees on each transaction (e.g., 0.07\% in Aave V2, 0.05\% in Aave V3 \cite{aave2025flashloans}). These fees allow the protocol to generate revenue, while maintaining a low-risk environment for its liquidity providers.

    Arbitrageurs utilize flash loans to exploit price discrepancies across different platforms or markets. For example, if a cryptocurrency is listed at different prices on two exchanges, they can first borrow a large amount of that cryptocurrency through a flash loan, then purchase the cryptocurrency where the price is lower, immediately sell it in the venue where the price is higher, repay the loan (and the interests) within the same transaction, and retain the profit generated from the arbitrage strategy.
    Since everything occurs within a single transaction, there is no need for the arbitrageurs to have their own capital or provide collateral. Additionally, the risk is minimal for both parties: the lender is assured of getting their funds back, and the arbitrageur can operate without financial exposure.

    Malicious actors use flash loans to manipulate the price of the market, ultimately resulting in the theft of assets. Ever since their introduction, flash loans have become increasingly exploited in DeFi attacks. Cao et al (2021) \cite{cao2021flashot} identified nine separate instances between February and December 2020 in which attackers successfully siphoned a total of $\$49.58$ million from DeFi protocols through flash loan-assisted exploits. The largest of these, the Harvest Attack in October 2020, saw the attackers extract $\$26$ million from Harvest, using the Curve and Uniswap protocols while relying on a flash loan from Uniswap v2 \cite{carter2021defi}. 
    In 2021, a flash loan attack caused the value of the token Bunny to drop over $95\%$ -- from $\$146$ to $\$6.17$ --, and the attacker profited $ \$3$ million \cite{crawley2021}. These attacks are becoming a serious problem in cryptocurrency and are increasing yearly. In 2021, attackers gained over $ \$3.2$ billion in various attacks, hacks, and scams. In 2022, this value raised to $ \$3.7$ billion.

 \textit{   In systemic risk terms, market manipulation and flash loan attacks typically originate as highly localized events, targeting specific protocols or asset pools. However, their speed of contagion in DeFi is extremely high due to blockchain transparency, composability, and automated execution. The potential scale of risk may depend on the protocol or assets involved. In contrast, manipulation in TradFi tends to unfold more gradually, though it can still have widespread consequences when large institutions or critical markets are affected.
}

 \begin{tcolorbox}[colback=interyellow!10, colframe=interyellow!80!black, boxrule=0.5pt, arc=4pt, left=3pt, right=3pt, top=3pt, bottom=3pt]\textbf{Capital reinforcement loops}
\end{tcolorbox}
    Capital reinforcement loops refer to a self-perpetuating dynamic where an entity’s ability to raise or deploy capital feeds directly into asset price inflation, which in turn enhances that entity’s capacity to secure additional funding. 

   For instance, in traditional finance, Strategy’s inclusion in the NASDAQ 100 index prompted index-tracking funds to purchase its stock, driving up the share price. This higher valuation then enabled the company to raise additional capital through debt, equity, or bond offerings, which it used to purchase more Bitcoin -- further fueling upward pressure on BTC’s price \cite{reuters2024microstrategy, aufiero2025cryptocurrencies}.The simultaneous rise in both Strategy’s stock and its Bitcoin holdings inflated the firm’s perceived value, further attracting investment and reinforcing the cycle.

    A parallel phenomenon emerged in decentralized finance through FTX’s handling of the Serum (SRM) token \cite{bloomberg2022ftx}. 
    In that scenario, FTX and its affiliate Alameda Research created a token supply, allocated the majority of tokens to themselves, and sold only a small fraction (approximately $3\%$) into the public market. Because the token’s price -- determined by the relatively small publicly traded float -- was extrapolated to the entire token supply, the resulting fully diluted market capitalization\footnote{ \textit{Market cap} is the total market value of a cryptocurrency's circulating supply, while \textit{fully diluted market cap} is the market cap if the maximum supply was in circulation. As stated by Matt Levine in \cite{bloomberg2022ftx}: ``if for instance some company creates a token, and says that there can be $10$ billion of the token, and reserves them all for itself, and then sells $1$ million of them to outside investors for $\$1$ each, then the market cap of that token is $\$1$ million ($\$1$ times 1 million circulating tokens), while the fully diluted market cap is $\$10$ billion ($\$1$ times $10$ billion total tokens), and the issuer’s $9,999,000,000$ remaining tokens have a value, on this math, of $\$9.999$ billion.''} on paper was vastly inflated. Importantly, FTX and Alameda did not purchase these tokens on the open market using cash. Rather, as initial backers of the Serum protocol, they effectively received the vast bulk of their holdings for free (beyond some startup costs). 
    This arrangement generated a self-reinforcing loop: the large reported valuation of Serum (up to $\$5.4$ billion on FTX’s balance sheet at one point) provided an appearance of robust financial health. However, in practical terms, offloading this enormous stash of tokens -- around two-thirds of the $97\%$ of SRM that FTX and Alameda controlled -- would likely have crashed the market price. The discrepancy between the theoretical value of these locked-up tokens and their realizable market value illustrates the risk of such self-referential valuations in crypto markets: if market conditions reverse, the chain reaction can be severe, exposing the fragility of a financial edifice built primarily on reflexive demand, leverage, and speculative enthusiasm.

    \textit{This risk has a high contagion speed, as feedback loops can rapidly amplify asset price movements and market instability -- especially in volatile environments or when key players are involved. While it may originate locally (e.g., within one institution or protocol), its systemic repercussions can be far-reaching.}

 \begin{tcolorbox}[colback=interyellow!10, colframe=interyellow!80!black, boxrule=0.5pt, arc=4pt, left=3pt, right=3pt, top=3pt, bottom=3pt]\textbf{Leverage}
\end{tcolorbox}    DeFi is characterised by the high leverage that can be sourced from lending and trading platforms. While loans are typically overcollateralized, funds borrowed in one instance can be re-used to serve as collateral in other transactions, allowing investors to build increasingly large exposure chains for a given amount of collateral. Derivatives trading on DEXs also involves leverage, as the agreed payments take place only in the future. The maximum permitted margin in DEXs is higher than in regulated exchanges in the established financial system \cite{aramonte2021defi}. And unregulated crypto centralized exchanges allow even higher leverage.

    High leverage in crypto markets exacerbates procyclicality. By enabling investors to control larger positions with limited initial capital, leverage amplifies gains in rising markets -- but also magnifies losses when prices fall. When debt must be reduced, whether due to investment losses or depreciating collateral, investors are forced to liquidate positions, exerting further downward pressure on prices. This self-reinforcing dynamic is often stronger in early-stage markets, where speculative trading dominates and liquidity is thinner. In DeFi, the problem is compounded by the high degree of interconnectedness between protocols, which can transmit stress rapidly across the ecosystem.

    Financial intermediation in DeFi relies exclusively on collateral to mitigate risk and enable transactions when participants cannot trust each other. Thus, there are no shock absorbers in DeFi that can cut in during stress periods. By contrast, in traditional finance, banks are elastic nodes that can expand their balance sheets (extending loans or purchasing distressed assets) via the issuance of bank deposits, which are a widely accepted medium of exchange \cite{borio2019money}.
    The destabilising role of leverage came to the fore in a cryptoasset crash in September 2021 \cite{ponciano2021}. Forced liquidations of derivatives positions and loans on DeFi platforms accompanied sharp price falls and spikes in volatility.

    \textit{Leverage risks in DeFi can lead to rapid contagion, as margin calls and forced liquidations escalate quickly when collateral values drop, though not instantaneously. In a volatile crypto environment, unwinding leverage can trigger a rapid spiral. These risks have systemic potential, as excessive leverage impacts entire markets, especially when multiple protocols use the same collateral and are interconnected via yield strategies. A significant deleveraging event can create downward pressure on prices across the ecosystem, leading to a system-wide crisis.}

 \begin{tcolorbox}[colback=marketblue!10, colframe=marketblue!80!black, boxrule=0.5pt, arc=4pt, left=3pt, right=3pt, top=3pt, bottom=3pt]\textbf{Market Crashes and Volatility Spikes}
\end{tcolorbox}
    Market crashes in traditional finance are characterized by significant declines in asset prices, often triggered by economic downturns, geopolitical events, or the bursting of financial bubbles. These crashes can lead to substantial losses for investors and can have long-term effects on economic stability. In trading, sharp price declines trigger widespread sell-offs as traders rush to exit positions, creating downward pressure on prices. Additionally, margin calls can force leveraged traders to liquidate their holdings, further amplifying the sell-off and exacerbating market instability.
    DeFi markets, given the inherent high volatility of cryptocurrencies, are particularly susceptible to sharp price swings. Automated liquidation mechanisms on DeFi lending platforms can exacerbate market downturns; as asset prices fall, collateralized positions may be liquidated, further driving down prices in a feedback loop. This self-reinforcing mechanism can lead to more pronounced and rapid market crashes compared to TradFi, underscoring the need for robust risk management strategies in DeFi platforms.
    
  \textit{  Because rapid price swings can trigger forced liquidations that feed back into deeper selling pressure -- especially in DeFi -- market crashes and volatility spikes tend to escalate quickly and impact many market participants simultaneously.}

 \begin{tcolorbox}[colback=marketblue!10, colframe=marketblue!80!black, boxrule=0.5pt, arc=4pt, left=3pt, right=3pt, top=3pt, bottom=3pt]\textbf{ Liquidity crises }
\end{tcolorbox}
    Liquidity crises occur in traditional finance when financial institutions face a shortage of liquid assets to meet short-term obligations, often leading to a credit crunch and broader economic instability. In the DeFi context, liquidity can evaporate rapidly if liquidity providers withdraw their funds or if there is a sudden spike in demand for withdrawals. Limited liquidity can result in high slippage for traders and can compromise the stability of DeFi platforms. The decentralized nature of DeFi means that liquidity is often provided by individual users rather than institutions, making the system more susceptible to sudden liquidity fluctuations.

  \textit{  Given that liquidity may evaporate extremely fast (e.g., when liquidity providers collectively pull funds or there is a rush to withdraw in DeFi) and that the impact spread throughout the system (leading to broader market instability or credit-style crunches), liquidity crises are placed toward the upper-right portion of the speed of contagion/ scale of risk space.}

 \begin{tcolorbox}[colback=interyellow!10, colframe=interyellow!80!black, boxrule=0.5pt, arc=4pt, left=3pt, right=3pt, top=3pt, bottom=3pt]\textbf{Protocol Interdependence }
\end{tcolorbox}
    The interconnectedness of financial institutions in traditional finance can lead to systemic risks, where the failure of one institution can trigger cascading failures, as seen during the 2008 financial crisis. In DeFi, protocols often build upon each other ( following the ``composability'' principle). While this allows for the creation of innovative financial products and services, it also means that a failure in one protocol can have ripple effects across others. For example, if a widely used stablecoin loses its peg, it can destabilize multiple platforms and protocols that rely on it, highlighting the need for careful risk assessment and diversification.

\textit{    Protocol interdependence is intrinsically a  systemic vulnerability, as it directly reflects the interconnected nature of both TradFi and DeFi systems. Failures in one protocol or institution often propagate through dependent entities, creating cascading risks. The speed of contagion, however, is typically higher in DeFi due to the composability of smart contracts and the automated execution of dependencies.}

    \begin{tcolorbox}[colback=marketblue!10, colframe=marketblue!80!black, boxrule=0.5pt, arc=4pt, left=3pt, right=3pt, top=3pt, bottom=3pt]\textbf{Runs and Liquidity Pool Drains }
\end{tcolorbox}
    
    In traditional finance, a bank run occurs when a large number of customers withdraw their deposits simultaneously due to concerns about the bank's solvency, potentially leading to the bank's collapse. This phenomenon reflects a loss of confidence in financial institutions, and can have widespread economic repercussions. Within the DeFi ecosystem, a similar event manifests when users rapidly withdraw their assets from liquidity pools or lending protocols. Such mass withdrawals are often triggered by fears of smart contract vulnerabilities, hacks, or rapid declines in asset values. For example, if rumors spread about a potential exploit in a DeFi protocol, users may rush to withdraw their funds to mitigate losses, thereby draining liquidity and destabilizing the protocol. This reaction mirrors the panic-induced withdrawals in TradFi, highlighting the psychological factors influencing investor behavior across both systems.

\textit{    Bank runs and liquidity pool drains tend to escalate quickly once fear sets in -- whether in a TradFi bank run or a DeFi protocol facing a sudden rush of withdrawals. If enough participants are involved, the fallout can spread broadly, undermining confidence and triggering secondary failures across the system.}

\vspace{1cm}
To summarize, the mapping in Fig. \ref{fig:mapping} shows clear groupings of risks, each with distinct patterns of contagion speed and systemic impact.

The \textbf{blue cluster} covers market risks, which exhibit varying speeds of contagion but share a systemic nature when they escalate. Liquidity crises and bank runs are positioned at the upper-right corner, reflecting their rapid escalation and far-reaching systemic implications. Market crashes and volatility spikes, while slightly slower, similarly propagate widely due to their impact on investor behavior and market stability. Interest rate risk, on the other hand, is more slowly-moving and predominantly localized, emphasizing its more contained, yet critical, role in financial ecosystems.

The \textbf{red cluster} -- regulatory and macroeconomic risks -- typically evolves slowly but may reshape entire financial ecosystems over time, placing them firmly in the upper-left ``slow but systemic” quadrant.

The \textbf{green cluster} represents operational and governance risks. These risks are positioned closer to the local and moderate-contagion region, reflecting their typically isolated origins. However, as discussed, risks such as counterparty default may occasionally propagate more broadly if critical protocols or smart contracts are involved. 

The \textbf{orange cluster}, representing technological risks,falls in the mid-range for both speed and scale, with moderate-to-high speeds of contagion depending on the scope of the vulnerability or infrastructure affected, and a moderate scale of risk. 

The \textbf{grey cluster} represents manipulation and exploitation risks, and is positioned in the lower right part of the scheme, reflecting their moderate-to-high speed of contagion which remains predominantly localized nature. 

Finally, \textbf{the yellow cluster} -- representing interconnectedness risks -- occupies the middle-to-upper-right quadrant, reflecting both their systemic nature and moderate-to-high speed of contagion. 

Taken together, the figure highlights how localized risks can serve as triggers, cascading into systemic crises through amplifiers. Moreover, the plot highlights how the dynamics of slow-moving systemic risks (e.g., regulatory changes) contrast with fast-moving but localized threats (e.g., manipulation).

\section{DeFi-TradFi Contagion: \textit{Crosstagion}}\label{sec:results2}

\begin{tcolorbox}[colback=cyan!10, colframe=cyan!80!black, boxrule=0.5pt, arc=4pt, left=6pt, right=6pt, top=6pt, bottom=6pt]
\textbf{Definition: \textit{Crosstagion}} \vspace{0.5cm}\\
\textit{Crosstagion} refers to the bidirectional transmission of financial instability between decentralized finance (DeFi) markets and traditional finance (TradFi) systems. This phenomenon captures how disruptions in DeFi -- such as liquidity crises, protocol failures, or stablecoin depegging -- can propagate into TradFi through interconnected assets, shared intermediaries, and overlapping markets. Conversely, disturbances originating in TradFi -- such as regulatory shifts, macroeconomic shocks, or central bank policy changes -- can trigger cascading effects within DeFi ecosystems. This two-way transmission highlights the growing interdependence between the two systems, where vulnerabilities in one sphere can amplify risks in the other.
\end{tcolorbox} 
Until 2021, DeFi operated largely in isolation from the traditional financial system.  Conservative regulatory frameworks limited banks' crypto participation, restricting their activities primarily to modest equity investments in crypto-related firms \cite{basel2021crypto}. Even as late as mid-2022, the integration between cryptocurrencies and traditional banking institutions remained minimal. Andrea Enria, then Chair of the ECB Supervisory Board, emphasized to the European Parliament in June 2022, that there were ``still very limited'' connections between cryptocurrencies and banks \cite{nasdaq2022crypto}. Supporting this observation, Jeff Berman from US Financial Services Regulatory Group noted that ``banks do not hold crypto, and they have been very careful about lending against crypto. In fact, most crypto lending has been conducted by crypto specialists, resulting in overall low exposure for banks'' \cite{nasdaq2022crypto}. Consequently, at this stage, the crypto ecosystem was deemed unlikely to pose systemic risk to traditional financial institutions or the broader financial system.

The first U.S. banks that have actively pursued clients in the DeFi space are Silvergate Bank, Signature Bank, and Metropolitan Community Bank \cite{carter2021defi}.
Among them, Silvergate Bank became a key nexus connecting traditional banking and the digital currency industry, boasting $\$5.5$ billion in total assets and $\$5.03$ billion in cryptocurrency deposits as of Q4 2020. By the end of 2021, Silvergate's total assets had surged to over $\$16$ billion \cite{federalreserve2023silvergate}. Yet, this rapid expansion was not free from vulnerabilities; in Q4 2022, Silvergate's deposits from digital asset clients plummeted from $\$11.9$ billion in September to $\$3.8$ billion by December, driven by market turmoil and declining confidence. Ultimately, in March 2023, Silvergate announced its intention to voluntarily liquidate \cite{federalreserve2023silvergate},  highlighting the volatility and inherent risks of the sector.

Since 2022, the interaction between DeFi and the traditional financial system has expanded markedly, driven by a series of significant developments. The New York Stock Exchange (NYSE), the world's largest stock exchange, had an equity market capitalization exceeding $\$30$ trillion as of September 2024 \cite{statista2025largestexchanges}. By comparison, the global cryptocurrency market, with a capitalization of $\$3.44$ trillion by the end of 2024, now represents $12\%$ of the NYSE's size. This valuation surpasses the GDP of major economies such as Italy ($\$2.38$ trillion), Canada ($\$2.21$ trillion), and Brazil ($\$2.19$ trillion), and is just $\$0.06$ trillion shy of overtaking France’s GDP of $\$3.17$ trillion \cite{economictimes2025cryptomarket}.

The rising correlation between crypto asset prices and mainstream risky financial assets during periods of market stress challenges their effectiveness as tools for portfolio diversification. Notably, the correlation between crypto asset and stock returns increased significantly during the market turmoil of March 2020, as well as during the sell-offs in December 2021 and May 2022. This pattern suggests that during times of heightened risk aversion in broader financial markets, crypto assets have become more interconnected with traditional risk assets -- a trend likely influenced, at least partly, by the growing participation of institutional investors in crypto markets \cite{iyer2022cryptic, mungo2024cryptocurrency}.
Numerous influential studies have employed various models to explore the interactions and relationships between cryptocurrencies and traditional financial assets. Recent empirical investigations and analytical approaches, including those by Cao and Xie (2022) \cite{xie4769838dynamic}, Zhang et al. (2021) \cite{zhang2021risk}, Wang et al. (2022) \cite{wang2022dynamic}, Scagliarini et al. (2022) \cite{scagliarini2022pairwise}, Bendob et al. (2022) \cite{fund2022understanding}, and Aufiero et al. (2025) \cite{aufiero2025cryptocurrencies} emphasize the growing interconnectedness between cryptocurrency markets and traditional financial systems, pointing to a complex interdependence between the two.
These studies suggest that the relationship between cryptocurrency and the stock market is not consistent over time. For instance, a significant positive correlation between Bitcoin and risk assets has been observed during extreme market events, such as the Covid-19 pandemic, while in the case of Middle East countries, the relationship varies across regions. Studies also reveal that this relationship is highly complex, driven by asymmetric and time-varying risk spillovers. The analysis in \cite{xie4769838dynamic} of China’s financial market highlights that the spillover effect from cryptocurrencies to the financial market is stronger than the reverse.
Pi\~neiro et al. (2022) \cite{pineiro2022preliminary} measure the relationship among the returns of DeFi tokens and traditional assets, finding that DeFi acts, similar to other crypto assets, as a safe haven. 
In the study by Ugolini et al. (2023) \cite{ugolini2023connectedness}, they find that DeFi and cryptocurrency asset markets demonstrate strong returns spillovers both within their own markets and between each other, whereas stock and safe-haven asset markets exhibit relatively weak interconnectedness. Safe-haven assets are identified as minor participants in spillover effects, acting as limited receivers and transmitters of such dynamics.
Pacelli et al. (2024) \cite{pacelli2024cryptocurrencies} investigates the relationship between the cryptocurrency market and the global equity indexes considering the bearish and bullish market conditions in both markets.
Zhou (2024) \cite{zhou2024cryptocurrency} emphasize the dual role of cryptocurrencies in being vulnerable to risk contagion from traditional markets while also acting as a source of risk for those markets. Mainstream cryptocurrencies such as Bitcoin and Ethereum demonstrate strong influence and relative stability but also carry significant risks, whereas stablecoins such as Tether and USD Coin provide hedging functions but remain tied to fiat currency constraints. Emerging cryptocurrencies such as Dogecoin and Solana are marked by high innovation and growth potential, but are heavily influenced by speculation and market sentiment.

Interconnectedness with the wider financial system has also been growing. Major payment networks have also stepped up their support of crypto asset services, leveraging their retail networks and making crypto assets more easily accessible to consumers and businesses alike \cite{ecb2022stability}. 
Traditional financial institutions have shown a growing interest in DeFi and crypto assets. Major banks and asset management firms have started exploring blockchain technology for asset tokenization, settlement processes, and offering crypto-related services to clients. For instance, some banks have begun to provide custody services for digital assets, while asset managers are considering or actually including crypto assets as part of diversified portfolios.
Regulatory bodies worldwide have intensified their focus on DeFi and the broader crypto ecosystem to mitigate potential systemic risks. The U.S. Securities and Exchange Commission (SEC) and the Commodity Futures Trading Commission (CFTC) have increased enforcement actions and proposed new regulations targeting crypto exchanges, stablecoins, and DeFi platforms. In the European Union, the Markets in Crypto Assets (MiCA) regulation aims to create a regulatory framework for crypto assets and related services.
Following incidents such as the collapse of TerraUSD (UST) in May 2022, stablecoins have come under heightened scrutiny. Regulators are now more concerned about the reserves backing these digital assets, and their potential impact on financial stability \cite{ecb2022stability}. This has ultimately led to discussions about implementing stricter reserve requirements and transparency standards for stablecoin issuers.

Central banks have accelerated research and pilot programs for Central Bank Digital Currencies (CBDCs) \cite{auer2020rise}, which could further bridge the gap between traditional finance and digital assets. 
As of September 2024, 134 countries, representing $98\%$ of the global economy, are exploring CBDCs, with 44 nations in advanced development stages \cite{reuters2024cbdc}. China continues to lead in CBDC development with its digital yuan (e-CNY). By June 2024, e-CNY transactions reached 7 trillion yuan (approximately \$986 billion), a substantial increase from the previous year \cite{atlanticcouncil2025cbdctracker}. The People's Bank of China (PBOC) has expanded trials to 26 areas across 17 provinces, integrating the digital yuan into sectors such as education, healthcare, and tourism \cite{krasia2025digitalyuan}. 
In Europe, the European Central Bank is progressing with its digital euro initiative. The ECB has entered a multi-year pilot phase, focusing on both retail and wholesale CBDC models to enhance the efficiency and security of digital payments across the Eurozone \cite{atlanticcouncil2025cbdctracker}. Similarly, the Bank of England is exploring a digital pound, considering its potential for large-scale transactions between financial institutions \cite{fnlondon2025cbdc}. 
In contrast, the United States has taken a more cautious approach, because political developments have slowed progress toward a digital dollar.
The International Monetary Fund (IMF) has noted that while interest in CBDCs remains strong, central banks are increasingly aware of the complexities involved. Issues such as cybersecurity, data privacy, and the implications for monetary policy operations are central to ongoing discussions \cite{ft2025cbdc}. 
There has been a greater degree of collaboration between international regulatory bodies to address the global nature of DeFi and crypto markets. Organizations such as the Financial Stability Board (FSB) and the International Organization of Securities Commissions (IOSCO) are working on frameworks to manage cross-border risks and standardize regulations.

Traditional non-bank investors, such as family offices and hedge funds, started to be increasingly interested in DeFi and the broader crypto space, and their crypto assets have grown significantly. Funds with meaningful crypto exposure, some of which focus exclusively on DeFi while others are more diversified, have increased their assets from about $\$5$ billion in 2018 to about $\$50$ billion in 2021 \cite{aramonte2021defi}. 
Frolov et al. (2024) \cite{frolov2024interaction} considered how DeFi interacts with the traditional banking system and assessed the impact of DeFi on financial stability, regulatory policy, and the global financial ecosystem. Their comprehensive analysis of financial markets in the United States, India, and the United Kingdom provides a detailed understanding of interdependencies and dynamics in these economies.
The findings emphasize the significant influence of monetary policy, credit conditions, inflation, and Total Value Locked (TVL) in DeFi on shaping the financial systems of these nations. While each country’s economic landscape and regulatory structure differ, the analysis reveals common threads in how these factors drive market behavior. Despite variations in economic contexts, the core mechanisms governing financial dynamics are remarkably interconnected across different global markets. For policymakers, analysts, and investors, the study stresses the necessity for a multifaceted approach.

In a recent review of November 2024 \cite{azar2024digitalassets}, the Federal Reserve Bank of New York highlighted the growing risk of overflow within markets, warning that as crypto assets become more entangled with TradFi, price collapses could create stronger spillovers, exacerbating systemic risk.
European Central Bank officials echoed this sentiment, comparing Bitcoin to a speculative bubble destined to burst \cite{bindseil2024distributional}.
Adding to this dynamic is the paradox of crypto’s growing institutionalization. An industry originally built on the ethos of decentralization and resistance to state control now finds itself bolstered by state-backed institutional adoption, such as endorsements from the US government. This has created an ironic situation where the prices of crypto assets are driven upward by the very establishment they sought to avoid. This institutional embrace, while providing legitimacy, also integrates crypto markets with the very systemic vulnerabilities they were designed to escape, deepening the risks to global financial stability \cite{ft2025article}.

\subsection{The Covid-19 Crisis: a Case Study}

The Covid-19 crisis had profound repercussions on global financial markets, sparking extensive research into both TradFi and DeFi sectors. Given its unprecedented scale, the pandemic serves as an ideal case study for examining contagion dynamics across diverse markets.

In the following, we first examine the effects of Covid-19 on traditional finance, summarizing the most significant findings from existing research. We then turn to its interconnectedness with the DeFi space, capturing the full extent of market contagion triggered by this global shock.

Rizwan et al (2020) \cite{rizwan2020systemic} find that Covid-19 has slowed down the global economy, leading financial institutions to experience heightened liquidity risks, increased loan defaults, and reduced intermediation revenues. Due to the interconnectedness among financial institutions, individual issues have spread across the network, amplifying stress within the financial system. They show a significant rise in systemic risk during the Covid-19 period for the analyzed countries (Canada, China, France, Germany, Italy, Spain, UK, USA).
Duan et al (2021) \cite{duan2021bank}, using a sample of $1,584$ listed banks from 64 countries during the Covid-19 pandemic, showed that systemic risk increased across countries as well. The study reveals that this heightened risk operates through channels such as government policy responses and bank default risks. Furthermore, the adverse effects on systemic stability were particularly pronounced for large, highly leveraged, riskier banks with high loan-to-asset ratios, low capitalization, and low network centrality. However, the negative impact was mitigated by factors such as formal bank regulation (e.g., deposit insurance), ownership structure (e.g., foreign and government ownership), and informal institutions (e.g., culture and trust).
Lai et al. (2021) \cite{lai2021study} estimate the Granger causalities of stock markets for 20 different countries from August 2019 to March 2020 and establish a network for global stock markets based on this data. The results show that Covid-19 strengthened financial connections between countries, causing the impact to spread over shorter distances and accelerating the transmission of the crisis. By comparing the topology of the financial network and centrality during stabilization and fluctuation periods, financial risks can be identified: topological structure and centrality analysis of networks can effectively measure and warn of systemic risks.

In March 2020, Ethereum (ETH) experienced a significant crash coinciding with the global financial turmoil induced by the Covid-19 pandemic. Between March 12 and March 13, crypto assets registered a price drop across the board up to 60\%. ETH, in particular, dropped from 193\$ all the way down to 95\$ in less than 24 hours. This drastic decline was exacerbated by a liquidity crisis in the DeFi sector, particularly affecting platforms like MakerDAO. The rapid drop in ETH prices led to mass liquidations of collateralized debt positions, causing systemic instability within the DeFi ecosystem.

Caferra et al. (2020) \cite{caferra2021raised} found that both cryptocurrency and stock markets experienced sharp declines in March during the pandemic, indicating the emergence of financial contagion. However, while cryptocurrencies quickly rebounded, stock markets remained in a bear phase. In other words, the price dynamics during the pandemic depended on the type of the market. 
Corbet (2020) \cite{corbet2020contagion} examine the potential contagion effects of the Covid-19 pandemic on gold and cryptocurrencies and consider that cryptocurrencies may play a role similar to that of precious metals during economic crises.
Umar and Gubareva (2020) \cite{umar2020time} and Umar et al. (2021) \cite{umar2021impact} analyze the potential interdependences between foreign exchange and cryptocurrency markets from the perspective of contagion and their possible role as safe havens during periods of economic turbulence, such as the SARS-CoV-2 outbreak.
Yousaf et al. (2023) \cite{yousaf2023connectedness}  explores the spillover between DeFi assets and US sector indices between January 2, 2019, and October 8, 2021, and they find high levels of the total connectedness index.
Akhtaruzzaman et al. (2022) \cite{akhtaruzzaman2022systemic} use the Conditional Value-at-Risk (CoVaR) model to develop a systemic contagion index (SCI) for cryptocurrencies and analyze their spillover effects. The results show that the SCI reached its highest value during the Covid-19 period, providing evidence of pandemic-driven contagion channels. Additionally, their analysis of cryptocurrency systemic networks reveals increased interconnections during this period, indicating a higher number of systemic contagion channels caused by the pandemic.

Elsayed et al. (2022) \cite{elsayed2022risk} examine the return and risk transmissions between Bitcoin, traditional financial assets (such as Crude Oil, Gold, Stocks, Bonds, and the US Dollar), and global uncertainty measures from April 2013 to June 2020. The results show that return and volatility spillovers reached unprecedented levels during the Covid-19 period and have remained elevated since. Gold is identified as the center of return spillovers, demonstrating its role as a safe haven asset, while Bitcoin acts as a net transmitter of volatility to other markets, particularly during the Covid-19 period. The study further confirms unidirectional volatility spillovers from Bitcoin to Gold, Stocks, Bonds, VIX, and Crude Oil, providing important implications for market dynamics.
Yousaf et al. (2022) \cite{yousaf2022linkages} study the spillovers between four DeFi assets (Chainlink, Maker, Basic Attention Token, and Synthetix) and four major conventional currencies (Chinese Yuan, Japanese Yen, Euro, and Pound Sterling). The results show a low static connectedness between DeFi and traditional currency markets. However, dynamic analysis reveals a significant increase in connectedness in early 2020, coinciding with the onset of the Covid-19 pandemic. Interestingly, the spillover from the Chinese Yuan did not increase during this period, indicating a pandemic-induced decoupling of the Chinese financial system from other centralized and decentralized markets. Despite heightened spillovers to DeFi markets at the pandemic's start, DeFi assets primarily acted as net shock transmitters to conventional currencies in 2020.

\section{TradFi Meets DeFi: Opportunities and Pitfalls}\label{sec:opportuinities}

The ongoing convergence of traditional finance and decentralized finance has sparked significant debate about the virtues of disintermediation, transparency, and innovation versus the stability and regulatory protections afforded by established institutions. On one hand, DeFi protocols promise lower transaction costs, near-instant settlements, and automated services -- features supported by smart contracts and often open-source development. These characteristics could offer TradFi a roadmap for reducing overheads and increasing efficiency, for instance by shifting certain lending or market-making functions onto ``semi-decentralized'' platforms. On the other hand, a historical pattern emerges in which ecosystems initially hailed as peer-to-peer or trustless end up centralizing over time, driven by legal, commercial, or practical pressures.
A clear illustration of this pattern can be found in non-financial examples. \emph{Napster}, for instance, began as a peer-to-peer file-sharing network but faced major copyright lawsuits and was forced to relaunch as a centrally managed music service, thus surrendering its pure P2P model. \emph{Skype}, similarly, relied on a distributed ``supernode'' architecture at its inception, but later evolved into a more centralized, server-based platform after being acquired by Microsoft, in part to meet user expectations and regulatory demands. Meanwhile, \emph{BitTorrent} remains decentralized in its underlying protocol, yet many users depend on large indexing sites or aggregator websites -- effectively creating centralized choke points. Within the blockchain sphere itself, several projects that aspired to total decentralization, such as \emph{Steem} (a blockchain-based social media platform, later forked to Hive due to a major stakeholder takeover) or \emph{EOS} (a platform for decentralized applications), have encountered centralization tied to governance decisions or an uneven token distribution.

From a financial perspective, these examples suggest that while peer-to-peer technology can help eliminate intermediaries and reduce costs, it also introduces new forms of risk. In DeFi, a protocol vulnerable to malicious exploits, poor governance safeguards, or liquidity runs may experience a rapid bank run scenario where users withdraw funds en masse. In response, the community or external investors may intervene to recapitalize the platform, often implementing stricter controls and governance measures to restore confidence. Ironically, these interventions can result in the platform adopting structures that closely resemble traditional, centralized financial systems, thereby diluting the foundational principles of decentralization.
A regulatory shift is currently unfolding in the United States, though. The Securities and Exchange Commission (SEC) has announced the creation of a cryptocurrency-focused task force (21 January 2025, \cite{bloomberg2025cryptotaskforce}). Acting Chair Mark Uyeda emphasized that this task force will work to establish a ``comprehensive and clear'' regulatory framework for crypto assets. This move highlights how governance challenges within DeFi ecosystems and regulatory uncertainty at the institutional level are both driving forces behind the convergence of decentralized platforms toward more centralized structures.
A parallel can be drawn to TradFi during crises, such as the collapse of Silicon Valley Bank (SVB) in March 2023. SVB faced a sudden bank run, with depositors withdrawing $\$42$ billion in a single day, leading to its insolvency. In the aftermath, regulatory authorities facilitated the acquisition of SVB's UK arm by HSBC for a nominal fee of $\pounds 1$, effectively stabilizing the situation and protecting depositors. This swift action by regulators and a larger financial institution prevented broader systemic contagion and ensured continuity for SVB UK's clients \cite{hsbc2023acquisition}. 

The notion of privately issued tokens circulating as money is not novel; it echoes the free banking experiments of the nineteenth century, when commercial banks in Scotland, Canada, and parts of the United States issued their own redeemable banknotes backed by specie or other assets \cite{dowd1992experience, selgin1987evolution}. Much like today’s stablecoins or governance tokens, a free bank’s notes derived their value from the issuing institution’s balance sheet strength; secondary markets quickly discounted, or refused, notes from weaker banks, providing market driven quality control. The historical record therefore reminds us that decentralized or privately supplied media of exchange can flourish -- and also fail -- depending on collateral backing, liquidity, and the legal environment. 

Over time, repeated bailouts or consolidations can erode the decentralized principles, effectively returning governance and oversight to a handful of dominant players \cite{eichengreen2019commodity}.
Nonetheless, TradFi has much to learn from DeFi’s design principles. The use of smart contracts can streamline lending and settlement procedures, and real-time auditability can enhance market transparency. The open-source feature has led to continuous experimentation -- liquidity pools, stablecoin mechanisms, and yield strategies -- that could help incumbents innovate at lower costs. Moreover, DeFi’s global accessibility holds lessons for advancing financial inclusion. 

Finally, it is worth emphasizing that the trigger for a crisis -- whether a hack, an oracle manipulation, or a sharp decline in asset prices -- tends to matter less than the underlying conditions that magnify vulnerabilities. If a DeFi ecosystem is over-leveraged or dependent on concentrated sources of liquidity, a minor shock can spiral into a system-wide disruption. By contrast, a protocol with robust risk buffers and diversified governance may weather the same shock with minimal disruption. Looking ahead, the interplay between DeFi’s capacity for innovation and TradFi’s demand for order and reliability will likely produce a range of hybrid models -- some retaining decentralized elements, others reverting to centralized ownership.

\section{Conclusions}\label{sec:concl}

In this paper, we (i) propose a framework for understanding the process of systemic risk formation in traditional finance (TradFi), (ii) extend this analysis to decentralized finance (DeFi) to account for its structural and technological characteristics, (iii) provide a detailed mapping of how core risks manifest in both environments (e.g. trading vulnerabilities, liquidity crises), and (iv) introduce the concept of \textit{crosstagion}, which refers to the bidirectional transmission of instability between TradFi and DeFi. 

This paper emphasizes how systemic risks, while rooted in similar underlying principles, manifest and propagate differently across these two ecosystems, and highlights the increasing entanglement between TradFi and DeFi as their boundaries blur.
We show that leverage, liquidity stress, interconnected balance sheets and governance failures are universal drivers of instability; what differs is the machinery through which each architecture amplifies and transmits those shocks.
A central contribution is summarised in Fig. \ref{fig:mapping}. By locating each risk along the axes of contagion speed and local‑to‑systemic scale of risk, the scheme provides an immediate visual diagnosis of where TradFi and DeFi are most vulnerable, and how quickly a local fault can translate into a sector‑wide crisis. The emergent clusters in the speed-scale of contagion risks (i.e., market, technological, regulatory, operational and interconnectedness risks) highlight which problems migrate faster from protocol level to full‑system meltdown.

On the basis of this mapping we introduce the notion of \textit{crosstagion}: a bidirectional shock channel through which a DeFi disruption can destabilise TradFi markets, and conversely, how abrupt monetary policy moves or regulatory shocks in TradFi can trigger DeFi liquidations. Recognising this mechanism is critical for regulators and central banks, because it calls for supervisory desks that track on‑chain and off‑chain data in real time rather than maintaining partitioned oversight of crypto and traditional windows. 

The framework thus offers a policy maker toolkit, highlighting where enhanced disclosure, or technical standards are most urgently needed, and it suggests how cross‑sector coordination can mute crosstagion before it may become systemic.

This work also lays the foundation for a broader, fully quantitative research program. By integrating blockchain transaction records with high-frequency market data and granular, bank-level exposures, researchers can calibrate the proposed two-dimensional risk map, develop agent-based models, and conduct network stress tests under scenarios of simultaneous TradFi and DeFi stress. Such analyses could reveal nonlinear amplification channels, feedback loops, and the thresholds at which localized shocks escalate into systemic crises. As tokenization advances and mainstream financial institutions adopt crypto-infrastructure, these data-driven extensions will be indispensable for both market participants and regulators aiming to safeguard financial stability and design effective macroprudential policies.

\section*{Acknowledgments}
P.V. acknowledges support from UKRI FLF Scheme (No. MR/X023028/1). F.C. acknowledges support of the Economic and Social Research Council (ESRC) in funding the Systemic Risk Centre at the LSE (ES/Y010612/1).


\begin{thebibliography}{100}
\expandafter\ifx\csname url\endcsname\relax
  \def\url#1{\texttt{#1}}\fi
\expandafter\ifx\csname urlprefix\endcsname\relax\def\urlprefix{URL }\fi
\expandafter\ifx\csname href\endcsname\relax
  \def\href#1#2{#2} \def\path#1{#1}\fi

\bibitem{maker2017whitepaper}
{Maker Foundation}, The Dai stablecoin system, Tech. rep., MakerDAO, white paper (2017).
\newline\urlprefix\url{https://makerdao.com/whitepaper/DaiDec17WP.pdf}

\bibitem{arrow1954existence}
K.~J. Arrow, G.~Debreu, Existence of an equilibrium for a competitive economy, Econometrica: Journal of the Econometric Society (1954) 265--290.

\bibitem{fama1970efficient}
E.~F. Fama, Efficient capital markets, Journal of finance 25~(2) (1970) 383--417.

\bibitem{muth1961rational}
J.~F. Muth, Rational expectations and the theory of price movements, Econometrica: Journal of the Econometric Society (1961) 315--335.

\bibitem{silva2017analysis}
W.~Silva, H.~Kimura, V.~A. Sobreiro, An analysis of the literature on systemic financial risk: A survey, Journal of Financial Stability 28 (2017) 91--114.

\bibitem{grilli2015markets}
R.~Grilli, G.~Tedeschi, M.~Gallegati, Markets connectivity and financial contagion, Journal of Economic Interaction and Coordination 10 (2015) 287--304.


\bibitem{rochet1996interbank}
J. C. Rochet and J. Tirole,
\newblock Interbank lending and systemic risk,
\newblock {Journal of Money, Credit and Banking} 28(4) (1996) 733--762.


\bibitem{allen2000financial}
F.~Allen, D.~Gale, Financial contagion, Journal of Political Economy 108~(1) (2000) 1--33.

\bibitem{oort1990banks}
C.~J. Oort, Banks and the stability of the international financial system, De Economist 138~(4) (1990) 451--463.

\bibitem{de2000systemic}
O.~De~Bandt, P.~Hartmann, \href{https://ssrn.com/abstract=258430}{Systemic risk: a survey}, Available at SSRN 258430 (2000).
\newline\urlprefix\url{https://ssrn.com/abstract=258430}

\bibitem{summer2003banking}
M.~Summer, Banking regulation and systemic risk, Open Economies Review 14 (2003) 43--70.

\bibitem{lehar2005measuring}
A.~Lehar, Measuring systemic risk: A risk management approach, Journal of Banking \& Finance 29~(10) (2005) 2577--2603.

\bibitem{ecb2009fsr}
{European Central Bank}, \href{https://www.ecb.europa.eu/pub/pdf/fsr/financialstabilityreview200912en.pdf}{Financial stability review}, Tech. rep., European Central Bank (2009).
\newline\urlprefix\url{https://www.ecb.europa.eu/pub/pdf/fsr/financialstabilityreview200912en.pdf}

\bibitem{95a16e2e-027f-345c-a203-76b01010c3b9}
T.~Adrian, M.~K. Brunnermeier, CoVaR, The American Economic Review 106~(7) (2016) 1705--1741.

\bibitem{billio2012econometric}
M.~Billio, M.~Getmansky, A.~W. Lo, L.~Pelizzon, Econometric measures of connectedness and systemic risk in the finance and insurance sectors, Journal of Financial Economics 104~(3) (2012) 535--559.

\bibitem{patro2013simple}
D.~K. Patro, M.~Qi, X.~Sun, A simple indicator of systemic risk, Journal of Financial Stability 9~(1) (2013) 105--116.

\bibitem{bandt2012systemic}
O.~de~Bandt, P.~Hartmann, J.~L. Peydr\'o, Systemic risk in banking: An update, in: A.~N. Berger, P.~Molyneux, J.~O.~S. Wilson (Eds.), The Oxford Handbook of Banking, Oxford University Press, Oxford, 2012, pp. 633--672, online edition, 18 September 2012.

\bibitem{benoit2017risks}
S.~Benoit, J.-E. Colliard, C.~Hurlin, C.~P{\'e}rignon, Where the risks lie: A survey on systemic risk, Review of Finance 21~(1) (2017) 109--152.

\bibitem{colander2009financial}
D.~Colander, M.~Goldberg, A.~Haas, K.~Juselius, A.~Kirman, T.~Lux, B.~Sloth, The financial crisis and the systemic failure of the economics profession, Critical Review 21~(2-3) (2009) 249--267.

\bibitem{bardoscia2017pathways}
M.~Bardoscia, S.~Battiston, F.~Caccioli, G.~Caldarelli, Pathways towards instability in financial networks, Nature communications 8~(1) (2017) 14416.

\bibitem{haldane2011systemic}
A.~G. Haldane, R.~M. May, Systemic risk in banking ecosystems, Nature 469~(7330) (2011) 351--355.

\bibitem{jackson2021systemic}
M.~O. Jackson, A.~Pernoud, Systemic risk in financial networks: A survey, Annual Review of Economics 13~(1) (2021) 171--202.

\bibitem{gale2007financial}
D.~M. Gale, S.~Kariv, Financial networks, American Economic Review 97~(2) (2007) 99--103.

\bibitem{eisenberg2001systemic}
L.~Eisenberg, T.~H. Noe, Systemic risk in financial systems, Management Science 47~(2) (2001) 236--249.

\bibitem{elliott2014financial}
M.~Elliott, B.~Golub, M.~O. Jackson, Financial networks and contagion, American Economic Review 104~(10) (2014) 3115--3153.

\bibitem{corsi2018measuring}
F.~Corsi, F.~Lillo, D.~Pirino, L.~Trapin, Measuring the propagation of financial distress with Granger-causality tail risk networks, Journal of Financial Stability 38 (2018) 18--36.

\bibitem{forer2025financial}
P.~Forer, B.~Budnick, P.~Vivo, S.~Aufiero, S.~Bartolucci, F.~Caccioli, Financial instability transition under heterogeneous investments and portfolio diversification, arXiv preprint arXiv:2501.19260 (2025).

\bibitem{acharya2007too}
V.~V. Acharya, T.~Yorulmazer, Too many to fail — an analysis of time-inconsistency in bank closure policies, Journal of Financial Intermediation 16~(1) (2007) 1--31.

\bibitem{allen2012asset}
F.~Allen, A.~Babus, E.~Carletti, Asset commonality, debt maturity and systemic risk, Journal of Financial Economics 104~(3) (2012) 519--534.

\bibitem{diebold2014network}
F.~X. Diebold, K.~Y{\i}lmaz, On the network topology of variance decompositions: Measuring the connectedness of financial firms, Journal of Econometrics 182~(1) (2014) 119--134.

\bibitem{bardoscia2019multiplex}
M.~Bardoscia, G.~Bianconi, G.~Ferrara, Multiplex network analysis of the UK over-the-counter derivatives market, International Journal of Finance \& Economics 24~(4) (2019) 1520--1544.

\bibitem{bargigli2014interaction}
L.~Bargigli, G.~Tedeschi, Interaction in agent-based economics: A survey on the network approach, Physica A: Statistical Mechanics and its Applications 399 (2014) 1--15.

\bibitem{bardoscia2021physics}
M.~Bardoscia, P.~Barucca, S.~Battiston, F.~Caccioli, G.~Cimini, D.~Garlaschelli, F.~Saracco, T.~Squartini, G.~Caldarelli, The physics of financial networks, Nature Reviews Physics 3~(7) (2021) 490--507.

\bibitem{axtell2025agent}
R.~L. Axtell, J.~D. Farmer, Agent-based modeling in economics and finance: Past, present, and future, Journal of Economic Literature 63~(1) (2025) 197--287.

\bibitem{thurner2011systemic}
S.~Thurner, \href{https://citeseerx.ist.psu.edu/document?repid=rep1&type=pdf&doi=433fb0508b3a498bf875cb15a6f05cca7517a6a9}{Systemic financial risk: agent based models to understand the leverage cycle on national scales and its consequences}, IfP/fgS Working Paper 14 (2011).
\newline\urlprefix\url{https://citeseerx.ist.psu.edu/document?repid=rep1&type=pdf&doi=433fb0508b3a498bf875cb15a6f05cca7517a6a9}

\bibitem{samanidou2007agent}
E.~Samanidou, E.~Zschischang, D.~Stauffer, T.~Lux, Agent-based models of financial markets, Reports on Progress in Physics 70~(3) (2007) 409.

\bibitem{cpmi2003glossary}
Committee~on~Payment and Settlement Systems, \href{https://www.bis.org/cpmi/glossary_030301.pdf}{A glossary of terms used in payments and settlement systems} (2003).
\newline\urlprefix\url{https://www.bis.org/cpmi/glossary_030301.pdf}

\bibitem{almarzoqi2015does}
R.~Almarzoqi, M.~S.~B. Naceur, A.~Scopelliti, \href{https://www.imf.org/external/pubs/ft/wp/2015/wp15210.pdf}{How does bank competition affect solvency, liquidity and credit risk? Evidence from the MENA countries}, International Monetary Fund, (2015).
\newline\urlprefix\url{https://www.imf.org/external/pubs/ft/wp/2015/wp15210.pdf}

\bibitem{bhattacharya1985preference}
S.~Bhattacharya, D.~Gale, W.~Barnett, K.~Singleton, Preference shocks, liquidity, and central bank policy, Liquidity and crises 35~(1) (1985) 69--88.

\bibitem{kelly2014tail}
B.~Kelly, H.~Jiang, Tail risk and asset prices, The Review of Financial Studies 27~(10) (2014) 2841--2871.

\bibitem{bhansali2008tail}
V.~Bhansali, Tail risk management, Journal of Portfolio Management 34~(4) (2008) 68.

\bibitem{thurner2012leverage}
S.~Thurner, J.~D. Farmer, J.~Geanakoplos, Leverage causes fat tails and clustered volatility, Quantitative Finance 12~(5) (2012) 695--707.

\bibitem{mcrandal2012primer}
R.~McRandal, A.~Rozanov, A primer on tail risk hedging, Journal of Securities Operations \& Custody 5~(1) (2012) 29--36.

\bibitem{bernanke1986agency}
B.~S. Bernanke, M.~Gertler, \href{http://www.nber.org/papers/w2015}{Agency costs, collateral, and business fluctuations}, Working Paper 2015, National Bureau of Economic Research (1986).
\newblock \href {https://doi.org/10.3386/w2015} {\path{doi:10.3386/w2015}}.
\newline\urlprefix\url{http://www.nber.org/papers/w2015}

\bibitem{kiyotaki1997credit}
N.~Kiyotaki, J.~Moore, Credit cycles, Journal of Political Economy 105~(2) (1997) 211--248.

\bibitem{holmstrom1997financial}
B.~Holmstrom, J.~Tirole, Financial intermediation, loanable funds, and the real sector, Quarterly Journal of economics 112~(3) (1997) 663--691.

\bibitem{allen2000bubbles}
F.~Allen, D.~Gale, Bubbles and crises, The Economic Journal 110~(460) (2000) 236--255.

\bibitem{allen2013systemic}
F.~Allen, E.~Carletti, Systemic risk from real estate and macro-prudential regulation, International Journal of Banking, Accounting and Finance 5~(1-2) (2013) 28--48.

\bibitem{adrian2010liquidity}
T.~Adrian, H.~S. Shin, Liquidity and leverage, Journal of Financial Intermediation 19~(3) (2010) 418--437.

\bibitem{caccioli2009eroding}
F.~Caccioli, M.~Marsili, P.~Vivo, Eroding market stability by proliferation of financial instruments, The European Physical Journal B 71~(4) (2009) 467--479.

\bibitem{brock2009more}
W.~A. Brock, C.~H. Hommes, F.~O. Wagener, More hedging instruments may destabilize markets, Journal of Economic Dynamics and Control 33~(11) (2009) 1912--1928.

\bibitem{amihud2012market}
Y.~Amihud, H.~Mendelson, L.~H. Pedersen, Market liquidity: asset pricing, risk, and crises, Cambridge University Press, 2012.

\bibitem{allen2004financial}
F.~Allen, D.~Gale, Financial intermediaries and markets, Econometrica 72~(4) (2004) 1023--1061.

\bibitem{krishnamurthy2010amplification}
A.~Krishnamurthy, Amplification mechanisms in liquidity crises, American Economic Journal: Macroeconomics 2~(3) (2010) 1--30.

\bibitem{bryant1980model}
J.~Bryant, A model of reserves, bank runs, and deposit insurance, Journal of Banking \& Finance 4~(4) (1980) 335--344.

\bibitem{diamond1983bank}
D.~W. Diamond, P.~H. Dybvig, Bank runs, deposit insurance, and liquidity, Journal of Political Economy 91~(3) (1983) 401--419.

\bibitem{brunnermeier2009market}
M.~K. Brunnermeier, L.~H. Pedersen, Market liquidity and funding liquidity, The Review of Financial Studies 22~(6) (2009) 2201--2238.

\bibitem{acharya2013precautionary}
V.~V. Acharya, O.~Merrouche, Precautionary hoarding of liquidity and interbank markets: Evidence from the subprime crisis, Review of Finance 17~(1) (2013) 107--160.

\bibitem{acharya2009theory}
V.~V. Acharya, A theory of systemic risk and design of prudential bank regulation, Journal of Financial Stability 5~(3) (2009) 224--255.

\bibitem{acharya2008information}
V.~V. Acharya, T.~Yorulmazer, Information contagion and bank herding, Journal of Money, Credit and Banking 40~(1) (2008) 215--231.

\bibitem{tirole2012overcoming}
J.~Tirole, Overcoming adverse selection: How public intervention can restore market functioning, American Economic Review 102~(1) (2012) 29--59.

\bibitem{anginer2014does}
D.~Anginer, A.~Demirguc-Kunt, M.~Zhu, How does deposit insurance affect bank risk? Evidence from the recent crisis, Journal of Banking \& Finance 48 (2014) 312--321.

\bibitem{cubillas2014financial}
E.~Cubillas, F.~Gonz{\'a}lez, Financial liberalization and bank risk-taking: International evidence, Journal of Financial Stability 11 (2014) 32--48.

\bibitem{chen1999banking}
Y.~Chen, Banking panics: The role of the first-come, first-served rule and information externalities, Journal of Political Economy 107~(5) (1999) 946--968.

\bibitem{dasgupta2004financial}
A.~Dasgupta, Financial contagion through capital connections: A model of the origin and spread of bank panics, Journal of the European Economic Association 2~(6) (2004) 1049--1084.

\bibitem{cespa2014illiquidity}
G.~Cespa, T.~Foucault, Illiquidity contagion and liquidity crashes, The Review of Financial Studies 27~(6) (2014) 1615--1660.

\bibitem{farhi2012collective}
E.~Farhi, J.~Tirole, Collective moral hazard, maturity mismatch, and systemic bailouts, American Economic Review 102~(1) (2012) 60--93.

\bibitem{bardoscia2019full}
M.~Bardoscia, G.~Ferrara, N.~Vause, M.~Yoganayagam, \href{https://ssrn.com/abstract=3344580}{Full payment algorithm}, Available at SSRN 3344580 (2019).
\newline\urlprefix\url{https://ssrn.com/abstract=3344580}

\bibitem{freixas2000systemic}
X.~Freixas, B.~M. Parigi, J.-C. Rochet, Systemic risk, interbank relations, and liquidity provision by the Central Bank, Journal of Money, Credit and Banking (2000) 611--638.

\bibitem{gai2010contagion}
P.~Gai, S.~Kapadia, Contagion in financial networks, Proceedings of the Royal Society A: Mathematical, Physical and Engineering Sciences 466~(2120) (2010) 2401--2423.

\bibitem{acemoglu2015systemic}
D.~Acemoglu, A.~Ozdaglar, A.~Tahbaz-Salehi, Systemic risk and stability in financial networks, American Economic Review 105~(2) (2015) 564--608.

\bibitem{elliott2021systemic}
M.~Elliott, C.-P. Georg, J.~Hazell, Systemic risk shifting in financial networks, Journal of Economic Theory 191 (2021) 105157.

\bibitem{caccioli2014stability}
F.~Caccioli, M.~Shrestha, C.~Moore, J.~D. Farmer, Stability analysis of financial contagion due to overlapping portfolios, Journal of Banking \& Finance 46 (2014) 233--245.

\bibitem{duarte2021fire}
F.~Duarte, T.~M. Eisenbach, Fire-sale spillovers and systemic risk, The Journal of Finance 76~(3) (2021) 1251--1294.

\bibitem{cont2017fire}
R.~Cont, E.~Schaanning, \href{https://ssrn.com/abstract=2541114}{Fire sales, indirect contagion and systemic stress testing}, Tech. Rep. 02/2017, Norges Bank Working Paper (Jun 13 2017).
\newline\urlprefix\url{https://ssrn.com/abstract=2541114}

\bibitem{barucca2021common}
P.~Barucca, T.~Mahmood, L.~Silvestri, Common asset holdings and systemic vulnerability across multiple types of financial institution, Journal of Financial Stability 52 (2021) 100810.

\bibitem{bouchaud2009markets}
J.-P. Bouchaud, J.~D. Farmer, F.~Lillo, How markets slowly digest changes in supply and demand, in: Handbook of financial markets: dynamics and evolution, Elsevier, 2009, pp. 57--160.

\bibitem{watts2002simple}
D.~J. Watts, A simple model of global cascades on random networks, Proceedings of the National Academy of Sciences 99~(9) (2002) 5766--5771.

\bibitem{capponi2015price}
A.~Capponi, M.~Larsson, Price contagion through balance sheet linkages, The Review of Asset Pricing Studies 5~(2) (2015) 227--253.

\bibitem{greenwood2015vulnerable}
R.~Greenwood, A.~Landier, D.~Thesmar, Vulnerable banks, Journal of Financial Economics 115~(3) (2015) 471--485.

\bibitem{cifuentes2005liquidity}
R.~Cifuentes, G.~Ferrucci, H.~S. Shin, Liquidity risk and contagion, Journal of the European Economic Association 3~(2-3) (2005) 556--566.

\bibitem{caccioli2015overlapping}
F.~Caccioli, J.~D. Farmer, N.~Foti, D.~Rockmore, Overlapping portfolios, contagion, and financial stability, Journal of Economic Dynamics and Control 51 (2015) 50--63.

\bibitem{poledna2021quantification}
S.~Poledna, S.~Mart{\'\i}nez-Jaramillo, F.~Caccioli, S.~Thurner, Quantification of systemic risk from overlapping portfolios in the financial system, Journal of Financial Stability 52 (2021) 100808.

\bibitem{caccioli2024modelling}
F.~Caccioli, G.~Ferrara, A.~Ramadiah, Modelling fire sale contagion across banks and non-banks, Journal of Financial Stability 71 (2024) 101231.

\bibitem{reinhart2009aftermath}
C.~M. Reinhart, K.~S. Rogoff, The aftermath of financial crises, American Economic Review 99~(2) (2009) 466--472.

\bibitem{bebchuk2011self}
L.~A. Bebchuk, I.~Goldstein, Self-fulfilling credit market freezes, The Review of Financial Studies 24~(11) (2011) 3519--3555.

\bibitem{jackson2024credit}
M.~O. Jackson, A.~Pernoud, Credit freezes, equilibrium multiplicity, and optimal bailouts in financial networks, The Review of Financial Studies 37~(7) (2024) 2017--2062.

\bibitem{roukny2018interconnectedness}
T.~Roukny, S.~Battiston, J.~E. Stiglitz, Interconnectedness as a source of uncertainty in systemic risk, Journal of Financial Stability 35 (2018) 93--106.

\bibitem{corsi2016micro}
F.~Corsi, S.~Marmi, F.~Lillo, When micro prudence increases macro risk: The destabilizing effects of financial innovation, leverage, and diversification, Operations Research 64~(5) (2016) 1073--1088.

\bibitem{may2008ecology}
R.~M. May, S.~A. Levin, G.~Sugihara, Ecology for bankers, Nature 451~(7181) (2008) 893--894.

\bibitem{bormetti2015modelling}
G.~Bormetti, L.~M. Calcagnile, M.~Treccani, F.~Corsi, S.~Marmi, F.~Lillo, Modelling systemic price cojumps with Hawkes factor models, Quantitative Finance 15~(7) (2015) 1137--1156.

\bibitem{flannery1996financial}
M.~J. Flannery, Financial crises, payment system problems, and discount window lending, Journal of Money, Credit and Banking 28~(4) (1996) 804--824.

\bibitem{acharya2011rollover}
V.~V. Acharya, D.~Gale, T.~Yorulmazer, Rollover risk and market freezes, The Journal of Finance 66~(4) (2011) 1177--1209.

\bibitem{caballero2013fire}
R.~J. Caballero, A.~Simsek, Fire sales in a model of complexity, The Journal of Finance 68~(6) (2013) 2549--2587.

\bibitem{brunnermeier2013liquidity}
M.~Brunnermeier, G.~Gorton, A.~Krishnamurthy, Liquidity mismatch measurement, in: Risk topography: Systemic risk and macro modeling, University of Chicago Press, 2013, pp. 99--112.

\bibitem{battiston2012default}
S.~Battiston, D.~D. Gatti, M.~Gallegati, B.~Greenwald, J.~E. Stiglitz, Default cascades: When does risk diversification increase stability?, Journal of Financial Stability 8~(3) (2012) 138--149.

\bibitem{cai2018syndication}
J.~Cai, F.~Eidam, A.~Saunders, S.~Steffen, Syndication, interconnectedness, and systemic risk, Journal of Financial Stability 34 (2018) 105--120.

\bibitem{de2011systemic}
G.~De~Nicol{\`o}, M.~Lucchetta, et~al., \href{https://ssrn.com/abstract=1555441}{Systemic risks and the macroeconomy}, Tech. rep., National Bureau of Economic Research Cambridge, Massachusetts, US (2011).
\newline\urlprefix\url{https://ssrn.com/abstract=1555441}

\bibitem{glasserman2015likely}
P.~Glasserman, H.~P. Young, How likely is contagion in financial networks?, Journal of Banking \& Finance 50 (2015) 383--399.

\bibitem{arinaminpathy2012size}
N.~Arinaminpathy, S.~Kapadia, R.~M. May, Size and complexity in model financial systems, Proceedings of the National Academy of Sciences 109~(45) (2012) 18338--18343.

\bibitem{banulescu2015sifis}
G.-D. Banulescu, E.-I. Dumitrescu, Which are the SIFIs? A component expected shortfall approach to systemic risk, Journal of Banking \& Finance 50 (2015) 575--588.

\bibitem{kaufman2014too}
G.~G. Kaufman, Too big to fail in banking: What does it mean?, Journal of Financial Stability 13 (2014) 214--223.

\bibitem{upper2004estimating}
C.~Upper, A.~Worms, Estimating bilateral exposures in the german interbank market: Is there a danger of contagion?, European Economic Review 48~(4) (2004) 827--849.

\bibitem{elsinger2006risk}
H.~Elsinger, A.~Lehar, M.~Summer, Risk assessment for banking systems, Management Science 52~(9) (2006) 1301--1314.

\bibitem{drehmann2011systemic}
M.~Drehmann, N.~A. Tarashev, \href{https://ssrn.com/abstract=1785264}{Systemic importance: some simple indicators}, BIS Quarterly Review, March (2011).
\newline\urlprefix\url{https://ssrn.com/abstract=1785264}

\bibitem{iyer2011interbank}
R.~Iyer, J.-L. Peydro, Interbank contagion at work: Evidence from a natural experiment, The Review of Financial Studies 24~(4) (2011) 1337--1377.

\bibitem{jamilov2024two}
R.~Jamilov, T.~K{\"o}nig, K.~M{\"u}ller, F.~Saidi, \href{https://ssrn.com/abstract=4924699}{Two centuries of systemic bank runs}, Tech. rep. ((2024)).
\newline\urlprefix\url{https://ssrn.com/abstract=4924699}

\bibitem{amini2016resilience}
H.~Amini, R.~Cont, A.~Minca, Resilience to contagion in financial networks, Mathematical Finance 26~(2) (2016) 329--365.

\bibitem{bardoscia2015debtrank}
M.~Bardoscia, S.~Battiston, F.~Caccioli, G.~Caldarelli, Debtrank: A microscopic foundation for shock propagation, PloS one 10~(6) (2015) e0130406.

\bibitem{bardoscia2016distress}
M.~Bardoscia, F.~Caccioli, J.~I. Perotti, G.~Vivaldo, G.~Caldarelli, Distress propagation in complex networks: the case of non-linear debtrank, PloS one 11~(10) (2016) e0163825.

\bibitem{fischer2014no}
T.~Fischer, No-arbitrage pricing under systemic risk: Accounting for cross-ownership, Mathematical Finance: An International Journal of Mathematics, Statistics and Financial Economics 24~(1) (2014) 97--124.

\bibitem{barucca2020network}
P.~Barucca, M.~Bardoscia, F.~Caccioli, M.~D'Errico, G.~Visentin, G.~Caldarelli, S.~Battiston, Network valuation in financial systems, Mathematical Finance 30~(4) (2020) 1181--1204.

\bibitem{soramaki2007topology}
K.~Soram{\"a}ki, M.~L. Bech, J.~Arnold, R.~J. Glass, W.~E. Beyeler, The topology of interbank payment flows, Physica A: Statistical Mechanics and its Applications 379~(1) (2007) 317--333.

\bibitem{bech2010topology}
M.~L. Bech, E.~Atalay, The topology of the federal funds market, Physica A: Statistical Mechanics and its Applications 389~(22) (2010) 5223--5246.

\bibitem{craig2014interbank}
B.~Craig, G.~Von~Peter, Interbank tiering and money center banks, Journal of Financial Intermediation 23~(3) (2014) 322--347.

\bibitem{blasques2018dynamic}
F.~Blasques, F.~Br{\"a}uning, I.~Van~Lelyveld, A dynamic network model of the unsecured interbank lending market, Journal of Economic Dynamics and Control 90 (2018) 310--342.

\bibitem{glasserman2016contagion}
P.~Glasserman, H.~P. Young, Contagion in financial networks, Journal of Economic Literature 54~(3) (2016) 779--831.

\bibitem{antonopoulos2021mastering}
A.~M. Antonopoulos, O.~Osuntokun, R.~Pickhardt, \href{https://www.oreilly.com/library/view/mastering-the-lightning/9781492054856/}{Mastering the lightning network}, O'Reilly Media, Inc., (2021).
\newline\urlprefix\url{https://www.oreilly.com/library/view/mastering-the-lightning/9781492054856/}

\bibitem{nakamoto2008bitcoin}
S.~Nakamoto, \href{https://bitcoin.org/bitcoin.pdf}{Bitcoin: A peer-to-peer electronic cash system}, Bitcoin 4~(2) (2008) 15.
\newline\urlprefix\url{https://bitcoin.org/bitcoin.pdf}

\bibitem{makarov2022cryptocurrencies}
I.~Makarov, A.~Schoar, Cryptocurrencies and decentralized finance (defi), Brookings Papers on Economic Activity 2022~(1) (2022) 141--215.

\bibitem{buterin2013ethereum}
V.~Buterin, et~al., \href{https://ethereum.org/en/whitepaper/}{Ethereum white paper}, GitHub repository 1 (2013) 22--23.
\newline\urlprefix\url{https://ethereum.org/en/whitepaper/}

\bibitem{buterin2022proof}
V.~Buterin, \href{https://onlinebooks.library.upenn.edu/webbin/book/lookupid?key=olbp100520}{Proof of stake: The making of Ethereum and the philosophy of blockchains}, Seven Stories Press (2022).
\newline\urlprefix\url{https://onlinebooks.library.upenn.edu/webbin/book/lookupid?key=olbp100520}

\bibitem{singh2020sidechain}
A.~Singh, K.~Click, R.~M. Parizi, Q.~Zhang, A.~Dehghantanha, K.-K.~R. Choo, Sidechain technologies in blockchain networks: An examination and state-of-the-art review, Journal of Network and Computer Applications 149 (2020) 102471.

\bibitem{bartolucci2020percolation}
S.~Bartolucci, F.~Caccioli, P.~Vivo, A percolation model for the emergence of the Bitcoin lightning network, Scientific reports 10~(1) (2020) 4488.

\bibitem{bartolucci2020model}
S.~Bartolucci, A.~Kirilenko, A model of the optimal selection of crypto assets, Royal Society Open Science 7~(8) (2020) 191863.

\bibitem{mungo2024cryptocurrency}
L.~Mungo, S.~Bartolucci, L.~Alessandretti, Cryptocurrency co-investment network: token returns reflect investment patterns, EPJ Data Science 13~(1) (2024) 11.

\bibitem{binance2025post}
Binance, \href{https://www.binance.com/en/square/post/18496299612402}{Dex futures trading reaches \$285b in 2024, hyperliquid dominates} (2025).
\newline\urlprefix\url{https://www.binance.com/en/square/post/18496299612402}

\bibitem{schar2021decentralized}
F.~Sch{\"a}r, \href{https://ssrn.com/abstract=3571335}{Decentralized finance: On blockchain-and smart contract-based financial markets}, FRB of St. Louis Review (2021).
\newline\urlprefix\url{https://ssrn.com/abstract=3571335}

\bibitem{weingartner2023deciphering}
T.~Weing{\"a}rtner, F.~Fasser, P.~Reis S{\'a}~da Costa, W.~Farkas, Deciphering defi: A comprehensive analysis and visualization of risks in decentralized finance, Journal of Risk and Financial Management 16~(10) (2023) 454.

\bibitem{ibba2023preliminary}
G.~Ibba, S.~Khullar, E.~Tesfai, R.~Neykova, S.~Aufiero, M.~Ortu, S.~Bartolucci, G.~Destefanis, A preliminary analysis of software metrics in decentralised applications, in: Proceedings of the Fifth ACM International Workshop on Blockchain-enabled Networked Sensor Systems, 2023, pp. 27--33.

\bibitem{aufiero2024dapps}
S.~Aufiero, G.~Ibba, S.~Bartolucci, G.~Destefanis, R.~Neykova, M.~Ortu, Dapps ecosystems: Mapping the network structure of smart contract interactions, EPJ Data Science 13~(1) (2024) 60.

\bibitem{caldarelli2020understanding}
G.~Caldarelli, Understanding the blockchain oracle problem: A call for action, Information 11~(11) (2020) 509.

\bibitem{carter2021defi}
N.~Carter, L.~Jeng, \href{https://ssrn.com/abstract=3866699}{Defi protocol risks: The paradox of defi}, Regtech, suptech and beyond: innovation and technology in financial services 3 (2021).
\newline\urlprefix\url{https://ssrn.com/abstract=3866699}

\bibitem{forbes_stablecoins_2024}
{Forbes}, \href{https://www.forbes.com/digital-assets/categories/stablecoins/?sh=393f32ad1cd0}{Stablecoins: What you need to know} (2024).
\newline\urlprefix\url{https://www.forbes.com/digital-assets/categories/stablecoins/?sh=393f32ad1cd0}

\bibitem{aramonte2021defi}
S.~Aramonte, W.~Huang, A.~Schrimpf, Defi risks and the decentralisation illusion, BIS Quarterly Review 6 (2021).

\bibitem{mcleay2014money}
M.~McLeay, A.~Radia, R.~Thomas, Money creation in the modern economy, Bank of England Quarterly Bulletin (2014) Q1.

\bibitem{chainalysis2023crypto}
Chainalysis, \href{https://go.chainalysis.com/2023-Crypto-Crime-Report.html}{The 2023 crypto crime report} (2023).
\newline\urlprefix\url{https://go.chainalysis.com/2023-Crypto-Crime-Report.html}

\bibitem{ronin2022hack}
S.~Mavis, \href{https://therecord.media/hackers-return-12-million-taken-from-ronin-network}{Ronin validator security breach} (2022).
\newline\urlprefix\url{https://therecord.media/hackers-return-12-million-taken-from-ronin-network}

\bibitem{certik2022h1}
CertiK, \href{https://www.certik.com/resources/blog/7fuXtbfo4CXEXcwy5Pqijp-hack3d-the-web3-security-quarterly-report-q2-2022}{Web3 security quarterly report q2 2022} (2022).
\newline\urlprefix\url{https://www.certik.com/resources/blog/7fuXtbfo4CXEXcwy5Pqijp-hack3d-the-web3-security-quarterly-report-q2-2022}

\bibitem{bekemeier2021deceptive}
F.~Bekemeier, Deceptive assurance? A conceptual view on systemic risk in decentralized finance (DeFi), in: Proceedings of the 2021 4th International Conference on Blockchain Technology and Applications, 2021, pp. 76--87.

\bibitem{meyer2022decentralized}
E.~Meyer, I.~M. Welpe, P.~G. Sandner, \href{https://ssrn.com/abstract=4016497}{Decentralized finance - a systematic literature review and research directions}, ECIS, (2022).
\newline\urlprefix\url{https://ssrn.com/abstract=4016497}

\bibitem{werner2022sok}
S.~Werner, D.~Perez, L.~Gudgeon, A.~Klages-Mundt, D.~Harz, W.~Knottenbelt, Sok: Decentralized finance (DeFi), in: Proceedings of the 4th ACM Conference on Advances in Financial Technologies, 2022, pp. 30--46.

\bibitem{zhou2023sok}
L.~Zhou, X.~Xiong, J.~Ernstberger, S.~Chaliasos, Z.~Wang, Y.~Wang, K.~Qin, R.~Wattenhofer, D.~Song, A.~Gervais, Sok: Decentralized finance (DeFi) attacks, in: 2023 IEEE Symposium on Security and Privacy (SP), IEEE, 2023, pp. 2444--2461.

\bibitem{li2022survey}
W.~Li, J.~Bu, X.~Li, H.~Peng, Y.~Niu, Y.~Zhang, A survey of DeFi security: Challenges and opportunities, Journal of King Saud University-Computer and Information Sciences 34~(10) (2022) 10378--10404.

\bibitem{qin2021attacking}
K.~Qin, L.~Zhou, B.~Livshits, A.~Gervais, Attacking the DeFi ecosystem with flash loans for fun and profit, in: International Conference on Financial Cryptography and Data Security, Springer, 2021, pp. 3--32.

\bibitem{chang2022risk}
T.~Chang, J.~Ho, Z.~Tirrell, G.~Weng, J.~You, A risk classification framework for decentralized finance protocols: Exploring emerging risks for insurers and reinsurers, Tech. rep., Society of Actuaries (2022).

\bibitem{kaur2023risk}
S.~Kaur, S.~Singh, S.~Gupta, S.~Wats, Risk analysis in decentralized finance (DeFi): a fuzzy-AHP approach, Risk Management 25~(2) (2023) 13.

\bibitem{danielsson2016cyber}
J.~Danielsson, M.~Fouche, R.~Macrae, \href{https://cepr.org/voxeu/columns/cyber-risk-systemic-risk}{Cyber risk as systemic risk}, VOX CEPR (2016).
\newline\urlprefix\url{https://cepr.org/voxeu/columns/cyber-risk-systemic-risk}

\bibitem{barrera2018blockchain}
C.~Barrera, S.~Hurder, \href{https://ssrn.com/abstract=3192208}{Blockchain upgrade as a coordination game}, University of Cambridge (2018).
\newline\urlprefix\url{https://ssrn.com/abstract=3192208}

\bibitem{tikhomirov2018smartcheck}
S.~Tikhomirov, E.~Voskresenskaya, I.~Ivanitskiy, R.~Takhaviev, E.~Marchenko, Y.~Alexandrov, Smartcheck: Static analysis of ethereum smart contracts, in: Proceedings of the 1st international workshop on emerging trends in software engineering for blockchain, 2018, pp. 9--16.

\bibitem{luu2016making}
L.~Luu, D.-H. Chu, H.~Olickel, P.~Saxena, A.~Hobor, Making smart contracts smarter, in: Proceedings of the 2016 ACM SIGSAC conference on computer and communications security, 2016, pp. 254--269.

\bibitem{ron2013quantitative}
D.~Ron, A.~Shamir, Quantitative analysis of the full Bitcoin transaction graph, in: Financial Cryptography and Data Security: 17th International Conference, FC 2013, Okinawa, Japan, April 1-5, 2013, Revised Selected Papers 17, Springer, 2013, pp. 6--24.

\bibitem{destefanis2018smart}
G.~Destefanis, M.~Marchesi, M.~Ortu, R.~Tonelli, A.~Bracciali, R.~Hierons, Smart contracts vulnerabilities: a call for blockchain software engineering?, in: 2018 International Workshop on Blockchain Oriented Software Engineering (IWBOSE), IEEE, 2018, pp. 19--25.

\bibitem{ibba2024curated}
G.~Ibba, S.~Aufiero, R.~Neykova, S.~Bartolucci, M.~Ortu, R.~Tonelli, G.~Destefanis, A curated Solidity smart contracts repository of metrics and vulnerability, in: Proceedings of the 20th International Conference on Predictive Models and Data Analytics in Software Engineering, 2024, pp. 32--41.

\bibitem{cryptoslate2021aave}
CryptoSlate, \href{https://cryptoslate.com/data-shows-how-aave-overtook-compound-in-defi-lending/}{Data shows how aave overtook compound in DeFi lending} (2021).
\newline\urlprefix\url{https://cryptoslate.com/data-shows-how-aave-overtook-compound-in-defi-lending/}



\bibitem{saengchote2021defi}
K. Saengchote, Where do DeFi stablecoins go? A closer look at what DeFi composability really means, \newblock {PIER Discussion Papers}, 156, Puey Ungphakorn Institute for Economic Research, 2021.
\newblock URL: \url{https://www.pier.or.th/files/dp/pier_dp_156.pdf}.



\bibitem{gencer2018decentralization}
A.~E. Gencer, S.~Basu, I.~Eyal, R.~Van~Renesse, E.~G. Sirer, Decentralization in Bitcoin and Ethereum networks, in: Financial Cryptography and Data Security: 22nd International Conference, FC 2018, Nieuwpoort, Cura{\c{c}}ao, February 26--March 2, 2018, Revised Selected Papers 22, Springer, 2018, pp. 439--457.

\bibitem{sec_enforcement_2022}
{U.S. Securities and Exchange Commission}, \href{https://www.sec.gov/spotlight/cybersecurity-enforcement-actions}{{SEC} enforcement actions against crypto and {DeFi} projects} (2022).
\newline\urlprefix\url{https://www.sec.gov/spotlight/cybersecurity-enforcement-actions}

\bibitem{gensler2021remarks}
G.~Gensler, \href{https://www.sec.gov/newsroom/speeches-statements/gensler-remarks-crypto-markets-040422}{Remarks by chair Gary Gensler on crypto markets and {DeFi}}, u.S. SEC official speech (2021).
\newline\urlprefix\url{https://www.sec.gov/newsroom/speeches-statements/gensler-remarks-crypto-markets-040422}

\bibitem{mica_proposal_2022}
European~Commission, \href{https://www.europarl.europa.eu/legislative-train/theme-a-europe-fit-for-the-digital-age/file-crypto-assets-1}{Proposal for a regulation of the European Parliament and of the council on markets in crypto-assets ({MiCA})}, cOM(2020) 593 final, updated discussions in 2022-2023.
\newline\urlprefix\url{https://www.europarl.europa.eu/legislative-train/theme-a-europe-fit-for-the-digital-age/file-crypto-assets-1}

\bibitem{ecb_2022_crypto}
\href{https://eur-lex.europa.eu/legal-content/EN/TXT/?uri=OJ%3AC%3A2021%3A152%3ATOC}{ECB opinion on crypto-asset markets and {MiCA}}, European Central Bank, 28 September 2022.
\newline\urlprefix\url{https://eur-lex.europa.eu/legal-content/EN/TXT/?uri=OJ%3AC%3A2021%3A152%3ATOC}

\bibitem{mas2022guidelines}
Monetary~Authority of~Singapore, \href{https://www.mas.gov.sg/regulation/guidelines/ps-g02-guidelines-on-provision-of-digital-payment-token-services-to-the-public}{Guidelines on provision of digital payment token services to the public}, consultation Paper, 2022.
\newline\urlprefix\url{https://www.mas.gov.sg/regulation/guidelines/ps-g02-guidelines-on-provision-of-digital-payment-token-services-to-the-public}

\bibitem{china2021ban}
\href{https://www.weforum.org/stories/2022/01/what-s-behind-china-s-cryptocurrency-ban/}{China bans crypto transactions and mining}, People’s Bank of China Announcement, 2021.
\newline\urlprefix\url{https://www.weforum.org/stories/2022/01/what-s-behind-china-s-cryptocurrency-ban/}

\bibitem{fsb2022risks}
{Financial Stability Board (FSB)}, \href{https://www.fsb.org/2022/02/assessment-of-risks-to-financial-stability-from-crypto-assets/}{Assessment of risks to financial stability from crypto-assets and DeFi} (February 2022).
\newline\urlprefix\url{https://www.fsb.org/2022/02/assessment-of-risks-to-financial-stability-from-crypto-assets/}

\bibitem{li2020survey}
X.~Li, P.~Jiang, T.~Chen, X.~Luo, Q.~Wen, A survey on the security of blockchain systems, Future generation computer systems 107 (2020) 841--853.

\bibitem{deshmukh2021decentralized}
S.~Deshmukh, S.~Warren, K.~Warbach, Decentralized finance (DeFi) policy-maker toolkit, in: World economic forum, Vol.~8, 2021, p. 2021.

\bibitem{tasca2017taxonomy}
P.~Tasca, C.~J. Tessone, A taxonomy of blockchain technologies: Principles of identification and classification, Ledger 4 (2019) 1--39.

\bibitem{de2020blockchain}
P.~De~Filippi, M.~Mannan, W.~Reijers, Blockchain as a confidence machine: The problem of trust \& challenges of governance, Technology in Society 62 (2020) 101284.

\bibitem{vidal2023ftx}
D.~Vidal-Tom{\'a}s, A.~Briola, T.~Aste, FTX’s downfall and Binance’s consolidation: The fragility of centralised digital finance, Physica A: Statistical Mechanics and its Applications 625 (2023) 129044.

\bibitem{gudgeon2020defi}
L.~Gudgeon, S.~Werner, D.~Perez, W.~J. Knottenbelt, DeFi protocols for loanable funds: Interest rates, liquidity and market efficiency, in: Proceedings of the 2nd ACM Conference on Advances in Financial Technologies, 2020, pp. 92--112.

\bibitem{coinbase_impermanent_loss}
Coinbase, \href{https://www.coinbase.com/en-gb/learn/crypto-glossary/what-is-impermanent-loss}{What is impermanent loss?}
\newline\urlprefix\url{https://www.coinbase.com/en-gb/learn/crypto-glossary/what-is-impermanent-loss}

\bibitem{christidis2016blockchains}
K.~Christidis, M.~Devetsikiotis, Blockchains and smart contracts for the internet of things, IEEE Access 4 (2016) 2292--2303.

\bibitem{ezzat2022blockchain}
S.~K. Ezzat, Y.~N. Saleh, A.~A. Abdel-Hamid, Blockchain oracles: State-of-the-art and research directions, IEEE Access 10 (2022) 67551--67572.

\bibitem{zhang2024security}
M.~Zhang, X.~Zhang, Y.~Zhang, Z.~Lin, Security of cross-chain bridges: Attack surfaces, defenses, and open problems, in: Proceedings of the 27th International Symposium on Research in Attacks, Intrusions and Defenses, 2024, pp. 298--316.

\bibitem{auer2020rise}
R.~Auer, G.~Cornelli, J.~Frost, Rise of the central bank digital currencies: drivers, approaches and technologies, CEPR Discussion Paper No. DP15363 (2020).

\bibitem{dupont2017experiments}
Q.~DuPont, \href{https://www.taylorfrancis.com/chapters/oa-edit/10.4324/9781315211909-8/experiments-algorithmic-governance-quinn-dupont}{Experiments in algorithmic governance}, Bitcoin and beyond (2017) 157.
\newline\urlprefix\url{https://www.taylorfrancis.com/chapters/oa-edit/10.4324/9781315211909-8/experiments-algorithmic-governance-quinn-dupont}

\bibitem{yawrisk}
A.~Hinneh, Risk management in decentralised finance (DeFi), Academia (2024) 72.

\bibitem{bisias2012survey}
D.~Bisias, M.~Flood, A.~W. Lo, S.~Valavanis, A survey of systemic risk analytics, Annu. Rev. Financ. Econ. 4~(1) (2012) 255--296.

\bibitem{lowenstein2023svb}
Lowenstein Sandler LLP, \href{https://www.lowenstein.com/news-insights/publications/client-alerts/silicon-valley-bank-a-timeline-and-summary-of-events-debt-finance}{Silicon valley bank: A timeline and summary of events} ((2023)).
\newline\urlprefix\url{https://www.lowenstein.com/news-insights/publications/client-alerts/silicon-valley-bank-a-timeline-and-summary-of-events-debt-finance}

\bibitem{gorton2023taming}
G.~B. Gorton, J.~Y. Zhang, Taming wildcat stablecoins, U. Chi. L. Rev. 90 (2023) 909.

\bibitem{pontem2022celsius}
P.~Network, \href{https://blog.pontem.network/the-celsius-crash-explained-be91ef715cd9}{The celsius crash explained} (2022).
\newline\urlprefix\url{https://blog.pontem.network/the-celsius-crash-explained-be91ef715cd9}

\bibitem{briola2024deep}
A.~Briola, S.~Bartolucci, T.~Aste, Deep limit order book forecasting: a microstructural guide, Quantitative Finance (2025) 1--31.

\bibitem{park2021conceptual}
A.~Park, \href{https://kenaninstitute.unc.edu/rethinc/wp-content/uploads/2022/03/Park_-Andreas-Automated-Market-Makers.pdf}{The conceptual flaws of constant product automated market making}, Available at SSRN 3805750 (2021).
\newline\urlprefix\url{https://kenaninstitute.unc.edu/rethinc/wp-content/uploads/2022/03/Park_-Andreas-Automated-Market-Makers.pdf}

\bibitem{cai2003there}
F.~Cai, \href{https://www.federalreserve.gov/econres/ifdp/was-there-front-running-during-the-ltcm-crisis.htm}{Was there front running during the LTCM crisis?}, Available at SSRN 385560 (2003).
\newline\urlprefix\url{https://www.federalreserve.gov/econres/ifdp/was-there-front-running-during-the-ltcm-crisis.htm}

\bibitem{malinova2024learning}
K.~Malinova, A.~Park, \href{https://www.newyorkfed.org/medialibrary/media/research/conference/2023/FinTech/6_1030am_Saleh_slidestrueAMM_Discussion.pdf?sc_lang=en&hash=B6D9682B78CDCAA6031EC9EA983A8146}{Learning from DeFi: Would automated market makers improve equity trading?}, Available at SSRN 4531670 (2024).
\newline\urlprefix\url{https://www.newyorkfed.org/medialibrary/media/research/conference/2023/FinTech/6_1030am_Saleh_slidestrueAMM_Discussion.pdf?sc_lang=en&hash=B6D9682B78CDCAA6031EC9EA983A8146}

\bibitem{xu2019anatomy}
J.~Xu, B.~Livshits, The anatomy of a cryptocurrency Pump-and-Dump scheme, in: 28th USENIX Security Symposium (USENIX Security 19), 2019, pp. 1609--1625.

\bibitem{aave2025flashloans}
Aave, \href{https://aave.com/docs/concepts/flash-loans}{Flash loans documentation}.
\newline\urlprefix\url{https://aave.com/docs/concepts/flash-loans}

\bibitem{cao2021flashot}
Y.~Cao, C.~Zou, X.~Cheng, Flashot: a snapshot of flash loan attack on DeFi ecosystem, arXiv preprint arXiv:2102.00626 (2021).

\bibitem{crawley2021}
J.~Crawley, \href{https://www.coindesk.com/markets/2021/05/20/flash-loan-attack-causes-defi-token-bunny-to-crash-over-95/}{Flash loan attack causes DeFi token bunny to crash over 95\%} ((2021)).
\newline\urlprefix\url{https://www.coindesk.com/markets/2021/05/20/flash-loan-attack-causes-defi-token-bunny-to-crash-over-95/}

\bibitem{reuters2024microstrategy}
Reuters, \href{https://www.reuters.com/business/finance/bitcoin-buyer-microstrategy-jumps-nasdaq-100-entry-2024-12-16/}{Bitcoin buyer MicroStrategy jumps on Nasdaq 100 entry} ((2024)).
\newline\urlprefix\url{https://www.reuters.com/business/finance/bitcoin-buyer-microstrategy-jumps-nasdaq-100-entry-2024-12-16/}

\bibitem{aufiero2025cryptocurrencies}
S.~Aufiero, A.~Briola, T.~Salarin, F.~Caccioli, S.~Bartolucci, T.~Aste, Cryptocurrencies in the balance sheet: Insights from (micro) strategy--bitcoin interactions, arXiv preprint arXiv:2505.14655 (2025).

\bibitem{bloomberg2022ftx}
Bloomberg~Opinion, \href{https://www.bloomberg.com/opinion/articles/2022-11-14/ftx-s-balance-sheet-was-bad}{FTX’s balance sheet was bad} ((2022)).
\newline\urlprefix\url{https://www.bloomberg.com/opinion/articles/2022-11-14/ftx-s-balance-sheet-was-bad}

\bibitem{borio2019money}
C.~Borio, On money, debt, trust, and central banking, Cato J. 39 (2019) 267.

\bibitem{ponciano2021}
J.~Ponciano, \href{https://www.forbes.com/sites/jonathanponciano/2021/09/07/crypto-flash-crash-wipes-out-400-billion-in-market-value-on-el-salvadors-bitcoin-day/}{Crypto flash crash wipes out \$400 billion in market value on bitcoin day as El Salvador’s president ‘buys the dip’}, Forbes (2021).
\newline\urlprefix\url{https://www.forbes.com/sites/jonathanponciano/2021/09/07/crypto-flash-crash-wipes-out-400-billion-in-market-value-on-el-salvadors-bitcoin-day/}

\bibitem{basel2021crypto}
{Basel Committee on Banking Supervision}, \href{https://www.bis.org/bcbs/publ/d519.pdf}{Prudential treatment of cryptoasset exposures}, Tech. rep., Bank for International Settlements (June 2021).
\newline\urlprefix\url{https://www.bis.org/bcbs/publ/d519.pdf}

\bibitem{nasdaq2022crypto}
Nasdaq, \href{https://www.nasdaq.com/articles/the-crypto-market-is-not-immune-to-contagion-risk}{The crypto market is not immune to contagion risk} (2022).
\newline\urlprefix\url{https://www.nasdaq.com/articles/the-crypto-market-is-not-immune-to-contagion-risk}

\bibitem{federalreserve2023silvergate}
Federal~Reserve~Office of~Inspector~General, \href{https://oig.federalreserve.gov/reports/board-review-silvergate-summary-sep2023.pdf}{Review of the board’s supervision of silvergate bank} (September 2023).
\newline\urlprefix\url{https://oig.federalreserve.gov/reports/board-review-silvergate-summary-sep2023.pdf}

\bibitem{statista2025largestexchanges}
Statista, \href{https://www.statista.com/statistics/270126/largest-stock-exchange-operators-by-market-capitalization-of-listed-companies/}{Largest stock exchange operators worldwide as of september 2024, by market capitalization of listed companies}.
\newline\urlprefix\url{https://www.statista.com/statistics/270126/largest-stock-exchange-operators-by-market-capitalization-of-listed-companies/}

\bibitem{economictimes2025cryptomarket}
T.~E. Times, \href{https://economictimes.indiatimes.com/markets/cryptocurrency/crypto-market-capitalisation-hits-record-3-2-trillion/articleshow/115286813.cms?from=mdr}{Crypto market capitalisation hits record \$3.2 trillion}.
\newline\urlprefix\url{https://economictimes.indiatimes.com/markets/cryptocurrency/crypto-market-capitalisation-hits-record-3-2-trillion/articleshow/115286813.cms?from=mdr}

\bibitem{iyer2022cryptic}
T.~Iyer, \href{https://www.imf.org/en/Publications/global-financial-stability-notes/Issues/2022/01/10/Cryptic-Connections-511776}{Cryptic connections: spillovers between crypto and equity markets}, International Monetary Fund, 2022.
\newline\urlprefix\url{https://www.imf.org/en/Publications/global-financial-stability-notes/Issues/2022/01/10/Cryptic-Connections-511776}

\bibitem{xie4769838dynamic}
W.~Xie, G.~Cao, \href{https://ssrn.com/abstract=4769838}{Dynamic spillover effects between cryptocurrencies and China's financial markets: New evidence from a Tvp-Var extended joint connectedness approach}, Available at SSRN 4769838 (2024).
\newline\urlprefix\url{https://ssrn.com/abstract=4769838}

\bibitem{zhang2021risk}
Y.-J. Zhang, E.~Bouri, R.~Gupta, S.-J. Ma, Risk spillover between Bitcoin and conventional financial markets: An expectile-based approach, The North American Journal of Economics and Finance 55 (2021) 101296.

\bibitem{wang2022dynamic}
P.~Wang, X.~Liu, S.~Wu, Dynamic linkage between Bitcoin and traditional financial assets: A comparative analysis of different time frequencies, Entropy 24~(11) (2022) 1565.

\bibitem{scagliarini2022pairwise}
T.~Scagliarini, G.~Pappalardo, A.~E. Biondo, A.~Pluchino, A.~Rapisarda, S.~Stramaglia, Pairwise and high-order dependencies in the cryptocurrency trading network, Scientific Reports 12~(1) (2022) 18483.

\bibitem{fund2022understanding}
A.~M. Fund, \href{https://www.amf.org.ae/sites/default/files/publications/2022-03/Understanding%20the%20Dynamic%20Correlation%20betwee.pdf}{Understanding the dynamic correlation between Bitcoin, gold, oil, and stock market indices in the selected Arab countries: Application of the DCC-GARCH model to the banking and insurance sectors}, Arab Monetary Fund Economic Studies Series (March 2022).
\newline\urlprefix\url{https://www.amf.org.ae/sites/default/files/publications/2022-03/Understanding%20the%20Dynamic%20Correlation%20betwee.pdf}

\bibitem{pineiro2022preliminary}
J.~Pi{\~n}eiro-Chousa, M.~{\'A}. L{\'o}pez-Cabarcos, A.~Sevic, I.~Gonz{\'a}lez-L{\'o}pez, A preliminary assessment of the performance of DeFi cryptocurrencies in relation to other financial assets, volatility, and user-generated content, Technological Forecasting and Social Change 181 (2022) 121740.

\bibitem{ugolini2023connectedness}
A.~Ugolini, J.~C. Reboredo, W.~Mensi, Connectedness between DeFi, cryptocurrency, stock, and safe-haven assets, Finance Research Letters 53 (2023) 103692.

\bibitem{pacelli2024cryptocurrencies}
V.~Pacelli, C.~Di~Tommaso, M.~Foglia, S.~Ingannamorte, Cryptocurrencies and systemic risk. the spillover effects between cryptocurrency and financial markets, in: Systemic Risk and Complex Networks in Modern Financial Systems, Springer Nature Switzerland Cham, 2024, pp. 343--358.

\bibitem{zhou2024cryptocurrency}
F.~Zhou, Cryptocurrency: A new player or a new crisis in financial markets? -- evolutionary analysis of association and risk spillover based on network science, Physica A: Statistical Mechanics and its Applications 648 (2024) 129955.

\bibitem{ecb2022stability}
European Central Bank, \href{https://www.ecb.europa.eu/press/financial-stability-publications/fsr/special/html/ecb.fsrart202205_02~1cc6b111b4.en.html}{Decrypting financial stability risks in crypto-asset markets} (2022).
\newline\urlprefix\url{https://www.ecb.europa.eu/press/financial-stability-publications/fsr/special/html/ecb.fsrart202205_02~1cc6b111b4.en.html}

\bibitem{reuters2024cbdc}
Reuters, \href{https://www.reuters.com/markets/currencies/central-bank-digital-currency-momentum-growing-study-shows-2024-09-17/}{Central bank digital currency momentum growing, study shows} (2024).
\newline\urlprefix\url{https://www.reuters.com/markets/currencies/central-bank-digital-currency-momentum-growing-study-shows-2024-09-17/}

\bibitem{atlanticcouncil2025cbdctracker}
Atlantic~Council, \href{https://www.atlanticcouncil.org/cbdctracker/}{Central bank digital currency (CBDC) tracker}.
\newline\urlprefix\url{https://www.atlanticcouncil.org/cbdctracker/}

\bibitem{krasia2025digitalyuan}
KrAsia, \href{https://kr-asia.com/china-explores-cross-border-uses-for-digital-yuan-in-new-trial}{China explores cross-border uses for digital yuan in new trial} (2024).
\newline\urlprefix\url{https://kr-asia.com/china-explores-cross-border-uses-for-digital-yuan-in-new-trial}

\bibitem{fnlondon2025cbdc}
Financial News London, \href{https://www.fnlondon.com/articles/bank-of-england-eyes-cbdc-for-banks-5a993b2d}{Bank of England eyes CBDC for banks} (2024).
\newline\urlprefix\url{https://www.fnlondon.com/articles/bank-of-england-eyes-cbdc-for-banks-5a993b2d}

\bibitem{ft2025cbdc}
Financial~Times, \href{https://www.ft.com/content/e50d0d40-a670-4ee7-b734-5f0ee3375aeb}{Even central banks are losing faith in CBDCs} (2024).
\newline\urlprefix\url{https://www.ft.com/content/e50d0d40-a670-4ee7-b734-5f0ee3375aeb}

\bibitem{frolov2024interaction}
S.~Frolov, M.~Ivasenko, M.~Dykha, I.~Shalyhina, V.~Hrabar, V.~Fenyves, Interaction between decentralized financial services and the traditional banking system: A comparative analysis, Banks and Bank Systems 19~(2) (2024) 53--74.

\bibitem{azar2024digitalassets}
P.~D. Azar, et~al., \href{https://www.newyorkfed.org/medialibrary/media/research/epr/2024/EPR_2024_digital-assets_azar.pdf}{Digital assets: Risk and regulation}, Economic policy review, Federal Reserve Bank of New York ((2024)).
\newline\urlprefix\url{https://www.newyorkfed.org/medialibrary/media/research/epr/2024/EPR_2024_digital-assets_azar.pdf}

\bibitem{bindseil2024distributional}
U.~Bindseil, J.~Schaaf, \href{https://ssrn.com/abstract=4985877}{The distributional consequences of Bitcoin}, Available at SSRN (2024).
\newline\urlprefix\url{https://ssrn.com/abstract=4985877}

\bibitem{ft2025article}
Financial~Times, \href{https://www.ft.com/content/af23fffc-e560-42eb-84a0-f25ca8d693c0}{Crypto industry dreams of a golden era under Trump} (2025).
\newline\urlprefix\url{https://www.ft.com/content/af23fffc-e560-42eb-84a0-f25ca8d693c0}

\bibitem{rizwan2020systemic}
M.~S. Rizwan, G.~Ahmad, D.~Ashraf, Systemic risk: The impact of Covid-19, Finance Research Letters 36 (2020) 101682.

\bibitem{duan2021bank}
Y.~Duan, S.~El~Ghoul, O.~Guedhami, H.~Li, X.~Li, Bank systemic risk around Covid-19: A cross-country analysis, Journal of Banking \& Finance 133 (2021) 106299.

\bibitem{lai2021study}
Y.~Lai, Y.~Hu, A study of systemic risk of global stock markets under Covid-19 based on complex financial networks, Physica A: Statistical Mechanics and its Applications 566 (2021) 125613.

\bibitem{caferra2021raised}
R.~Caferra, D.~Vidal-Tom{\'a}s, Who raised from the abyss? A comparison between cryptocurrency and stock market dynamics during the Covid-19 pandemic, Finance Research Letters 43 (2021) 101954.

\bibitem{corbet2020contagion}
S.~Corbet, C.~Larkin, B.~Lucey, The contagion effects of the Covid-19 pandemic: Evidence from gold and cryptocurrencies, Finance Research Letters 35 (2020) 101554.

\bibitem{umar2020time}
Z.~Umar, M.~Gubareva, A time--frequency analysis of the impact of the Covid-19 induced panic on the volatility of currency and cryptocurrency markets, Journal of Behavioral and Experimental Finance 28 (2020) 100404.

\bibitem{umar2021impact}
Z.~Umar, F.~Jare{\~n}o, M.~de~la O~Gonz{\'a}lez, The impact of Covid-19-related media coverage on the return and volatility connectedness of cryptocurrencies and fiat currencies, Technological Forecasting and Social Change 172 (2021) 121025.

\bibitem{yousaf2023connectedness}
I.~Yousaf, F.~Jare{\~n}o, M.~Tolentino, Connectedness between DeFi assets and equity markets during Covid-19: A sector analysis, Technological Forecasting and Social Change 187 (2023) 122174.

\bibitem{akhtaruzzaman2022systemic}
M.~Akhtaruzzaman, S.~Boubaker, D.~K. Nguyen, M.~R. Rahman, Systemic risk-sharing framework of cryptocurrencies in the Covid--19 crisis, Finance Research Letters 47 (2022) 102787.

\bibitem{elsayed2022risk}
A.~H. Elsayed, G.~Gozgor, C.~K.~M. Lau, Risk transmissions between Bitcoin and traditional financial assets during the Covid-19 era: The role of global uncertainties, International Review of Financial Analysis 81 (2022) 102069.

\bibitem{yousaf2022linkages}
I.~Yousaf, R.~Nekhili, M.~Gubareva, Linkages between defi assets and conventional currencies: Evidence from the Covid-19 pandemic, International Review of Financial Analysis 81 (2022) 102082.

\bibitem{bloomberg2025cryptotaskforce}
Bloomberg~News, \href{https://www.bloomberg.com/news/articles/2025-01-21/sec-launches-crypto-task-force-led-by-crypto-mom-hester-peirce}{Sec launches crypto task force led by `crypto mom' Hester Peirce} (2025).
\newline\urlprefix\url{https://www.bloomberg.com/news/articles/2025-01-21/sec-launches-crypto-task-force-led-by-crypto-mom-hester-peirce}

\bibitem{hsbc2023acquisition}
HSBC, \href{https://www.hsbc.com/news-and-views/news/media-releases/2023/hsbc-acquires-silicon-valley-bank-uk-limited}{Hsbc acquires Silicon Valley bank UK limited} (2023).
\newline\urlprefix\url{https://www.hsbc.com/news-and-views/news/media-releases/2023/hsbc-acquires-silicon-valley-bank-uk-limited}

\bibitem{dowd1992experience}
K.~Dowd, et~al., \href{https://www.jstor.org/stable/1060588}{The experience of free banking}, Routledge London, 1992.
\newline\urlprefix\url{https://www.jstor.org/stable/1060588}

\bibitem{selgin1987evolution}
G.~A. Selgin, L.~H. White, The evolution of a free banking system, Economic Inquiry 25~(3) (1987) 439--457.

\bibitem{eichengreen2019commodity}
B.~Eichengreen, \href{https://www.nber.org/papers/w25426}{From commodity to fiat and now to crypto: what does history tell us?}, Tech. rep., National Bureau of Economic Research (2019).
\newline\urlprefix\url{https://www.nber.org/papers/w25426}

\end{thebibliography}
\end{document}